\documentclass[aps,pra,10pt,twocolumn,showpacs,superscriptaddress,amsmath,amssymb,nofootinbib]{revtex4-1}
\usepackage[utf8]{inputenc}
\usepackage{natbib}
\usepackage{graphicx}         
\usepackage{dcolumn}          
\usepackage{bm}               
\usepackage{upgreek}
\usepackage{booktabs} 				
\usepackage{tabularx}
\usepackage{array}
\usepackage[colorlinks=true,urlcolor=blue,breaklinks=true]{hyperref}
\usepackage[hang,flushmargin,symbol*]{footmisc}
\usepackage{color}
\usepackage{multirow}
\usepackage{microtype}
\usepackage[version=3]{mhchem}
\usepackage[dvipsnames]{xcolor}
\usepackage{extdash}
\usepackage{float}
\usepackage{bibunits}
\DefineFNsymbolsTM{otherfnsymbols}{
  \textdagger    \dagger
}

\begin{document}

\begin{bibunit}[unsrt]

\title{Shadow-wall lithography of ballistic superconductor--semiconductor quantum devices}
\footnotetext{These authors contributed equally.}
\setfnsymbol{otherfnsymbols}

\author{Sebastian~Heedt$^*$}
\email{Sebastian.Heedt@Microsoft.com}
\affiliation{QuTech and Kavli Institute of Nanoscience, Delft University of Technology, 2600 GA Delft, The Netherlands}
\affiliation{Microsoft Quantum Lab Delft, 2600 GA Delft, The Netherlands}

\author{Marina~Quintero-P\'erez$^*$}
\affiliation{Microsoft Quantum Lab Delft, 2600 GA Delft, The Netherlands}

\author{Francesco~Borsoi$^*$}
\affiliation{QuTech and Kavli Institute of Nanoscience, Delft University of Technology, 2600 GA Delft, The Netherlands}

\author{Alexandra~Fursina}
\affiliation{Microsoft Quantum Lab Delft, 2600 GA Delft, The Netherlands}

\author{Nick~van Loo}
\affiliation{QuTech and Kavli Institute of Nanoscience, Delft University of Technology, 2600 GA Delft, The Netherlands}

\author{Grzegorz~P.~Mazur}
\affiliation{QuTech and Kavli Institute of Nanoscience, Delft University of Technology, 2600 GA Delft, The Netherlands}

\author{Micha\l{}~P.~Nowak}
\affiliation{AGH University of Science and Technology, Academic Centre for Materials and Nanotechnology, al.\ A.\ Mickiewicza 30, 30-059 Krak\'ow, Poland}

\author{Mark~Ammerlaan}
\affiliation{QuTech and Kavli Institute of Nanoscience, Delft University of Technology, 2600 GA Delft, The Netherlands}

\author{Kongyi~Li}
\affiliation{QuTech and Kavli Institute of Nanoscience, Delft University of Technology, 2600 GA Delft, The Netherlands}

\author{Svetlana~Korneychuk}
\affiliation{QuTech and Kavli Institute of Nanoscience, Delft University of Technology, 2600 GA Delft, The Netherlands}

\author{Jie~Shen}
\affiliation{QuTech and Kavli Institute of Nanoscience, Delft University of Technology, 2600 GA Delft, The Netherlands}

\author{May~An~Y.~van de Poll}
\affiliation{QuTech and Kavli Institute of Nanoscience, Delft University of Technology, 2600 GA Delft, The Netherlands}

\author{Ghada~Badawy}
\affiliation{Department of Applied Physics, Eindhoven University of Technology, 5600 MB Eindhoven, The Netherlands}

\author{Sasa~Gazibegovic}
\affiliation{Department of Applied Physics, Eindhoven University of Technology, 5600 MB Eindhoven, The Netherlands}

\author{Kevin~van Hoogdalem}
\affiliation{Microsoft Quantum Lab Delft, 2600 GA Delft, The Netherlands}

\author{Erik~P.~A.~M.~Bakkers}
\affiliation{Department of Applied Physics, Eindhoven University of Technology, 5600 MB Eindhoven, The Netherlands}

\author{Leo~P.~Kouwenhoven}
\affiliation{QuTech and Kavli Institute of Nanoscience, Delft University of Technology, 2600 GA Delft, The Netherlands}
\affiliation{Microsoft Quantum Lab Delft, 2600 GA Delft, The Netherlands}

\hyphenation{InSb na-no-wire u-sing con-si-der-ing mo-ni-tored na-no-struc-tures Jo-seph-son}

\date{\today}

\begin{abstract}
\noindent The realization of a topological qubit calls for advanced techniques to readily and reproducibly engineer induced superconductivity in semiconductor nanowires. Here, we introduce an on-chip fabrication paradigm based on shadow walls that offers substantial advances in device quality and reproducibility. It allows for the implementation of novel quantum devices and ultimately topological qubits while eliminating many fabrication steps such as lithography and etching. This is critical to preserve the integrity and homogeneity of the fragile hybrid interfaces. The approach simplifies the reproducible fabrication of devices with a hard induced superconducting gap and ballistic normal\Hyphdash*/superconductor junctions. Large gate-tunable supercurrents and high-order multiple Andreev reflections manifest the exceptional coherence of the resulting nanowire Josephson junctions. Our approach enables, in particular, the realization of 3-terminal devices, where zero-bias conductance peaks emerge in a magnetic field concurrently at both boundaries of the one-dimensional hybrids.
\end{abstract}

\maketitle

\noindent Hybrid superconducting/semiconducting nanowires are a promising material platform for the formation of one-dimensional topological superconductors bounded by pairs of Majorana modes~\cite{Oreg2010, Lutchyn2010, Lutchyn2018}. Owing to their non-Abelian exchange statistics, these localized Majorana bound states (MBS) are the fundamental constituents for fault-tolerant topological quantum computing~\cite{Kitaev2001, Nayak2008}. Individual qubits comprise at least four MBS in several interconnected nanowire segments with a hard induced superconducting gap~\cite{Plugge2017, Karzig2017}. Residual fermionic states within the gap would compromise the topological protection of the Majorana modes. Hence, a fundamental challenge in the development of topological qubits is the engineering of complex, interconnected hybrid devices with hard superconducting gaps and clean, homogenous interfaces~\cite{Takei2013, Gul2017}.\\
Here, we introduce a novel fabrication technique that overcomes these challenges and provides high-quality hybrid quantum devices involving minimal nanofabrication steps compared with previously established methods~\cite{Krogstrup2015, Gazibegovic2017}. Our approach is based on the deposition of superconducting thin films at a shallow angle onto semiconducting nanowires, which have been selectively placed on substrates with pre-patterned gates and shadow-wall structures. It enables complex hybrid devices with individual (i.e.\ mutually isolated) contacts to the nanowire segments while eliminating lithography, etching, and other fabrication steps after the deposition of the superconductor, in the following referred to as \textit{post-interface} fabrication. While shadow-wall lithography is compatible with a large variety of materials, we utilize InSb nanowires coated with Al half-shells to induce superconducting correlations -- a suitable material combination to study Majorana physics~\cite{Gazibegovic2017, deMoor2018}. The homogeneity of the interface between InSb and Al ultimately determines the device quality, but it is known to have very limited chemical and thermal stability~\cite{Gul2017}. Therefore, the reduction or elimination of post-interface fabrication steps represents a paradigm shift that enables pristine hybrid interfaces. Similar advances in quality and reproducibility were made possible by the reverse fabrication process established for carbon nanotube devices~\cite{Cao2005}.
\begin{figure*}[htb!]
\includegraphics[width=\linewidth]{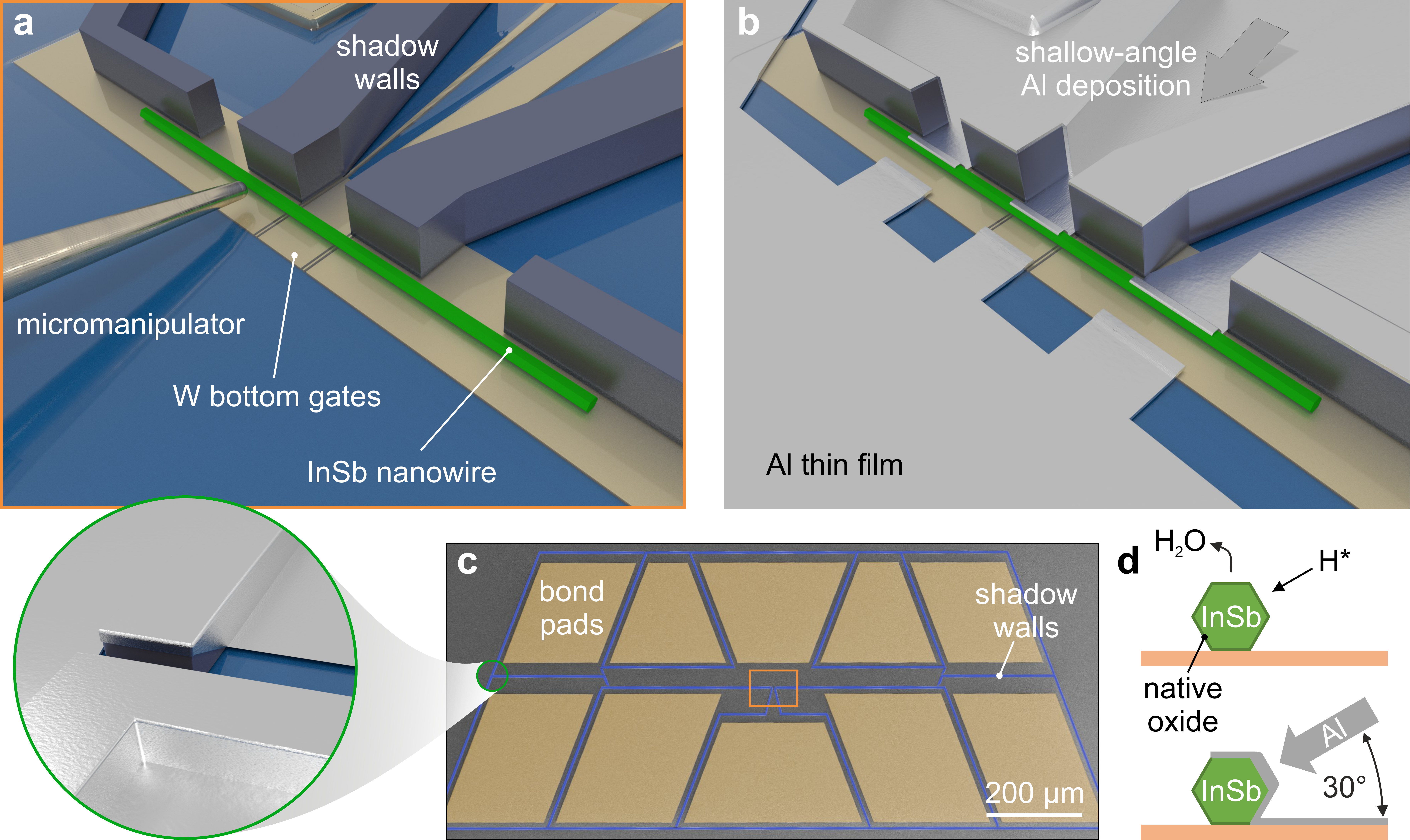}
\caption{\textbf{Illustration of the shadow-wall technique.} \textbf{a}~Micromechanical transfer of the nanowires onto local bottom gates (covered by Al$_2$O$_3$ dielectric) in the proximity of the Si$_3$N$_4$ shadow walls. \textbf{b}~Illustration of a final device following the H radical cleaning and Al deposition at a shallow angle. \textbf{c}~False-colour SEM image of an exemplary sample. Shadow walls are designated in blue and bond pads, which are enclosed by shadow walls, are shaded in dark yellow. Gaps are placed at critical locations along the shadow walls (cf.\ green circle and the illustration in the blow-up). This ensures that bond pads with leads are isolated from each other after the Al deposition. The area illustrated in panel~(\textbf{a}) is indicated by the orange box. \textbf{d}~Schematic of the InSb nanowire cross-section during H radical cleaning (top). The native oxide of the semiconductor is denoted by a dark green layer. The Al thin film deposited at a shallow angle of $30^{\circ}$ forms an electrical connection to the substrate (bottom).}
\label{fig:Figure 1}
\end{figure*}~\\
In this Article, we investigate the transport properties of hybrid nanowire shadow-wall devices. Initially, we examine Josephson junctions and detect subharmonic gap features that arise from multiple Andreev reflections~\cite{Octavio1983}. These junctions exhibit gate-tunable supercurrents of up to $90\,$nA, which is exceptionally large for InSb/Al nanowires~\cite{Nilsson2012, Li2016, Gul2017}. The shadow-wall method also facilitates 3-terminal hybrid devices with two normal-metal--superconductor (N--S) interfaces, which are crucial to corroborate earlier Majorana signatures~\cite{Mourik2012, Rosdahl2018}. We investigate the transport at a single N--S interface and observe a crossover between a hard induced gap and pronounced Andreev enhancement upon increasing the junction transparency, consistent with the expected behaviour for ballistic junctions~\cite{Blonder1982, Beenakker1992}. Finally, we report the emergence of discrete subgap states in the tunnelling conductance at both nanowire ends and detect stable zero-energy conductance peaks that coexist at certain magnetic fields and chemical potentials.\\
Our fabrication method paves the way for more advanced nanowire devices, including qubit implementations~\cite{Vijay2016, Plugge2017, Karzig2017} and other multi-terminal devices that are essential for fundamental research on topological superconductors~\cite{Beri2012, Rosdahl2018}. The versatility of the shadow-wall technique introduces a convenient and quick way to implement new device geometries with various combinations of semiconductor and superconductor materials.

\section*{Shadow-Wall Lithography}
\noindent A well-established approach to realize hybrid devices is based on the epitaxial growth of nanowires followed by the in-situ evaporation of a superconductor~\cite{Krogstrup2015, Bjergfelt2019}. This method requires a subsequent etching step to expose gate-tunable wire segments without metal. Nanowires have also been grown on opposite crystal facets of etched trenches~\cite{Rieger2016, Gazibegovic2017}, which enables the formation of shadowed junctions without the need to etch the superconductor~\cite{Gazibegovic2017}. The native oxide that forms during the ex-situ processing is removed prior to the deposition of the superconductor. Another recent study employed growth chips with bridges and trenches that act as selectively shadowing objects during the evaporation of a superconductor~\cite{Carrad2020}. However, common to those methods is that the hybrid nanowires are removed from the growth substrate following the evaporation and undergo several post-interface fabrication steps such as alignment via scanning electron microscopy (SEM), electron-beam lithography involving resist coating, or etching. However, hybrid devices are prone to degradation. High-temperature processing (e.g.\ certain dielectric deposition methods or resist baking) cannot be performed, as it would lead to chemical intermixing at the super-/semiconductor interface~\cite{Boscherini1987, Thomas2019}. The low chemical and thermal stability of the interface requires sample storage in vacuum at a temperature $T<0\,^{\circ}$C, which is hardly compatible with standard fabrication methods. The low thermal budget and the additional fabrication steps limit the achievable device performance in terms of electrical noise, lithographical alignment accuracy, contamination and disorder. The considerable variation from device to device imposes singular rather than standardized designs and results in a low yield and limited reproducibility of basic transport measurements.\\
In contrast, the core principle of our approach is to minimize or eliminate post-interface fabrication. We have engineered scalable substrates that comprise all desired functionality without being subject to any fabrication restrictions (e.g.\ thermal budget limitations) since the semiconductor nanowires are only introduced right before the superconductor deposition. As depicted in Fig.~\ref{fig:Figure 1}a, we transfer InSb nanowires~\cite{Badawy2019} onto these substrates on top of pre-patterned bottom gates covered by a continuous dielectric layer in the vicinity of shadow-wall structures. The nanowires are loaded into a customized evaporation chamber where the native oxide is removed at $T = 550\,$K by exposure to a directed flow of atomic hydrogen radicals. Without breaking the vacuum, Al is subsequently deposited onto the samples at $T = 140\,$K. The superconductor is evaporated at a shallow angle of $30^{\circ}$ with respect to the substrate plane, which creates a $3$-facet nanowire shell that is connected to the leads and bond pads on the substrate (Fig.~\ref{fig:Figure 1}d). As illustrated in Fig.~\ref{fig:Figure 1}b, the shadow walls enable selective deposition on both the nanowires and the substrate. Adding gaps at critical locations along the shadow walls (Fig.~\ref{fig:Figure 1}c) ensures that the leads are electrically isolated from one another while eliminating the need for post-interface fabrication such as lift-off patterning or Al etching. Fig.~\ref{fig:Figure 2}a shows an exemplary device that is directly bonded to a printed circuit board for low-temperature transport measurements. Here, the p$^+$-doped Si substrate enables back-gate control of the electron density in the nanowire (see Fig.~\ref{fig:Figure 2}b).

\section*{Material Analysis}
\begin{figure}[hbt!]
\includegraphics[width=\linewidth]{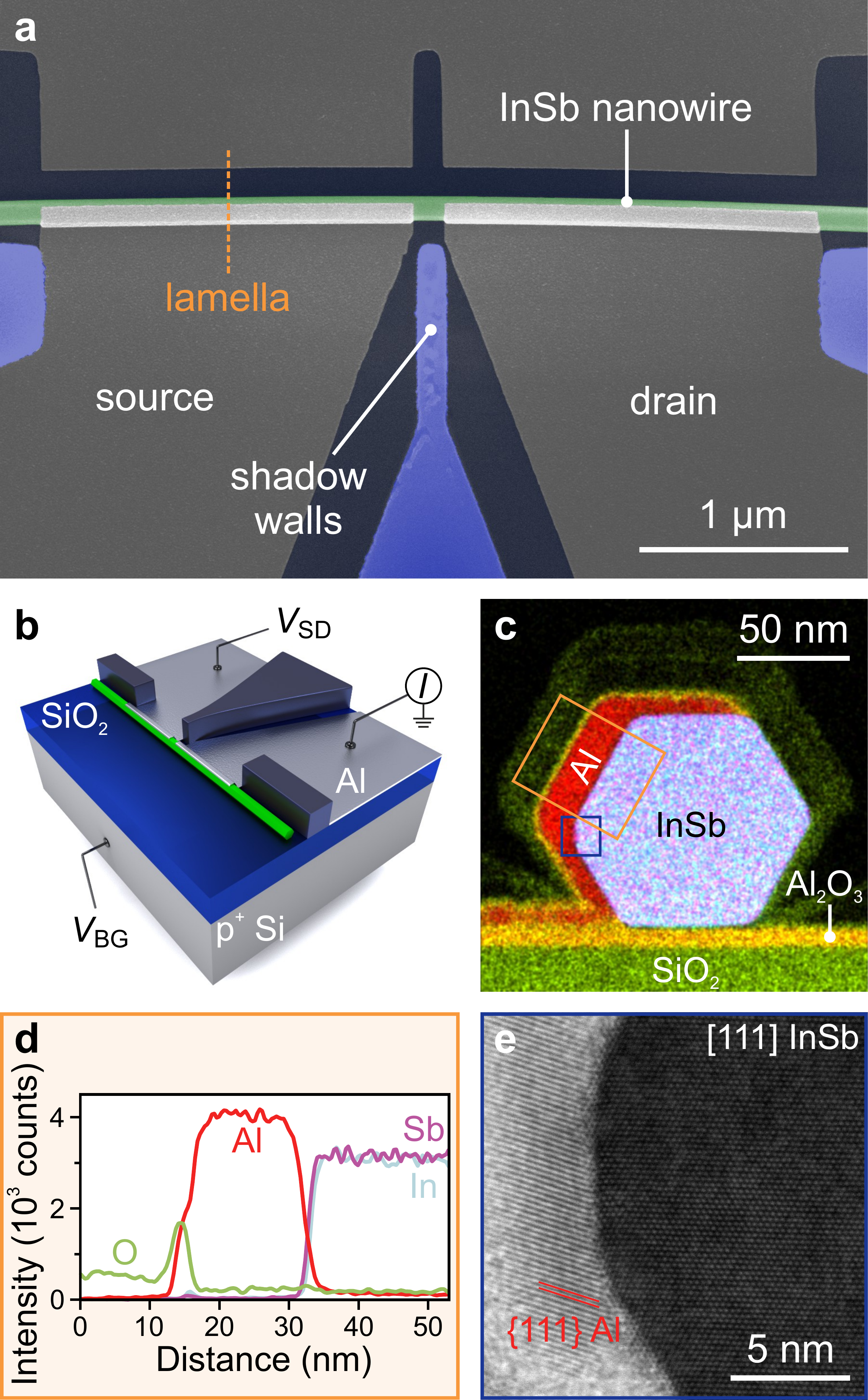}
\caption{\textbf{TEM analysis of the InSb/Al interface.} \textbf{a}~False-colour SEM image of an InSb nanowire Josephson junction. \textbf{b}~Schematic of the measurement setup. The back-gate voltage, $V_{\mathrm{BG}}$, is applied to the p$^+$-doped Si substrate to tune the electron density in the nanowire. \textbf{c}~Cross-sectional EDX elemental composite image of the $\left[111\right]$ InSb nanowire covered with the Al layer and a protective layer of SiN$_x$. \textbf{d}~Line-cuts of the integrated elemental counts within the orange box in panel~(\textbf{c}). \textbf{e}~High-resolution bright-field scanning TEM image of the InSb/Al interface at the location indicated by the blue box in panel~(\textbf{c}).}
\label{fig:Figure 2}
\end{figure}

\noindent The quality of the InSb nanowires, Al thin films, and InSb/Al interfaces is assessed by transmission electron microscopy (TEM) of cross-sectional lamellae prepared via focused ion beam (FIB). These lamellae are cut out from devices like the one depicted in Fig.~\ref{fig:Figure 2}a (cf.\ dashed line). The samples exhibit a sharp super\Hyphdash*/semiconductor interface and a continuous high-quality polycrystalline Al layer is formed on three facets of the InSb nanowires (see Figs.~\ref{fig:Figure 2}c,e and Supplementary Fig.~S1). No oxide formation is observed between the Al grains, which is evident in the elemental energy-dispersive X-ray spectroscopy (EDX) composite image (Fig.~\ref{fig:Figure 2}c). The middle facet has twice the Al layer thickness ($16\,$nm) compared to the top and bottom facets ($8\,$nm) due to the evaporation angle of $30^{\circ}$ with respect to the substrate plane. The InSb/Al interface is clean and there is no residual native oxide (see Figs.~\ref{fig:Figure 2}d,e), which confirms that our procedure of atomic hydrogen radical cleaning can effectively remove the nanowire oxide without damaging the InSb crystal structure. The nanowires are single-crystalline, defect-free, and exhibit a hexagonal geometry. The polycrystalline Al layer forms a continuous metallic connection from the nanowire to the substrate. This connection is crucial for the contact between the shell and the thin Al lead on the substrate and it is fundamental for more complex devices such as superconducting interferometers (see Supplementary Fig.~S25) and 3-terminal Majorana devices that can reveal the opening of a topological gap~\cite{Rosdahl2018}.

\section*{Highly Transparent Josephson Junctions}
\begin{figure}[hbt!]
\includegraphics{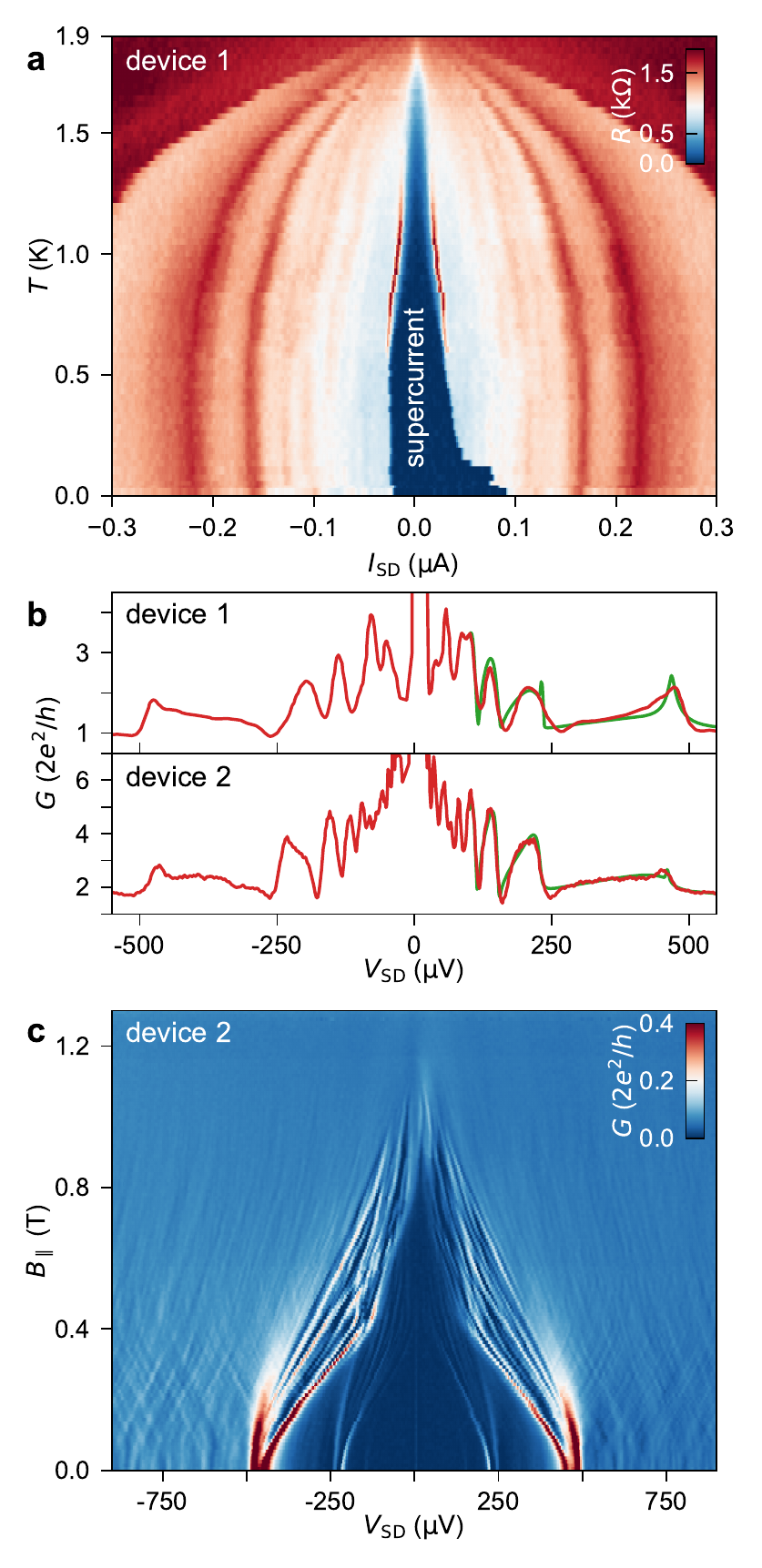} 
\caption{\textbf{Multiple Andreev reflections and supercurrent in InSb/Al Josephson junctions}. \textbf{a}~Differential resistance, $R$, as a function of $I_{\mathrm{SD}}$ (with upward sweep direction) and $T$ for device~1 at $V_{\mathrm{BG}} = 13.65\,$V. The switching current reaches a maximum of $\sim 90\,$nA at $T = 30\,$mK and persists up to $1.8\,$K. \textbf{b}~Conductance line traces (red) versus source--drain voltage for device~1 at $V_{\mathrm{BG}} = 5.1\,$V (top) and for device~2 at $V_{\mathrm{BG}} = 3.0\,$V (bottom). The theoretical fits (green) yield the transmissions, $T_n$, of the one-dimensional subbands with index $n$: $T_1 = 0.87$, $T_2 = 0.17$ (top) and $T_1 = 0.93$, $T_2 = 0.71$, $T_3 = 0.01$ (bottom). \textbf{c}~Differential conductance, $G$, as a function of $V_{\mathrm{SD}}$ and magnetic field, $B_{\parallel}$, which is oriented along the nanowire, for device~2 at $V_{\mathrm{BG}} = -0.9\,$V.}
\label{fig:Figure 3}
\end{figure}

\noindent We employ mesoscopic InSb/Al Josephson junctions like the one depicted in Fig.~\ref{fig:Figure 2}a to study the induced superconductivity in the nanowires. Each device comprises two Al contacts ($1.8\,\upmu$m wide) separated by a $110-150\,$nm long bare nanowire segment that is tunable by the back-gate voltage, $V_{\mathrm{BG}}$. The source--drain voltage, $V_{\mathrm{SD}}$, is applied or measured between the two Al electrodes (Fig.~\ref{fig:Figure 2}b). Fig.~\ref{fig:Figure 3}a shows the differential resistance, $R = \mathrm{d}V_{\mathrm{SD}}/\mathrm{d}I_{\mathrm{SD}}$, as a function of bias current, $I_{\mathrm{SD}}$, and temperature for a typical device. The blue region ($R = 0\,\Omega$) denotes the superconducting phase, which persists up to $\sim 1.8\,$K. At low temperatures ($T < 0.6\,$K), the hysteretic behaviour of the asymmetric $V_{\mathrm{SD}}$--$I_{\mathrm{SD}}$ traces indicates that the junction is in the underdamped regime according to the model of resistively and capacitively shunted junctions. Above $0.6\,$K, the thermal activation washes out the asymmetry of the traces. Remarkably, at $T = 30\,$mK the switching current, $I_{\mathrm{sw}}$, i.e.\ the observable supercurrent, ranges from $30$ to $90\,$nA across all devices in the open-channel regime. The magnitude of the intrinsic supercurrent, $I_{\mathrm{c}}$, in ballistic and short junctions, can be predicted via the Ambegaokar--Baratoff formula: $I_{\mathrm{c}} R_{\mathrm{N}} = \pi \Delta_{\mathrm{ind}}/2e$, with the normal-state resistance $R_{\mathrm{N}}$, the induced gap $\Delta_{\mathrm{ind}}$, and the electron charge $e$~\cite{Ambegaokar1963}. Here, the typical $I_{\mathrm{sw}} R_{\mathrm{N}}$ product is $\sim 110\,\upmu$V, i.e.\ only one-third of $\pi \Delta_{\mathrm{ind}}/2e \sim 360\,\upmu$V. The discrepancy between $I_{\mathrm{sw}}$ and $I_{\mathrm{c}}$ is in line with previous experiments~\cite{Doh2005, Nilsson2012, Li2016} and can be explained by premature switchings due to thermal activation and current fluctuations~\cite{Fulton1974, Tinkham1996}.
~\\
In Fig.~\ref{fig:Figure 3}b, we show the differential conductance, $G=\mathrm{d}I_{\mathrm{SD}}/\mathrm{d}V_{\mathrm{SD}}$, as a function of $V_{\mathrm{SD}}$ (red curves) for the same Josephson junction (top) and for a second device (bottom). The traces display subharmonic conductance peaks stemming from multiple Andreev reflection (MAR) processes~\cite{Octavio1983}. By fitting the conductance with a coherent scattering model (green curves), we can estimate the induced superconducting gap, $\Delta_{\mathrm{ind}}$ ($233\,\upmu$eV and $230\,\upmu$eV for device~1 and 2, respectively), and the gate-tunable tunnelling probability of the different subbands (see Supplementary Fig.~S6)~\cite{Scheer1997}.\\
In Fig.~\ref{fig:Figure 3}c, we report the evolution of the MAR pattern as a function of magnetic field, $B_{\parallel}$, parallel to the nanowire axis of device~2. Here, the presence of subgap states close to the gap edge alters the typical MAR pattern and gives rise to an intricate energy dispersion in magnetic field that is further discussed in Supplementary Section~III. Crucially, the magnetic field quenches the superconductivity at a critical value of $B_{\mathrm{c}} = 1.2-1.3\,$T. This limit can be enhanced to about $2\,$T by using a thinner Al shell (Supplementary Fig.~S11). These values are well above the magnetic field at which a topological phase transition should occur in hybrid InSb/Al nanowires~\cite{Nijholt2016}.
\begin{figure*}[htb]
\includegraphics[width=\linewidth]{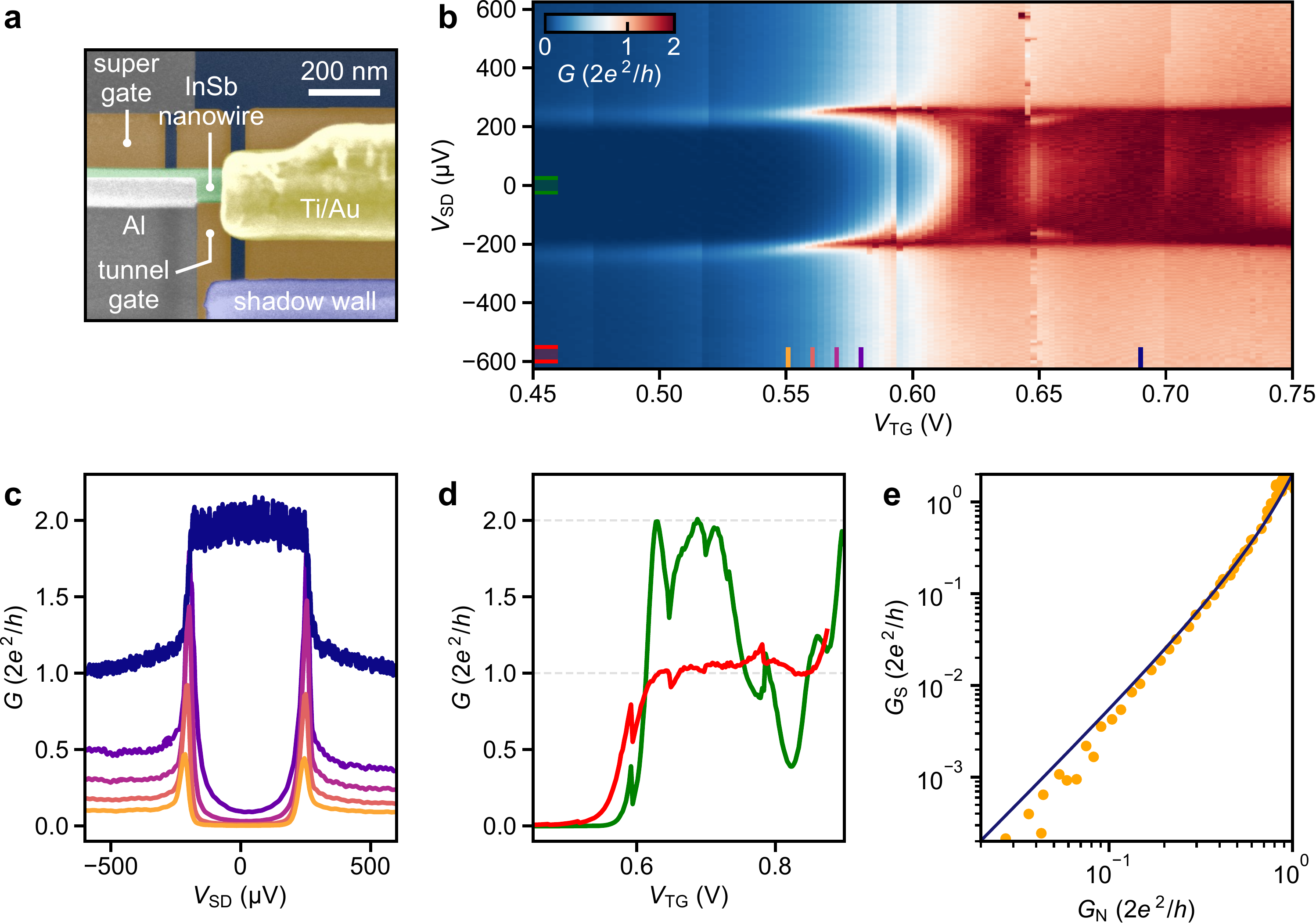}
\caption{\textbf{Ballistic Andreev transport.} \textbf{a}~False-colour SEM image of an exemplary N--S junction. The W bottom gates (brown) underneath the InSb nanowire (green) are covered by $18\,$nm of Al$_2$O$_3$ dielectric. \textbf{b}~Differential conductance, $G$, as a function of source--drain voltage, $V_{\mathrm{SD}}$, and bottom tunnel-gate voltage, $V_{\mathrm{TG}}$. The so-called \textit{super gate} which controls the chemical potential of the hybrid nanowire segment is grounded. \textbf{c}~$G$ versus $V_{\mathrm{SD}}$ line-cuts of the data in panel~(\textbf{a}) at the locations designated by the coloured lines. \textbf{d}~Subgap conductance (green) and above-gap conductance (red) averaged over the $V_{\mathrm{SD}}$ intervals designated in panel~(\textbf{b}). \textbf{e}~$G_{\mathrm{S}}$ (subgap conductance at zero bias) as a function of $G_{\mathrm{N}}$ (normal-state conductance at $V_{\mathrm{SD}}=650\,\upmu$eV) together with the theoretically predicted dependence which assumes Andreev-dominated transport in a single channel (blue line trace).}
\label{fig:Figure 4}
\end{figure*}~\\
In Fig.~\ref{fig:Figure 3}c, the out-of-gap conductance displays a dense pattern of faint peaks with an average spacing of about $30\,\upmu$V and an effective Land\'e $g$ factor of $\sim 20$ (extracted from the energy dispersion in magnetic field). This $g$ factor is larger than in Al ($|g| = 2$) but smaller than in InSb ($|g| = 30-50$), which indicates that these peaks stem from discrete states of the nanowire hybridized with the ones in the metal~\cite{Antipov2018}. The observation of this structure is a novelty and it might be correlated with our choice of nanowire surface treatment. In fact, the gentle atomic hydrogen cleaning preserves the pristine semiconductor crystal quality, unlike the invasive chemical or physical etching methods adopted in previous works~\cite{Doh2005, Nilsson2012, Abay2012, Li2016, Gul2017}.

\section*{Hard Induced Gap and Ballistic Superconductivity}
\noindent A common technique to detect Majorana bound states is tunnelling spectroscopy. Signatures of MBS in InSb-based N--S junctions are zero-bias peaks (ZBPs) in the differential conductance at moderately large magnetic fields~\cite{Mourik2012}. For Majorana zero modes, the ZBP height in the zero-temperature limit is predicted to be $G_0=2e^2/h$ due to resonant Andreev reflection, independent of the tunnel-coupling strength~\cite{Law2009}. Non-topological ZBPs may arise from disorder, which can mimic the subgap behaviour of MBS. A major challenge is to reduce the detrimental role of disorder at the semiconductor--superconductor interface, which determines the final device quality. The measure of success is a hard induced gap at a finite magnetic field and quantized Andreev transport as a signature of ballistic transport.\\
An exemplary N--S device is depicted in Fig.~\ref{fig:Figure 4}a. Here, the N contact to the InSb nanowire was formed in an optional post-interface fabrication step. In  Fig.~\ref{fig:Figure 4}b, we present voltage-bias spectroscopy at such an N--S junction where the transmission is tunable via a pre-fabricated bottom tunnel gate. The line-cuts in Fig.~\ref{fig:Figure 4}c at low tunnel-gate voltage, $V_{\mathrm{TG}}$, highlight the pronounced suppression of the subgap conductance, $G_{\mathrm{S}}$, by about two orders of magnitude compared with the normal-state conductance, $G_{\mathrm{N}}$ (cf.\ Supplementary Fig.~S18). As the first one-dimensional subband starts to conduct fully at $V_{\mathrm{TG}}>0.6\,$V, the above-gap conductance reaches the conductance quantum, $2e^2/h$, and the quantization manifests itself as a plateau in the tunnel-gate dependence (Fig.~\ref{fig:Figure 4}d). At the same time, the conductance below the gap edge reaches $4e^2/h$ owing to two-particle transport via Andreev reflection~\cite{Blonder1982}. This pronounced doubling of the normal-state conductance together with the quantization of $G_{\mathrm{N}}$ signifies a very low disorder strength in the junction and a strong coupling at the nanowire/Al interface~\cite{Zhang2017}. While the subgap conductance reaches up to $2G_0$, it drops again before the chemical potential reaches the bottom of the second confinement subband due to inter-subband scattering~\cite{Heedt2016,Zhang2017}. The plot of $G_{\mathrm{S}}$ versus $G_{\mathrm{N}}$ (Fig.~\ref{fig:Figure 4}e) follows the Beenakker model~\cite{Beenakker1992} without fitting parameter reasonably well, demonstrating that in the single-subband regime electrical transport below the gap edge is dominated by Andreev processes. The data are well-described by the BTK theory~\cite{Blonder1982} across the entire gate voltage range, demonstrating a hard induced gap (see Methods and Supplementary Fig.~S20). We determine an induced gap size of $\Delta_{\mathrm{ind}} \sim 230\,\upmu$eV using this model. Discrete subgap states and ZBPs appear at a finite magnetic field and field-dependent voltage-bias spectroscopy for this N--S device is presented in Supplementary Fig.~S21.

\begin{figure*}[htb]
\includegraphics[width=1.0\linewidth]{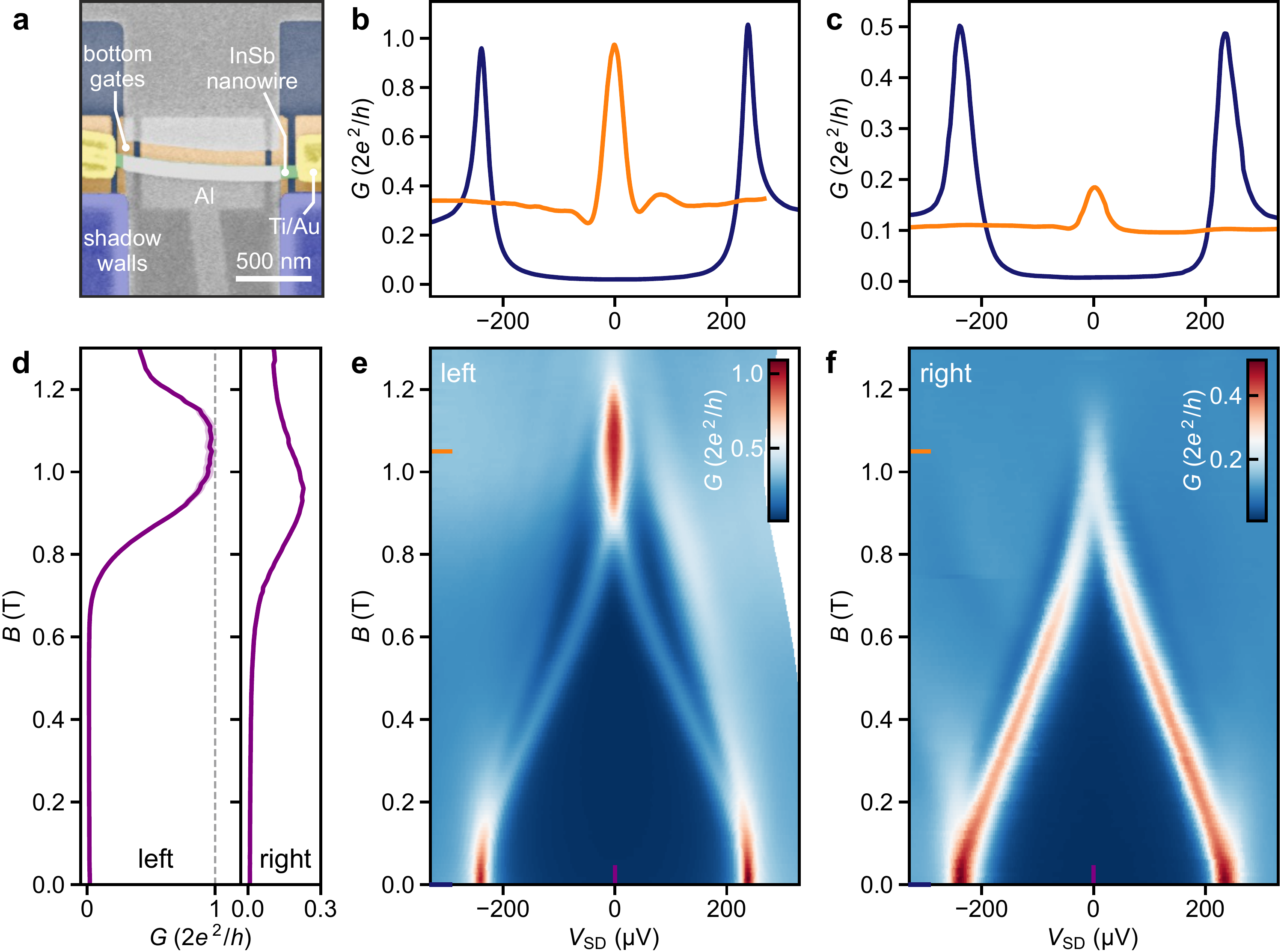}
\caption{\textbf{Zero-bias conductance peaks at two opposite N--S boundaries.} \textbf{a}~False-colour SEM image of the correlation device with a $1\,\upmu$m long hybrid nanowire segment. \textbf{b}, \textbf{c}~Line-cuts of the differential conductance at zero field (blue) and at $B=1.05\,$T (orange) taken from panel~(\textbf{e}) and (\textbf{f}), respectively. \textbf{d}~$G$ versus $B$ line-cut at $V_{\mathrm{SD}}=0\,\upmu$V taken from panel~(\textbf{e}) (left) and (\textbf{f}) (right). Shaded areas (light purple) illustrate the variation in conductance assuming an uncertainty of $\pm0.5\,$k$\Omega$ in the series resistance. For the line-cut at the right N--S junction, this variation is less than the line width. \textbf{e}, \textbf{f}~Differential conductance, $G=\mathrm{d}I_{\mathrm{SD}}/\mathrm{d}V_{\mathrm{SD}}$, as a function of bias voltage, $V_{\mathrm{SD}}$, and magnetic field, $B$, measured concurrently at the left and right junction, respectively. Here, the super gate underneath the hybrid nanowire segment is grounded.}
\label{fig:Figure 5}
\end{figure*}

\section*{Emergence of Correlated Zero-Bias Peaks}
\noindent The shadow-wall technique enables novel 3-terminal Majorana devices for nonlocal correlation experiments~\cite{Rosdahl2018, Lai2019} by harnessing the continuous connection of the Al shell to the substrate, as depicted in Fig.~\ref{fig:Figure 5}a. Here, the Al thin film serves as the superconducting drain lead. Established fabrication methods do not allow for the implementation of such devices since etching of Al causes disorder at the InSb surface and contacting the Al shell requires selective removal of the native oxide of Al, which affects the integrity of the thin film. As shown in Fig.~\ref{fig:Figure 5}a, optional Ti/Au contacts are again added at both sides of the hybrid segment. With this device type, we can study the simultaneous emergence of ZBPs at both N--S boundaries in a magnetic field oriented along the wire. Here, the hybrid nanowire segment is $1\,\upmu$m long and the chemical potential is controlled via a bottom gate (\textit{super gate}) at potential $V_{\mathrm{SG}}$. The differential conductance is measured concurrently at both N--S boundaries by alternating the $V_{\mathrm{SD}}$ sweep between the left and right N terminals. In Figs.~\ref{fig:Figure 5}e,f, we demonstrate the formation of zero-energy subgap states at both nanowire ends at $V_{\mathrm{SG}}=0\,$V. Another data set measured at $V_{\mathrm{SG}}=0.5\,$V is presented in Supplementary Fig.~S23. The concomitant behaviour of both ZBPs as a function of $V_{\mathrm{SG}}$ is shown in Supplementary Figs.~S23 and S24. The effective $g$ factor extracted from the linear energy dispersion at the two boundaries is $\sim 10$, albeit the values of $g$ can be strongly gate-dependent~\cite{deMoor2018}. Many experiments have demonstrated ZBPs in tunnelling spectroscopy, indicating the presence of a robust state at zero energy~\cite{Mourik2012, Nichele2017, Gul2018, Grivnin2019}. The robustness of ZBPs in the parameter space (i.e.\ chemical potential and magnetic field) has been used to substantiate the topological origin~\cite{Chen2017}. So far, no experiment has revealed the emergence of ZBPs concurrently at both boundaries of a long hybrid nanowire. While such an observation would corroborate the signatures of MBS, it cannot be regarded as conclusive evidence~\cite{Gul2018,Pan2020b}. Recent experimental works have highlighted accidental correlations between bound states at both ends of short (up to $400\,$nm long) hybrid nanowire devices~\cite{Anselmetti2019,Yu2020}. It is well known that ZBPs can originate from trivial Andreev bound states that arise from inhomogeneities in the chemical potential and random disorder, which emphasizes the need for long and pristine hybrids~\cite{Pan2020b}. Theoretical studies recently pointed out that ZBPs can be a generic feature in many N--S junctions~\cite{Vuik2019, Pan2020}. Additionally, the experimental ZBPs are in general substantially lower than the expected value of $2e^2/h$~\cite{Law2009}, which is a critical but not sufficient hallmark of MBS~\cite{Yu2020}. Figs.~\ref{fig:Figure 5}b,c show differential conductance line-cuts at zero field and at $1.05\,$T which reveal a zero-bias conductance close to $2e^2/h$ for the ZBP at the left boundary of the device, as highlighted in Fig.~\ref{fig:Figure 5}d. While ZBP conductance close to $2e^2/h$ can in principle be observed at both boundaries, it depends on the fine-tuning of the two tunnel barriers, which can be strongly affected by transmission resonances.

\begin{figure}[hbt!]
\includegraphics[width=1.0\linewidth]{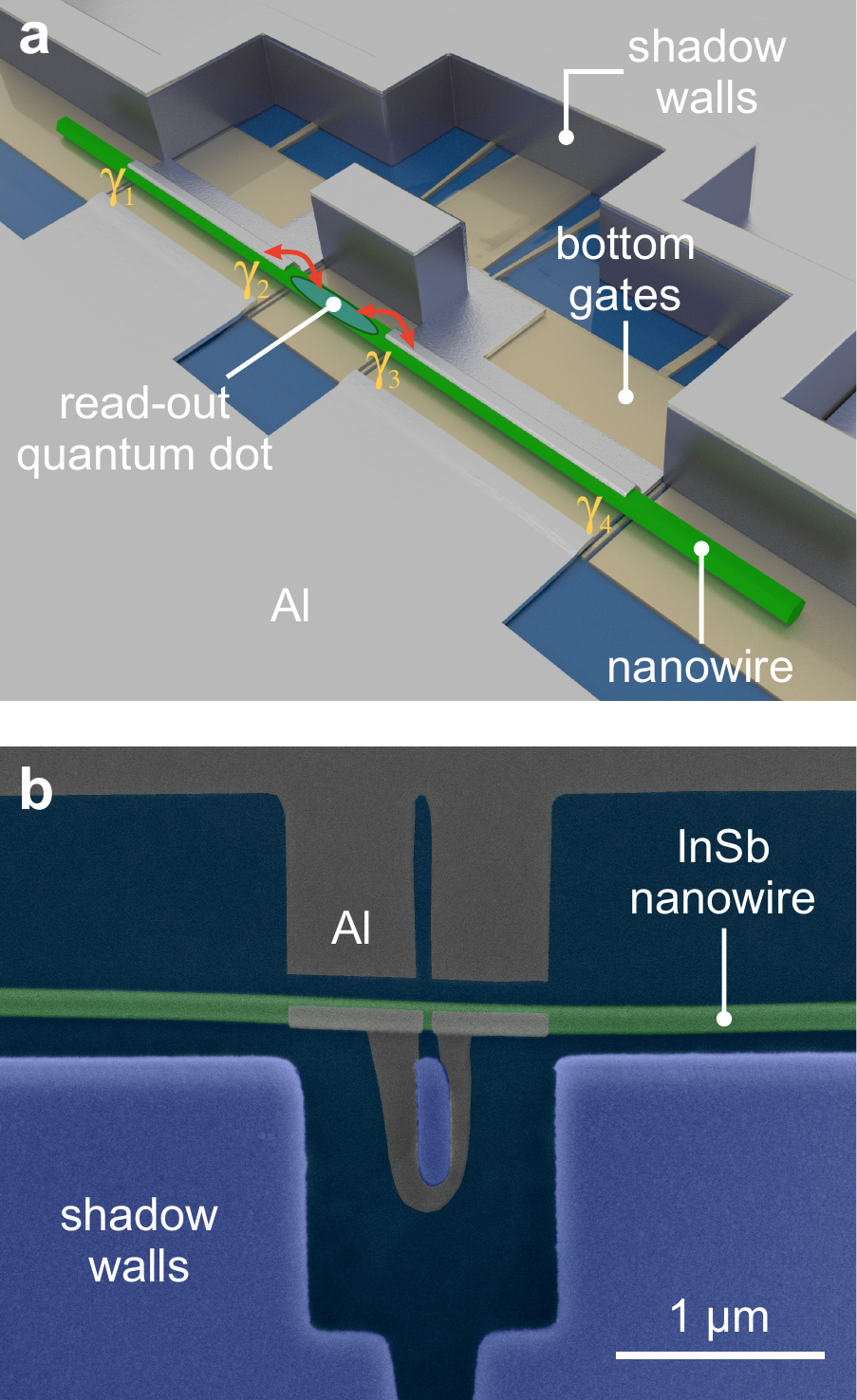}
\caption{\textbf{Illustration of the Majorana loop qubit.} \textbf{a}~Schematic of a single-nanowire loop-qubit device. The presumable locations of the MBS at the boundaries of the two hybrid segments are denoted by $\gamma_i$, where $i \in \left\{1,2,3,4\right\}$. The electron parity is fixed due to the finite charging energy of the loop qubit. This configuration offers the desired ground-state degeneracy for a single qubit and can provide information on decoherence and quasiparticle poisoning. \textbf{b}~False-colour SEM image of an InSb nanowire following the shadow-wall deposition. Two segments of the nanowire are covered with a superconducting 3-facet Al shell. These hybrid segments are interconnected via an Al loop running across the substrate.}
\label{fig:Figure 6}
\end{figure}

\section*{Outlook}
\noindent The 3-terminal hybrid nanowire devices provide a fundamental tool to study the evolution of the induced superconducting gap in the bulk of the hybrid, where electron- and hole-type bands become inverted at the topological phase transition. There, the closing and reopening of the induced gap are accompanied by the emergence of delocalized MBS, hallmarked by ZBPs at both boundaries of the hybrid nanowire. Here, we demonstrate hard-gap N--S junctions in a magnetic field where only discrete subgap states move to zero energy to form ZBPs at both boundaries and that respond similarly to variations in the chemical potential. While these are critical signatures of MBS, upcoming studies will attempt to correlate the local tunnelling conductance with the evolution of the induced bulk gap via the non-local conductance between the two N terminals~\cite{Rosdahl2018}.\\
Our approach promotes the development of intriguing nanowire-based quantum devices. The ballistic hard-gap N--S junctions together with the continuous thin Al connections across the substrate represent a vital starting point for realizing a topological qubit. A qubit implementation with a single read-out loop~\cite{Karzig2017} would allow for measuring the projection of the qubit state on one axis of the Bloch sphere. A schematic of the \textit{loop qubit} is presented in Fig.~\ref{fig:Figure 6}a. It is made from a single nanowire with two superconductor--semiconductor segments connected via a superconducting loop that encircles a central shadow-wall pillar. Bottom gates at the centre of the device are used to define a read-out quantum dot in the nanowire with tunable tunnel couplings to the MBS denoted as $\gamma_2$ and $\gamma_3$ in the schematic. Parity read-out will be performed by measuring the quantum capacitance via radio-frequency gate reflectometry~\cite{deJong2019,Plugge2017,Karzig2017}. In Fig.~\ref{fig:Figure 6}b we present an exemplary realization of the basic elements of such a device via the shadow-wall technique. It comprises a superconducting loop to provide a connection for the exchange of Cooper pairs that acts a blocker for quasiparticle transport between the two hybridized nanowire segments. The shadow-wall technique is ideally suited to facilitate the Al loop across the substrate without the need for any post-interface fabrication steps.

\section*{Methods}
\noindent\textbf{Nanowire growth}\\
The InSb nanowires are grown on InSb ($111$)B substrates covered with a pre-patterned SiN$_x$ mask via metalorganic vapour-phase epitaxy (MOVPE). These nanowires are not grown on top of InP stems but nucleate instead directly on the growth substrate at Au catalyst droplets~\cite{Badawy2019}. The investigated nanowires have an average diameter of $100\,$nm and a length in the order of $10\,\upmu$m controlled by the Au droplet size and the growth mask openings.\\

\noindent\textbf{Device fabrication}\\
Bottom gates are fabricated on Si/SiO$_2$ substrates via dry-etching of W thin films, which are subsequently covered by Al$_2$O$_3$ gate dielectric via atomic layer deposition (ALD). Shadow walls of $\sim 600\,$nm height are created via reactive-ion etching of thick layers of Si$_3$N$_4$ formed via plasma-enhanced chemical vapour deposition (PECVD). Using a micromanipulator, individual nanowires are placed deterministically next to the shadow walls. The native oxide of the nanowire is removed via hydrogen radical cleaning (see Supplementary Section~I) and Al thin films are deposited by evaporation under a shallow angle that forms continuous contacts from the nanowire to the substrate and creates segments on the chip which are electrically isolated from one another. This allows to immediately cool down the devices without the need for additional post-interface fabrication steps. Additional Ti/Au leads as depicted in Figs.~\ref{fig:Figure 4} and \ref{fig:Figure 5} are optional. Alternatively, Al leads that are defined by the shadow walls -- microns away from the N--S junction -- can serve as N contacts but require additional bottom gates to render all nanowire segments fully conducting (cf.\ Fig.~\ref{fig:Figure 1}b).\\

\noindent\textbf{TEM analysis}\\
The cross-sectional lamellae for TEM are prepared using the focused ion beam (FIB) technique with a Helios G4 UX FIB/SEM from Thermo Fisher Scientific after capping the devices with a protective layer of sputtered SiN$_x$. TEM analysis is carried out at an acceleration voltage of $200\,$kV with a Talos electron microscope from Thermo Fisher Scientific equipped with a Super-X energy-dispersive X-ray spectroscopy (EDX) detector.\\

\noindent\textbf{Transport measurements}\\
Electrical transport measurements are carried out in dilution refrigerators equipped with $3$-axes vector magnets. The base temperature is approximately $15\,$mK, corresponding to an electron temperature of about $30\,$mK measured with a metallic N--S tunnel junction thermometer. The sample space is evacuated for at least one day prior to the cool-down to remove surface adsorbates that limit the device performance. Conductance measurements are performed using a standard low-frequency lock-in technique. For voltage-bias measurements, the excitation voltage is $V_{\mathrm{AC}}\leq20\,\upmu$V at a lock-in frequency of at least $20\,$Hz. Current-driven measurements are carried out in a four-point configuration. After taking the data, we became aware of the relatively low bandwidth of the employed current-to-voltage amplifiers. Hence, we recalibrated the lock-in data via a mapping according to the measured DC conductance that does not suffer from any bandwidth limitations and is insensitive to the reactive response of the circuit (Supplementary Section~IV).\\

\noindent\textbf{Superconducting gap extraction}\\
The BCS--Dynes term is given by a smeared BCS density of states with the broadening parameter $\Gamma$~\cite{Dynes1978}:
\begin{equation}
\frac{dI_{\mathrm{SD}}}{dV_{\mathrm{SD}}}\left(V_{\mathrm{SD}}\right) = G_{\mathrm{N}}\operatorname{Re}\left[{\frac{eV_{\mathrm{SD}}-i\Gamma}{\sqrt{\left(eV_{\mathrm{SD}}-i\Gamma\right)^2-\Delta_{\mathrm{ind}}^2}}}\right].\nonumber
\end{equation}
For all of our N--S devices, the fit of the BCS--Dynes term yields typical broadening parameters of less than $10\,\upmu$eV. The model by Blonder, Tinkham, and Klapwijk (BTK) incorporates the transition between BCS tunnelling and Andreev reflection in the open channel~\cite{Blonder1982}. 
Fits of the BCS--Dynes term and of the BTK model to the N--S junction data (including the data in Fig.~\ref{fig:Figure 4}a) are presented in Supplementary Section~IV.\\
The subgap conductance for a ballistic N--S junction with a single subband, where the transport is dominated by Andreev processes, has been described by Beenakker~\cite{Beenakker1992}. At a large enough chemical potential~\cite{Liu2017}, it is given by 
\begin{equation}
G_{\mathrm{S}} = \frac{4e^2}{h} \frac{T^2}{\left(2-T\right)^2} = 2 \frac{G_{\mathrm{N}}^2}{\left(2G_0-G_{\mathrm{N}}\right)^2},\nonumber
\end{equation}
where the transmission probability, $T$, has been substituted with the normal-state conductance, $G_{\mathrm{N}}$, in units of $2e^2/h$. This function is plotted together with the measured data in Fig.~\ref{fig:Figure 4}e.

\section*{Data availability}
\noindent The data that support the plots within this paper and other findings of this study are available at \url{https://doi.org/10.5281/zenodo.3954465}.

\section*{Acknowledgments}
\noindent We are grateful to Olaf Benningshof for valuable technical support and to Emrah Y\"ucelen, Gijs de Lange, Bernard van Heck, Andrey E.\ Antipov, and Jay D.\ Sau for fruitful discussions. We thank Morteza Aghaee for support with dielectric deposition and TNO for providing access to their cleanroom facilities.
This work has been financially supported by the Dutch Organization for Scientific Research (NWO), the Foundation for Fundamental Research on Matter (FOM) and Microsoft Corporation Station Q. M.P.N.\ acknowledges support within the POIR.04.04.00-00-3FD8/17 project as part of the HOMING programme of the Foundation for Polish Science co-financed by the European Union under the European Regional Development Fund.

\section*{Author contributions}
\noindent S.H., M.Q.P., F.B., K.v.H.\ and L.P.K.\ conceived the experiment. S.H., M.Q.P., F.B., N.v.L., G.P.M., J.S.\ and M.A.Y.v.d.P.\ contributed to the fabrication and/or electrical transport measurements of the devices. S.H.\ and F.B.\ analysed the transport data. A.F.\ fabricated the substrates. M.P.N.\ performed numerical simulations of the MAR processes. M.A.\ made critical upgrades to the equipment and provided technical support. G.B., S.G.\ and E.P.A.M.B.\ carried out the nanowire synthesis. K.L.\ prepared the FIB lamellae. S.K.\ performed the TEM analysis. S.H., M.Q.P.\ and F.B.\ wrote the manuscript. All authors provided critical feedback. L.P.K.\ supervised the project.\\

\section*{Competing interests}
\noindent The authors declare no competing interests.

\end{bibunit}


\begin{bibunit}[unsrt]

\clearpage
\widetext

\begin{center}
\large\textbf{Supplementary Information: \linebreak Shadow-wall lithography of ballistic superconductor--semiconductor quantum devices}\\
\normalsize{~\\
Sebastian~Heedt$^*$,$^{1,2,\,\dagger}$ Marina~Quintero-P\'erez$^*$,$^{2}$ Francesco~Borsoi$^*$,$^{1}$ Alexandra~Fursina,$^2$\\
Nick~van Loo,$^1$ Grzegorz~P.~Mazur,$^1$ Micha\l{}~P.~Nowak,$^3$ Mark~Ammerlaan,$^1$ Kongyi~Li,$^1$\\
Svetlana~Korneychuk,$^1$ Jie~Shen,$^1$ May~An~Y.~van de Poll,$^1$ Ghada~Badawy,$^4$ Sasa~Gazibegovic,$^4$\\
Kevin~van Hoogdalem,$^2$ Erik~P.~A.~M.~Bakkers,$^4$ and Leo~P.~Kouwenhoven$^{1,2}$}
\small{~\\~\\
$^1$\textit{QuTech and Kavli Institute of Nanoscience, Delft University of Technology, 2600 GA Delft, The Netherlands}\\
$^2$\textit{Microsoft Quantum Lab Delft, 2600 GA Delft, The Netherlands}\\
$^3$\textit{AGH University of Science and Technology, Academic Centre for Materials\\
and Nanotechnology, al.\ A.\ Mickiewicza 30, 30-059 Krak\'ow, Poland}\\
$^4$\textit{Department of Applied Physics, Eindhoven University of Technology, 5600 MB Eindhoven, The Netherlands}}\\
$^{*}$ These authors contributed equally.\\
$^{\dagger}$ Sebastian.Heedt@Microsoft.com
\end{center}

\setcounter{equation}{0}
\setcounter{figure}{0}
\setcounter{table}{0}
\setcounter{page}{1}
\setcounter{section}{0}
\makeatletter
\renewcommand{\theequation}{S\arabic{equation}}
\renewcommand{\thefigure}{S\arabic{figure}}
\renewcommand{\bibnumfmt}[1]{[S#1]}
\renewcommand{\citenumfont}[1]{S#1}

\section{Fabrication recipes}

\subsection{Chips without Bottom Gates}
\noindent Chips that contain devices with a global back-gate like the ones presented in Figs.~2 and 3 of the main text are fabricated on p$^+$-doped Si wafers covered with $285\,$nm of thermal SiO$_2$. The first fabrication step consists of patterning the bond pads via electron-beam lithography (EBL), W sputtering and lift-off in acetone. Afterwards, plasma-enhanced chemical vapour deposition (PECVD) of $600\,$nm of Si$_3$N$_4$ is performed followed by EBL, reactive-ion etching (RIE) with CHF$_3$/$\mathrm{O}_2$ gases, resist lift-off and an oxygen plasma descum step to remove carbon residues. Eventually, nanowires are deposited under an optical microscope using a micromanipulator equipped with tungsten needles~\cite{Flohr2011}.

\subsection{Chips with Bottom Gates}
\noindent Chips with additional local bottom gates (used e.g.\ for the experiments in Figs.~4 and 5 of the main text) are fabricated by sputtering $17\,$nm of W on Si wafers covered with $285\,$nm of thermal SiO$_x$ (protected by a thin Al$_2$O$_3$ etch-stop layer), followed by EBL patterning and RIE of the W layer with SF$_6$ gas. Next, $18\,$nm of a high-quality Al$_2$O$_3$ layer are deposited by atomic layer deposition (ALD), acting as bottom-gate dielectric. Shadow walls on top of the bottom gates are created by first depositing $600\,$nm of Si$_3$N$_4$ by PECVD, followed by EBL patterning with precise alignment of the shadow walls with respect to the underlying fine bottom gates. Then, RIE with CHF$_3$/$\mathrm{O}_2$ gases is used to selectively etch Si$_3$N$_4$ while the Al$_2$O$_3$ gate dielectric acts as an etch-stop layer. Finally, after the resist lift-off an oxygen plasma descum step is used to remove carbon residues from the chips. The nanowires are then mechanically transferred on top of the bottom gates under an optical microscope using a micromanipulator equipped with tungsten needles~\cite{Flohr2011}.

\subsection{Additional Fabrication Steps for N--S Devices}
\noindent For devices with additional Ti/Au normal-metal contacts, such as the ones presented in Figs.~4 and 5 of the main text, an extra post-interface fabrication step is included. It consists of EBL patterning, $40\,$s of argon ion milling at $1.5 \cdot 10^{-3}\,$mbar with a commercial Kaufmann source in the load lock of an electron-beam evaporator, and in-situ evaporation of $10\,$nm/$150\,$nm of Ti/Au followed by lift-off in acetone. Note that this step is not essential and could have been omitted. Bottom gates underneath the nanowire can open up the channels and tune the conductance. Combining this electrostatic gate control with additional Al contacts that are defined by shadow walls microns away from the N--S junction allows to entirely avoid post-interface fabrication for these devices.

\subsection{Semiconductor Surface Treatment}
\noindent To obtain a pristine, oxide-free semiconductor surface, we accomplish a gentle oxygen removal via atomic hydrogen radical cleaning. For this purpose, a custom-made H radical generator is installed in the load lock of our aluminium electron-gun evaporator. It consists of a gas inlet for H$_2$ molecules connected to a mass-flow controller and a tungsten filament at a temperature of about $1700\,^{\circ}$C that dissociates a fraction of the molecules into hydrogen radicals~\cite{Webb2015}.\\
The cleaning process is assessed via the transport characteristics of InSb/Al nanowire junctions and TEM analysis of the same devices. In particular, we consider as critical indicators of the interface transparency the magnitude of the supercurrents and the amount of interface oxide measured via EDX. During the optimization, we vary the process duration and the hydrogen flow, and keep the substrate temperature constant at $550\,$K. It has been demonstrated in the literature that this temperature results in an efficient cleaning of InSb allowing for indium- and antimony-based oxides to be removed with similar efficiency~\cite{Haworth2000,Tessler2006}.\\
The optimal removal of the native oxide is achieved for a process duration of $30\,$mins and a hydrogen flow of $2\,$mln/min. During atomic hydrogen cleaning the H$_2$ pressure is $6.3\cdot10^{-5}\,$mbar. This recipe, which is used for all the devices shown in this paper, results in a constant EDX count of oxygen at the interface (i.e.\ the traces do not show oxygen peaks, see Fig.~2d of the main text) and yields the highest supercurrents in the Josephson junction devices ($\sim 90\,$nA).

\begin{figure*}[hbt!]
\includegraphics[width=1.0\linewidth]{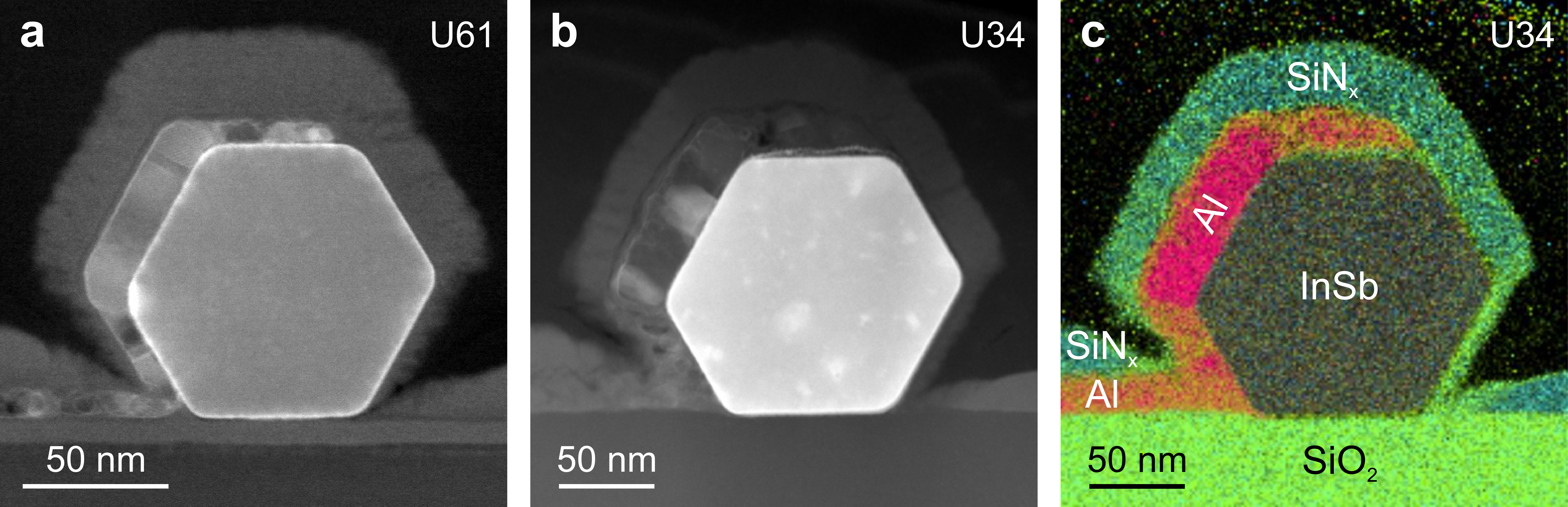}
\caption{Cross-sectional TEM images of InSb nanowires covered by a thin layer of Al. SiN$_x$ was sputter-coated as a protective layer before focused ion beam (FIB) lamella preparation. \textbf{a}~Annular dark-field (ADF) scanning TEM image of a nanowire cross-section. This sample is identical to the device presented in Figs.~2c-e of the main text (sample ID: U61). The Al is deposited at an angle of $30^{\circ}$ with respect to the substrate plane at $T=140\,$K. \textbf{b}~ADF scanning TEM image of another nanowire cross-section (sample ID: U34). Here, $30\,$nm of Al are deposited at an angle of $25^{\circ}$ with respect to the substrate plane at $T=80\,$K. \textbf{c}~EDX elemental composite image of the device in panel~(\textbf{b}) identifying the individual compounds and the Al thin film.}
\label{fig:S1_TEM}
\end{figure*}

\subsection{Superconductor Deposition}
\begin{figure}[H]
\centering
\includegraphics[width=1.0\linewidth]{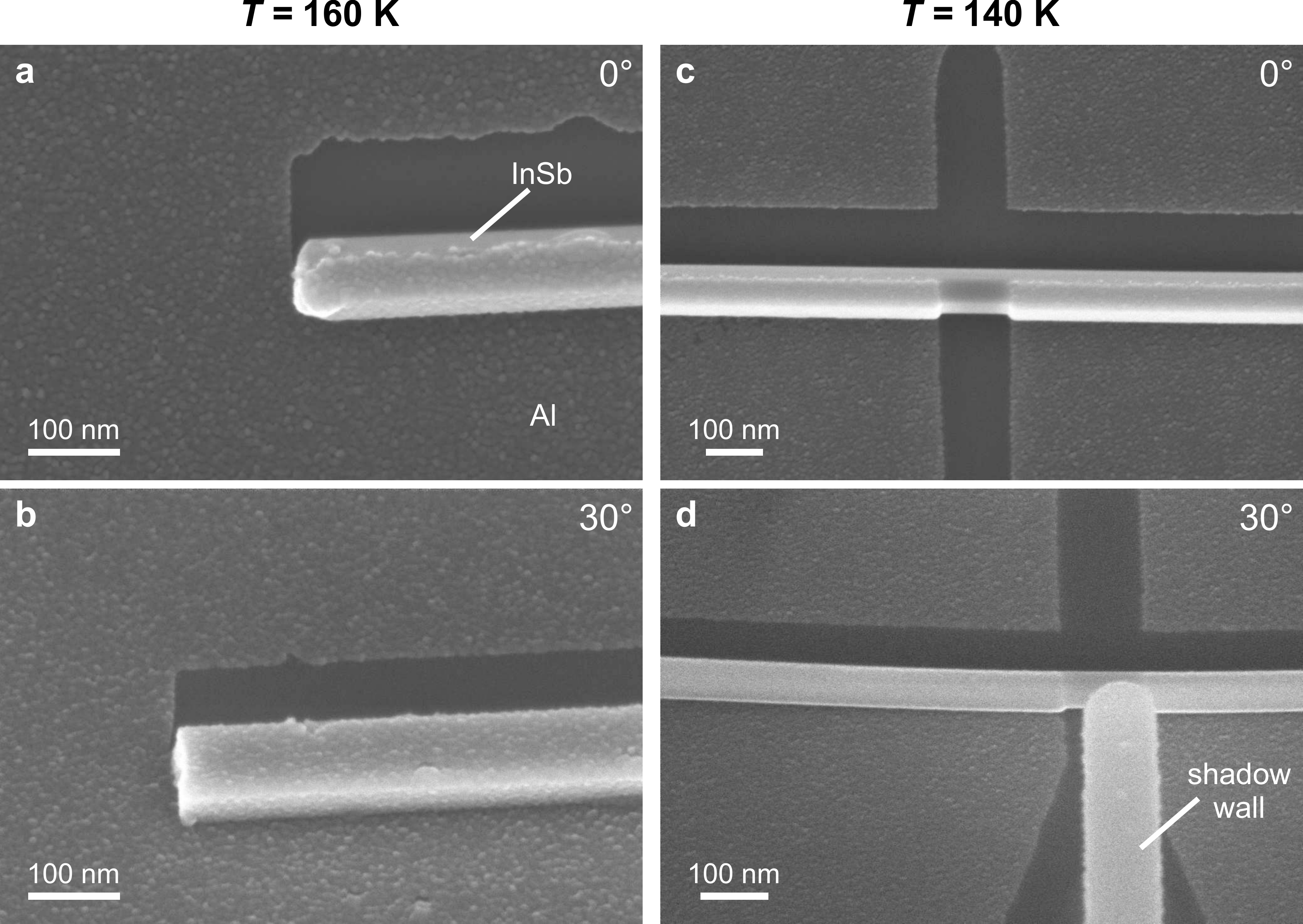}
\caption{SEM images of InSb nanowires with Al thin films deposited at different temperatures. For both samples the evaporation angle is $30^{\circ}$ relative to the substrate plane. \textbf{a}, \textbf{b}~Top-view and tilt-view (tilt angle: $30^{\circ}$) SEM images of InSb nanowires covered with an Al thin film deposited at $160\,$K. The maximum film thickness, which corresponds to the middle nanowire facet, is $20\,$nm. \textbf{c}, \textbf{d}~Top-view and tilt-view (tilt angle: $30^{\circ}$) SEM images of a nanowire Josephson junction. Here, the Al thin film is deposited at $140\,$K and the film thickness of the middle nanowire facet is $15\,$nm. Panel~(\textbf{d}) exhibits a featureless Al shell on the middle wire facet, whereas grains are visible on the middle facet in the case of Al grown at $160\,$K (panel~(\textbf{b})).}
\label{fig:RoughnessFacets}
\end{figure}

\noindent After the semiconducting surface cleaning, the chips are loaded into the evaporator main chamber and cooled down by actively circulating liquid nitrogen through the sample holder. After one hour of thermalization, aluminium is deposited by electron-beam evaporation at a typical rate of $0.2\,$nm/min.\\
The optimization of the aluminium growth is performed by studying the quality of thin films deposited on Si substrates -- sometimes containing scattered or deposited InSb nanowires -- at different evaporation rates, temperatures and angles. It is observed that evaporation angles close to $90^{\circ}$ with respect to the substrate plane are favourable for thin-film aluminium growth, whereas for shallower angles the self-shadowing effect of the Al atoms on the surface becomes more apparent, giving rise to columnar growth, possible voids in the film, and a larger roughness~\cite{Dong1996,BARRANCO201659}. To minimize this angle-dependent self-shadowing effect, the substrate temperature can be slightly increased to give the atoms arriving at the substrate enough momentum to rearrange into a crystal before the next atoms arrive at the substrate. Our results and the work by Dong~\emph{et al.}~\cite{Dong1996} indicate that, for a fixed deposition rate, the temperature optimum depends on the evaporation angle.\\
In this work, a temperature optimum of around $140\,$K is found for Al growth at $30^{\circ}$ with respect to the substrate plane, allowing for homogeneous 3-facet coverage of the hexagonal nanowires as well as a connection from the nanowires to the substrate. Josephson junctions made at this growth temperature exhibit roughly four times higher supercurrents than similar devices produced when Al was deposited at a substrate temperature of $\sim 80\,$K. Similarly, cross-sectional TEM images of FIB lamellae from nanowires with Al grown at $140\,$K and $80\,$K are presented in Fig.~\ref{fig:S1_TEM}a (as well as Figs.~2c,e of the main text) and Figs.~\ref{fig:S1_TEM}b,c, respectively. Comparing these figures, the superior quality of the deposition at $140\,$K is evident; nanowire facets are better covered and form a continuous film, the crystalline quality of Al is higher and the oxidation of the Al facets is much less prominent than in the case of the deposition at $80\,$K (in Fig.~\ref{fig:S1_TEM}b, the abundant oxide formation in the aluminium film at the the top and bottom-left nanowire facets is especially noticeable).\\
In addition, Fig.~\ref{fig:RoughnessFacets} illustrates a comparison between a higher Al growth temperature ($160\,$K) and Al grown at $140\,$K. The former leads to both grainy Al covering the middle nanowire facet, which can be better observed in the tilt-view picture Fig.~\ref{fig:RoughnessFacets}b, and a film on the substrate where the different grains are clearly distinguishable. Images corresponding to deposition at $140\,$K instead show a featureless Al film on the middle facet, where roughness is indiscernible at these SEM conditions (Fig.~\ref{fig:RoughnessFacets}d), and a grainy but more uniform Al structure on the substrate.

\subsection{Typical Yield of the Nanowire Transfer for Josephson Junction Devices}
\begin{figure}[H]
\centering
\includegraphics{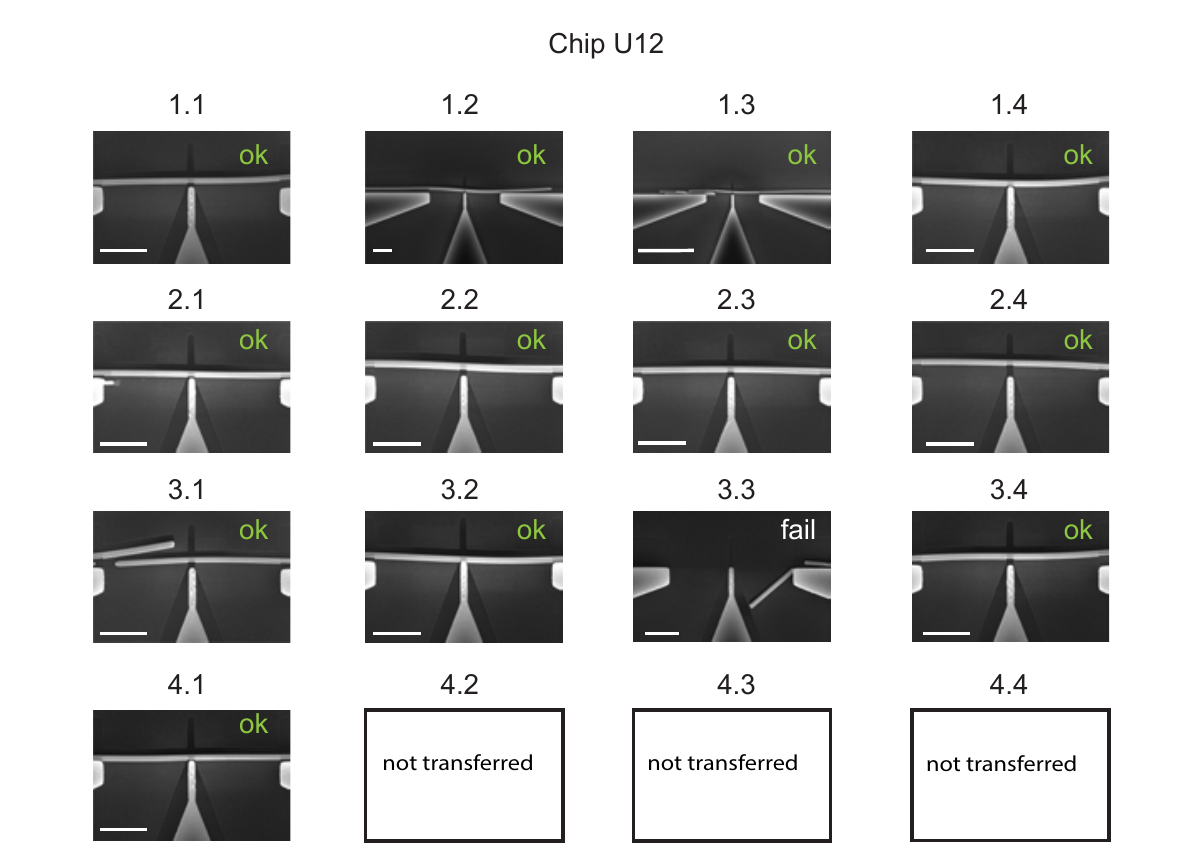}
\caption{Typical yield of the nanowire transfer for Josephson junction devices: Scanning electron micrographs of all Josephson junction devices on a typical chip (sample ID: U12) taken after the Al deposition. Out of $13$ nanowire transfer attempts, $12$ nanowires are perfectly positioned, and only in one case the transfer failed (device 3.3). The scale bars indicate $1\,\upmu$m.}
\label{fig:FigureS3}
\end{figure}

\begin{figure}[H]
\centering
\includegraphics{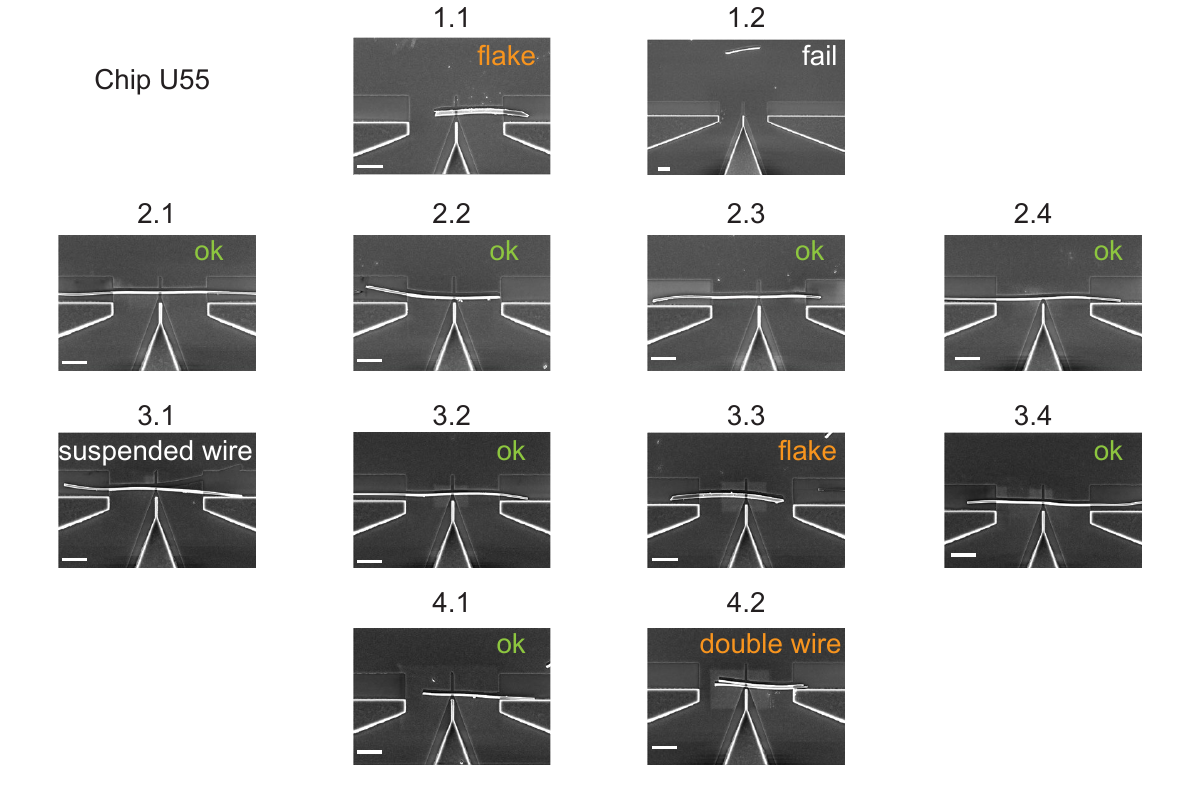}
\caption{Typical yield of the nanowire transfer for Josephson junction devices: Scanning electron micrographs of all Josephson junction devices on a typical chip (sample ID: U55) taken after the Al deposition. Out of $12$ nanowire transfer attempts, $7$ nanowires are perfectly positioned, $2$ of them are not (devices 1.2 and 3.1), $2$ narrow flakes -- rather than nanowires -- are accidentally transferred (devices 1.1 and 3.3), and in one case, two nanowires are transferred in the same location (device 4.2). The scale bars indicate $1\,\upmu$m.}
\label{fig:FigureS4}
\end{figure}

\noindent Depending on the layout, our pre-patterned chips typically accommodate up to $16$ nanowire devices. It is readily viable to have around $10$ devices on a single chip to consistently optimize the fabrication parameters. The yield per chip can be affected by the accidental transfer of multiple wires at once or by nanowires breaking during the transfer. In Figs.~\ref{fig:FigureS3} and \ref{fig:FigureS4}, we show scanning electron micrographs taken prior to the cool-down of the Josephson junctions. On the first chip (sample ID: U12) $13$ nanowires are transferred and result in $12$ working devices, i.e.\ where the junctions are well-defined. On the second chip (sample ID: U55) $12$ nanowires are positioned and yield $9$ working devices. Among those, $7$ are hexagonal-shaped nanowires and $2$ turned out to be narrow nano-flakes~\cite{deVries2019,Gazibegovic2019}, which can be indistinguishable from nanowires in optical microscopy.

\newpage
\section{Additional Transport Measurements of Josephson junctions}
\noindent In this section, we summarize the characteristics of the Josephson junction devices listed in Tab.~\ref{tab:JJ_devices}. All devices are fabricated by evaporating an Al thin film at $30^{\circ}$ with respect to the substrate plane. Device~3 differs from the other samples in the thickness of the Al shell. We note that despite such a low shell thickness, all nanowire devices on sample U59 are in electrical contact with the Al film on the substrate.

\begin{table}[hbt!]
\centering
\begin{tabular}{|c|c|c|c|c|c|}
\hline
\multirow{2}{*}{\textbf{Josephson junction}} & \textbf{Sample ID/} & \textbf{Evaporation} & \textbf{Channel} & \textbf{Maximum Al} & \multirow{2}{*}{\textbf{Oxidation}}\\
 & \textbf{device name} & \textbf{angle} & \textbf{width (nm)} & \textbf{thickness (nm)} & \\ \hline\hline
device 1  & U55/2.3  & $30^{\circ}$  & $100$  & $16$  & in O$_2$ atmosphere\\ \hline
device 2  & U51/1.2  & $30^{\circ}$  & $100$  & $16$  & in O$_2$ atmosphere\\ \hline
device 3  & U59/2.3  & $30^{\circ}$  & $100$  & $11$  & Al$_2$O$_3$ capping\\ \hline
device 4  & U55/3.3 & $30^{\circ}$  & $160$  & $16$  & in O$_2$ atmosphere\\ \hline
\end{tabular}
\caption{Summary of the Josephson junction devices presented in this study. Devices~1, 2, and 3 are all nominally identical in their geometries with a nanowire diameter of $100\,$nm and a separation between the Al contacts of $115\,$nm. Device~3 was made with a thinner Al shell thickness and capped with around $20\,$nm of Al$_2$O$_3$. Device~4 is a Josephson junction formed in an InSb nano-flake. Here, the channel width is $160\,$nm.}
\label{tab:JJ_devices}
\end{table}

\begin{figure}[H]
\centering
\includegraphics[width=0.8\linewidth]{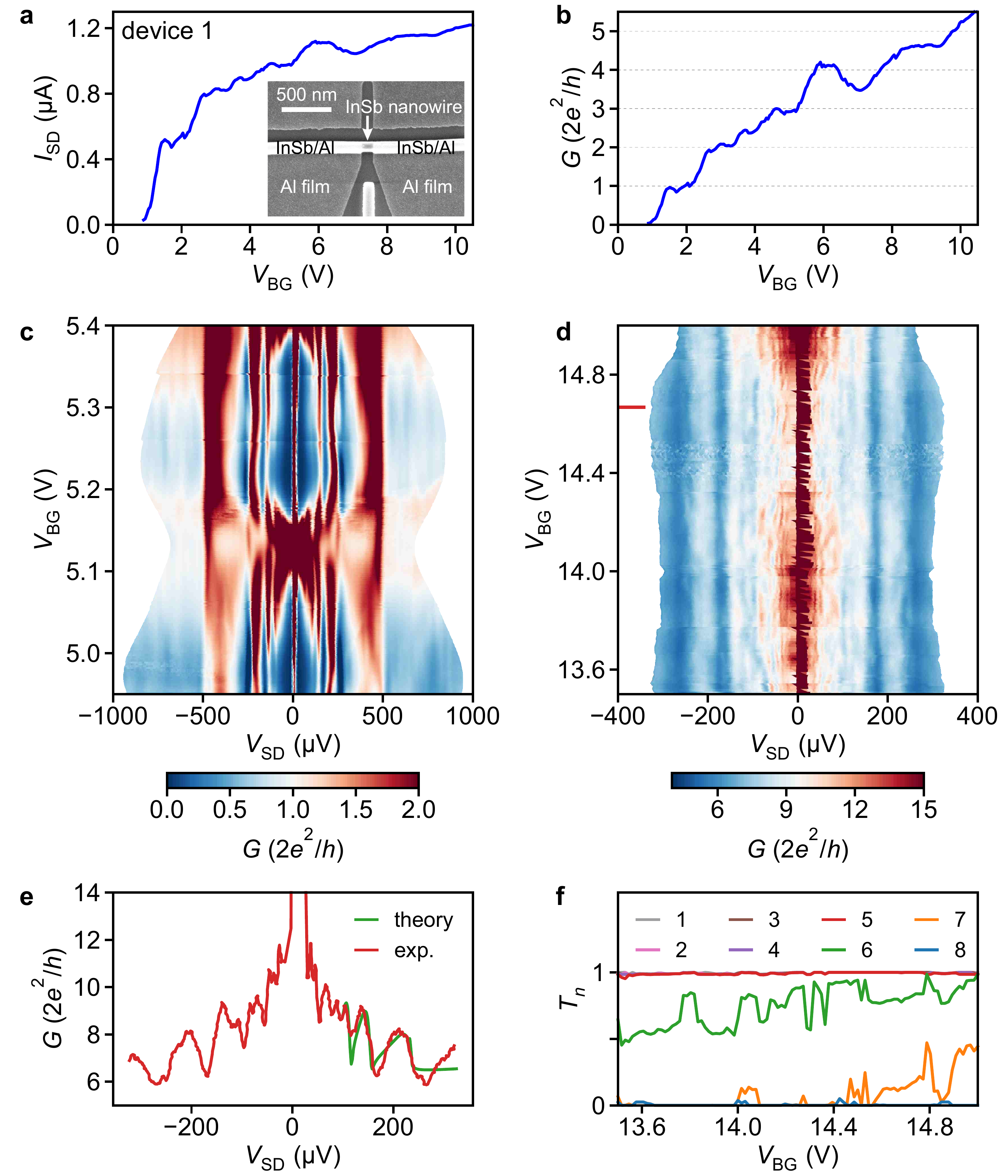}
\caption{Additional transport measurements of the first Josephson junction device. \textbf{a}~$I_{\mathrm{SD}}$ vs.\ $V_{\mathrm{BG}}$ sweep at $V_{\mathrm{SD}} = 10\,$mV, showing the field-effect tunability of the junction. Inset: scanning electron micrograph of the device. \textbf{b}~DC conductance, $G$, after subtracting the series resistance, as a function of $V_{\mathrm{BG}}$ at $10\,$mV bias voltage. \textbf{c}, \textbf{d}~$G$ vs.\ $V_{\mathrm{SD}}$ and $V_{\mathrm{BG}}$ in the few-subbands and many-subbands regime, respectively: vertical features in both scans at constant bias voltages are the characteristic peaks originating from MARs. \textbf{e}~Line-cut of (\textbf{d}) at $V_{\mathrm{BG}} = 14.67\,$V in red and best fit of the trace in green according to the coherent scattering model in Section~\ref{sec:MAR_model}. \textbf{f}~Extracted transmission probabilities, $T_n$, as a function of $V_{\mathrm{BG}}$ in the multi-subbands regime, with $n\in\left\{1, 2,\ldots,8\right\}$. In this back-gate voltage range, the transmission of the first five subbands is already saturated at $T_n = 1$.}
\label{fig:FigureSI_JJ1_01}
\end{figure}

\begin{figure}[H]
\centering
\includegraphics[width=0.8\linewidth]{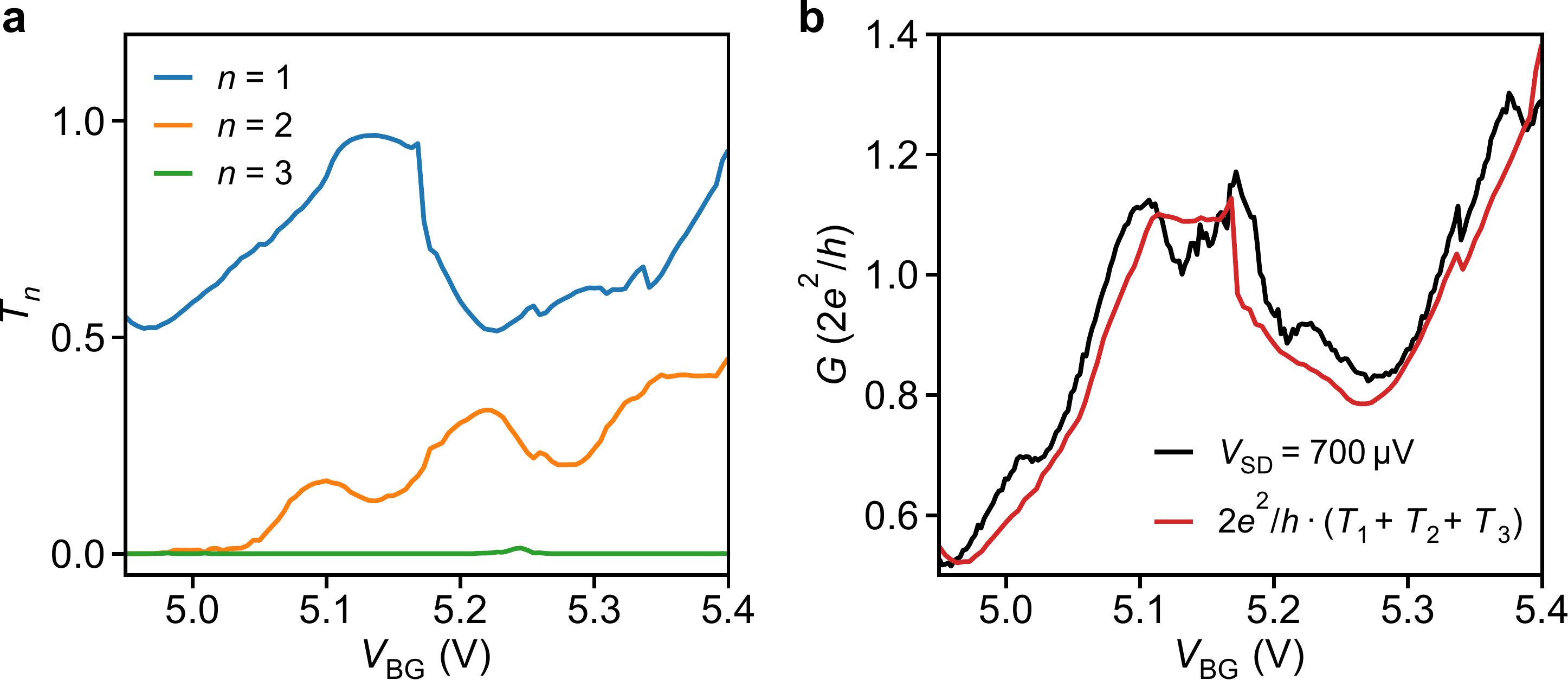}
\caption{Tunability of the subbands of the first Josephson junction device. \textbf{a}~Transmission probabilities, $T_n$, of the first three subbands as a function of $V_\mathrm{BG}$. The parameters are extracted by fitting the conductance map of Fig.~\ref{fig:FigureSI_JJ1_01}c with the coherent scattering model described in Section~\ref{sec:MAR_model}. \textbf{b}~Out-of-gap conductance as a function of $V_\mathrm{BG}$ in black (i.e.\ vertical line-cut of Fig.~\ref{fig:FigureSI_JJ1_01}c at $V_{\mathrm{SD}} = 700\,\upmu$V) together with the sum of the transmission probabilities in red.}
\label{fig:FigureSI_JJ1_02}
\end{figure}

\begin{figure}[H]
\centering
\includegraphics[width=0.8\linewidth]{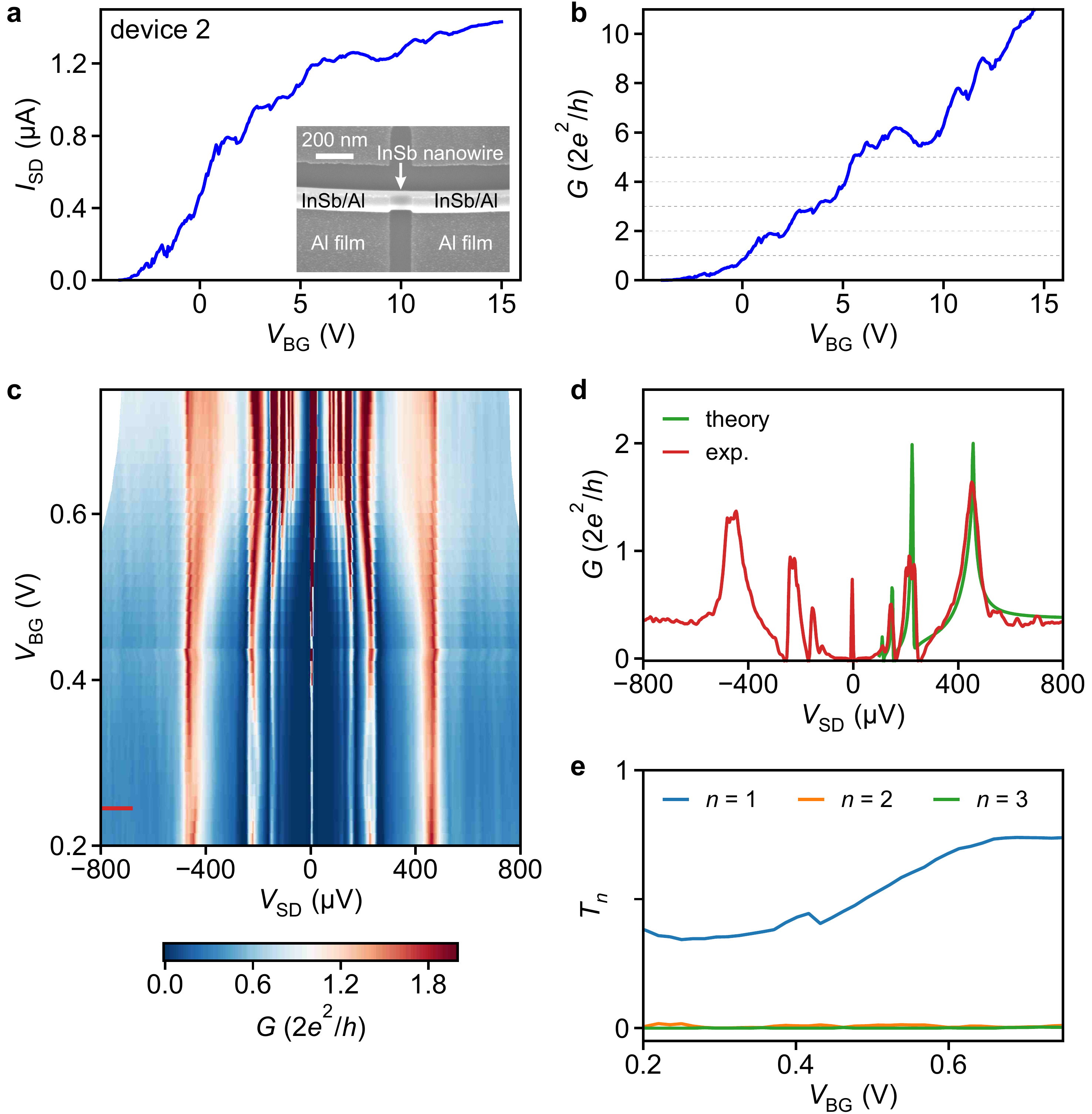}
\caption{Additional transport measurements of the second Josephson junction device. \textbf{a}~$I_{\mathrm{SD}}$ vs.\ $V_{\mathrm{BG}}$ sweep at $V_{\mathrm{SD}} = 10\,$mV, showing the field-effect tunability of the junction. Inset: scanning electron micrograph of the device. \textbf{b}~DC conductance, $G$, after subtracting the series resistance, as a function of $V_{\mathrm{BG}}$ at $10\,$mV bias voltage. \textbf{c}~$G$ vs.\ $V_{\mathrm{SD}}$ and $V_{\mathrm{BG}}$ in the weak-tunnelling regime: subharmonic gap features correspond to different orders of MARs. \textbf{d}~Line-cut of (\textbf{c}) at $V_{\mathrm{BG}} = 0.25\,$V in red and best fit of the trace in green according to the coherent scattering model in Section~\ref{sec:MAR_model}. \textbf{e}~Extracted transmission probabilities, $T_n$, depicted as a function of $V_{\mathrm{BG}}$ with $n\in\left\{1, 2,3\right\}$.}
\label{fig:FigureSI_JJ2_01}
\end{figure}

\begin{figure}[H]
\centering
\includegraphics[width=0.9\linewidth]{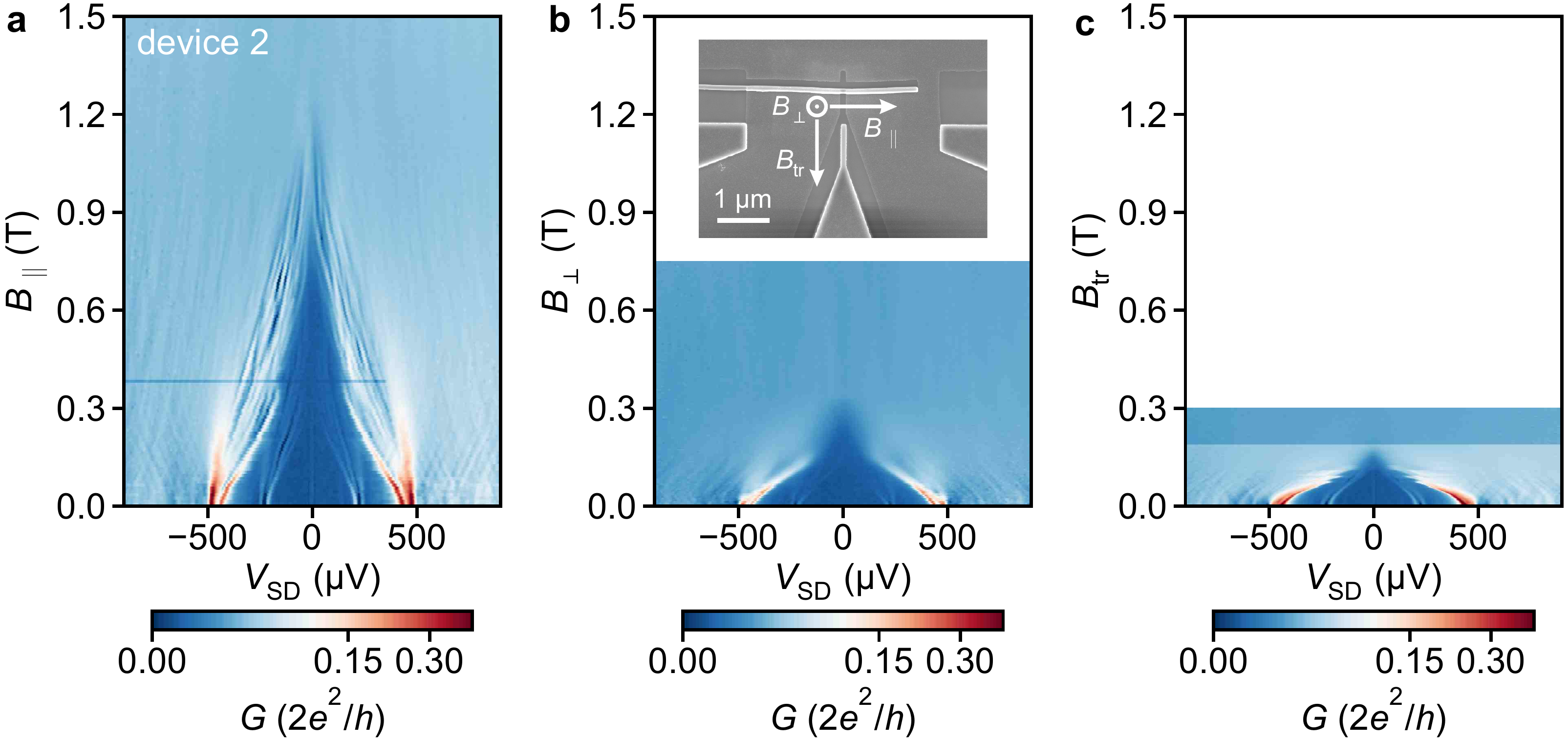}
\caption{Superconducting critical magnetic fields of the second Josephson junction device. Colour maps of $G$ vs.\ $V_{\mathrm{SD}}$ and $B$ taken at $V_{\mathrm{BG}} = 1.45\,$V for different magnetic field orientations: in \textbf{a} the field $B_{\parallel}$ is oriented parallel to the nanowire direction, in \textbf{b} $B_{\perp}$ is orthogonal to the plane of the substrate, and in \textbf{c} the transversal field $B_{\mathrm{tr}}$ is orthogonal to the nanowire direction but in the substrate plane. The inset in panel~(\textbf{b}) shows a scanning electron micrograph of the device together with the different magnetic field directions.}
\label{fig:FigureSI_JJ2_02}
\end{figure}

\begin{figure}[H]
\centering
\includegraphics[width=0.85\linewidth]{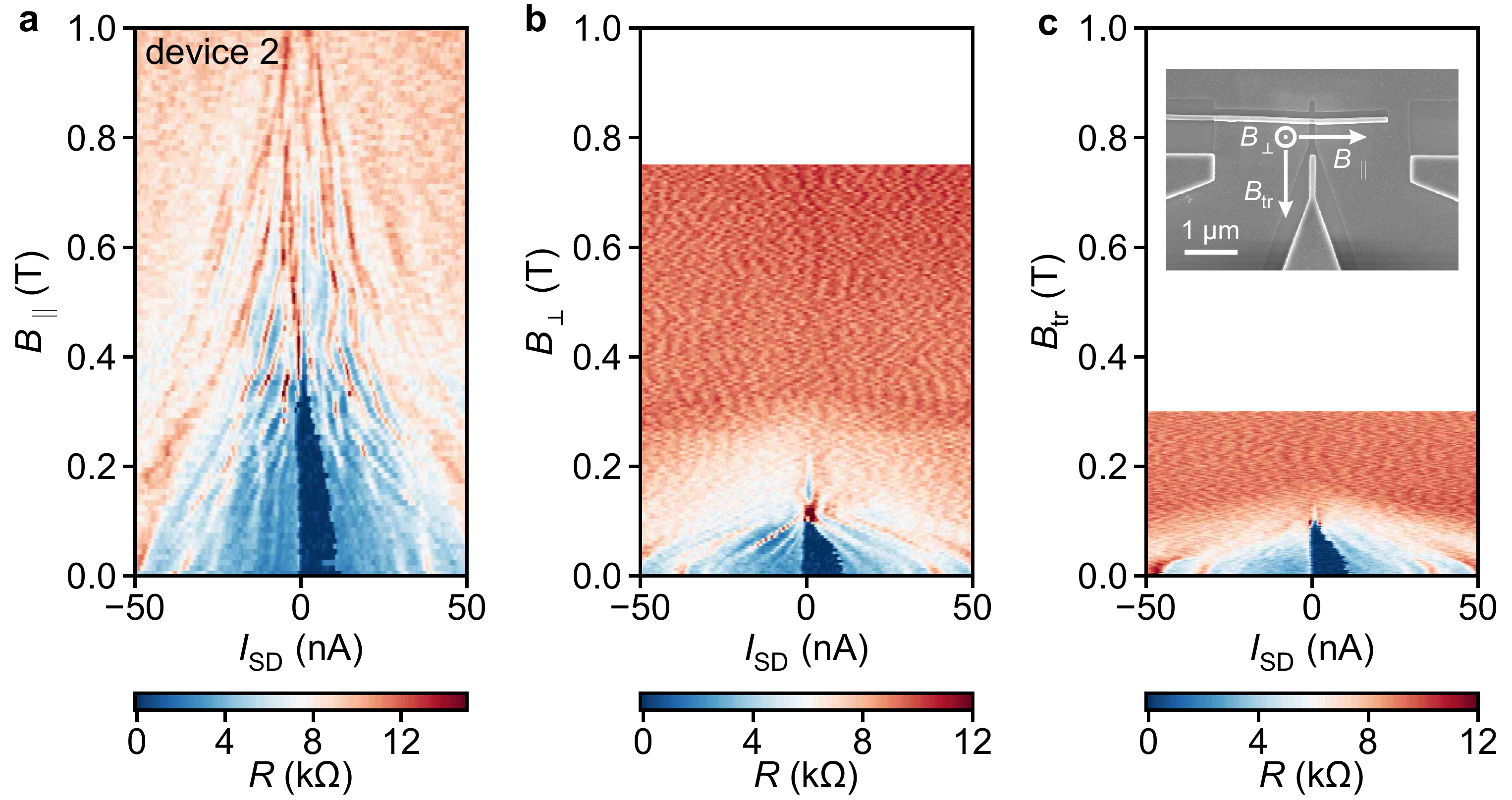}
\caption{Switching current of the second Josephson junction device in the open-channel regime ($V_{\mathrm{BG}} = 5.7\,$V). Differential resistance, $R$, as a function of $I_{\mathrm{SD}}$ and magnetic field in three different orientations: \textbf{a}~Magnetic field, $B_{\parallel}$, aligned parallel to the nanowire, \textbf{b}~magnetic field, $B_{\perp}$, oriented out-of-plane, and \textbf{c}~transversal in-plane magnetic field, $B_{\mathrm{tr}}$. The vectors in the inset of panel~(\textbf{c}) illustrate the three field orientations.}
\label{fig:FigureSI_JJ2_03}
\end{figure}

\begin{figure}[H]
\centering
\includegraphics[width=0.65\linewidth]{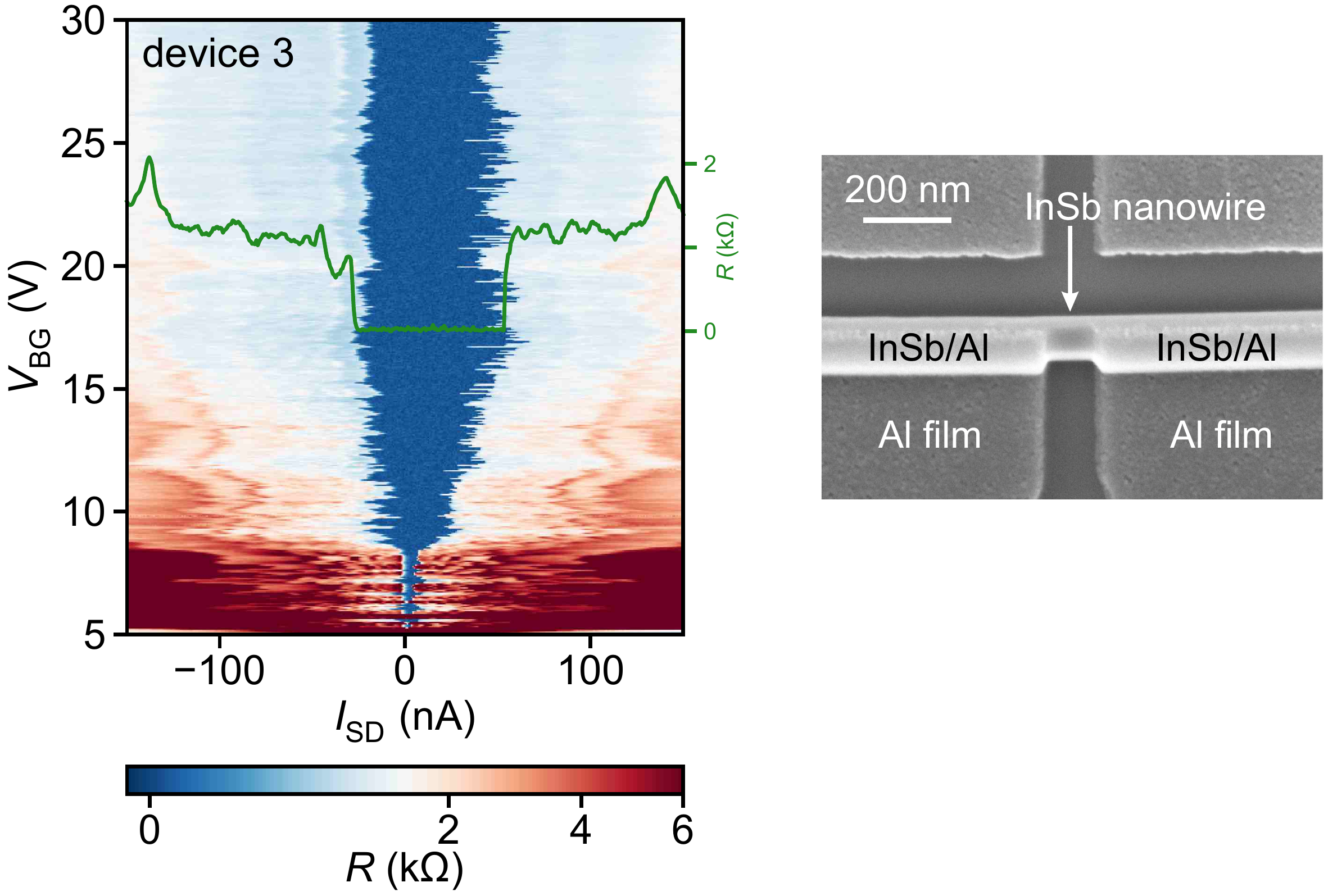}
\caption{Back-gate dependence of the switching current of the third Josephson junction device. Colour map of $R$ as a function of $I_{\mathrm{SD}}$ and $V_{\mathrm{BG}}$; the green trace is taken at $V_{\mathrm{BG}} = 13.82\,$V. The switching current (in dark blue) is suppressed in the low back-gate voltage regime. The inset on the right shows a scanning electron micrograph of the device. The Al segments are capped with a protective layer of Al$_2$O$_3$.}
\label{fig:FigureSI_JJ3_02}
\end{figure}

\begin{figure}[H]
\centering
\includegraphics[width=0.85\linewidth]{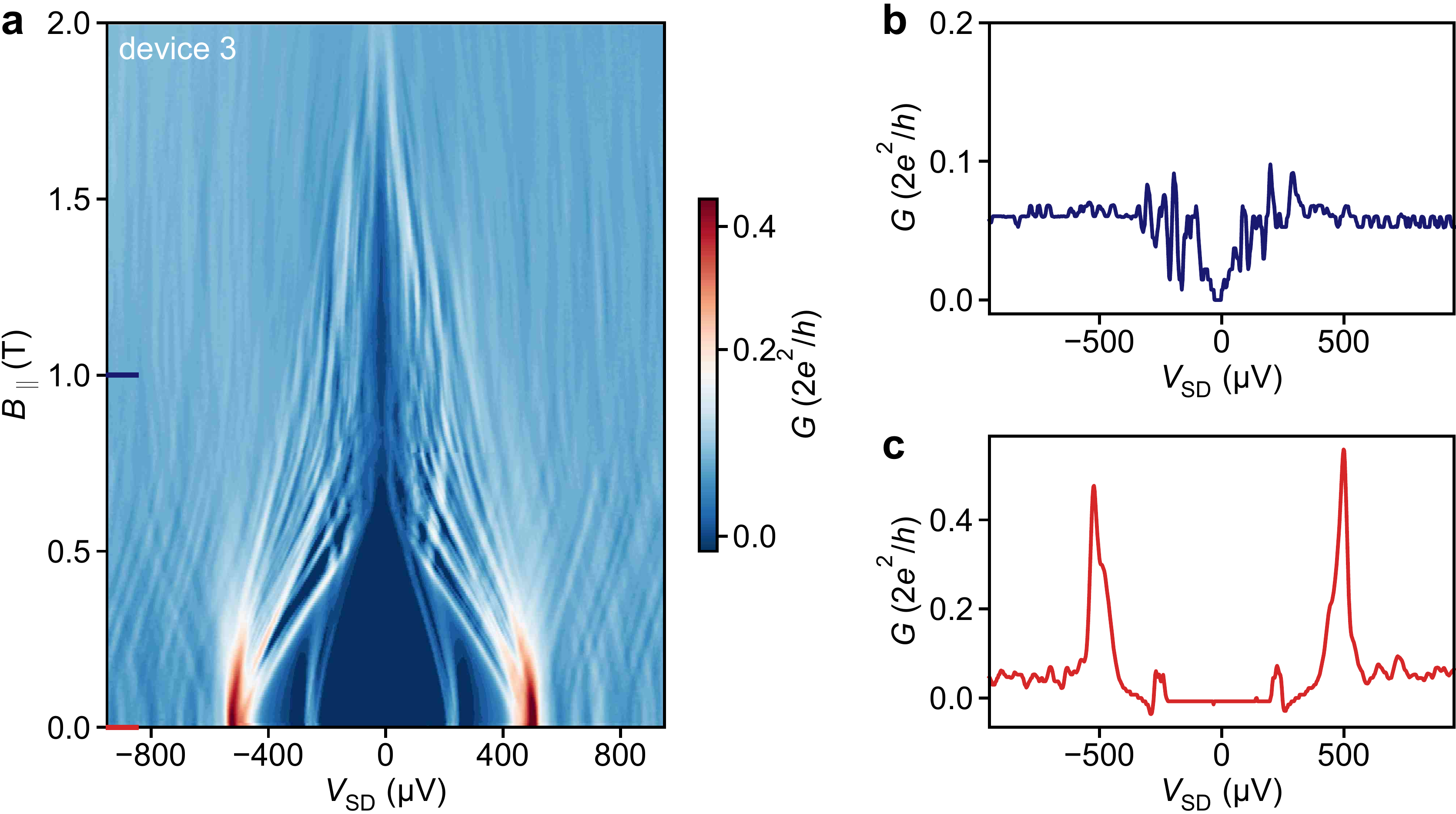}
\caption{Transport measurements of the third Josephson junction device. \textbf{a}~$V_{\mathrm{SD}}$ vs.\ $B_{\parallel}$ in the tunnelling regime. Owing to the thinner Al shell, the superconducting critical field is $B_{\mathrm{c}} \sim 2\,$T, much larger than for the previous two junctions. The tunnelling conductance peaks at $\pm 2 \Delta_{\mathrm{ind}}$ split into a manifold of resonances at a finite magnetic field due to the different $g$ factors of the discrete quasiparticle states at the gap edge. \textbf{b}, \textbf{c}~Line-cuts of (\textbf{a}) at the positions indicated by the two lines.}
\label{fig:FigureSI_JJ3_01}
\end{figure}

\begin{figure}[H]
\centering
\includegraphics[width=0.8\linewidth]{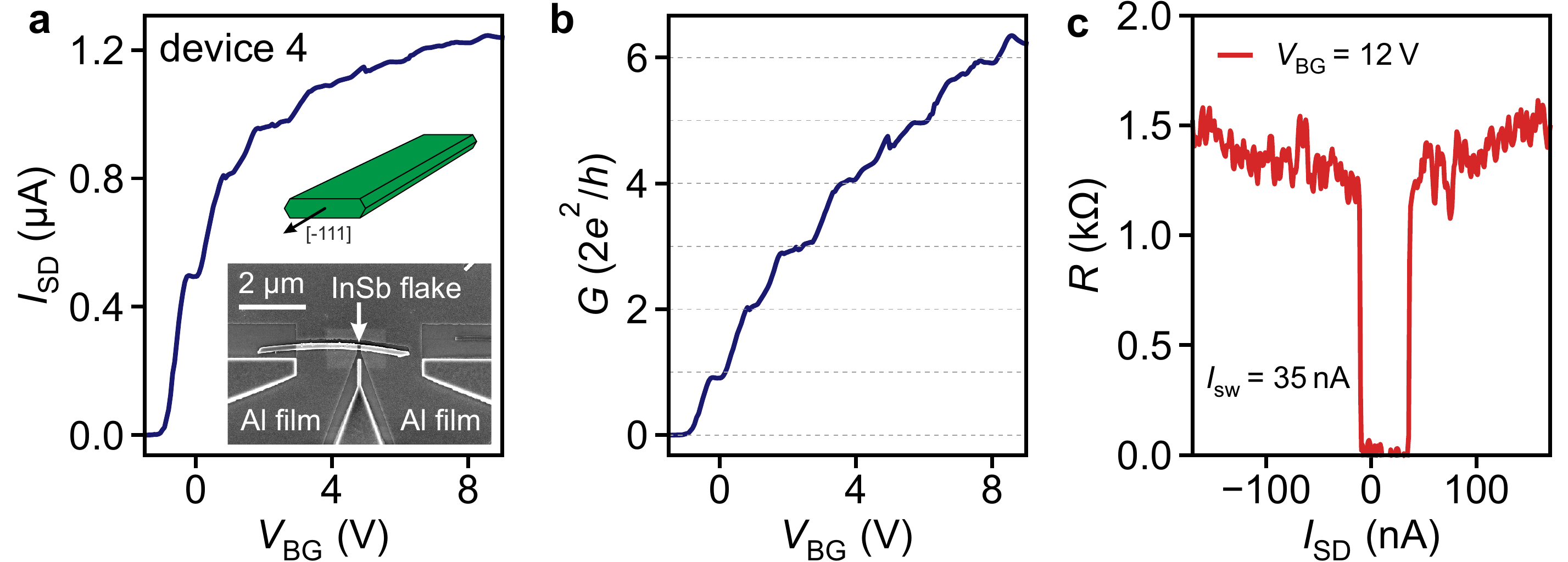}
\caption{Ballistic transport and supercurrent in an InSb flake Josephson Junction (device~4) at zero magnetic field. \textbf{a}~$I_{\mathrm{SD}}$ vs.\ $V_{\mathrm{BG}}$ at $V_{\mathrm{SD}} = 10\,$mV. Bottom inset: scanning electron micrograph of the nano-flake Josephson junction. Top inset: schematic of the cross-section of the nano-flake~\cite{Gazibegovic2019}. \textbf{b}~$G$ vs.\ $V_{\mathrm{BG}}$ at $V_{\mathrm{SD}} = 10\,$mV. The conductance displays distinct plateaus at multiples of $2e^2/h$, indicating the stepwise population of the one-dimensional subbands in the nanowire. \textbf{c}~$R$ vs.\ $I_{\mathrm{SD}}$ at $V_{\mathrm{BG}} = 12\,$V. The device exhibits a switching current of $\sim 35\,$nA in the open-channel regime.}
\label{fig:FigureSI_JJ4_01}
\end{figure}

\paragraph*{\textbf{Device 1}}
The current and differential conductance in the normal state ($V_{\mathrm{SD}} = 10\,$mV) display a steplike increase as a function of $V_{\mathrm{BG}}$ (Figs.~\ref{fig:FigureSI_JJ1_01}a,b), a signature of ballistic transport at zero magnetic field in device~1. At lower bias voltage, features of the induced superconductivity appear such as the conductance peaks due to multiple Andreev reflections and the zero-bias supercurrent peak (Figs.~\ref{fig:FigureSI_JJ1_01}c,d). A line-cut of Fig.~\ref{fig:FigureSI_JJ1_01}c is presented in Fig.~3b of the main text, whereas a line-cut of Fig.~\ref{fig:FigureSI_JJ1_01}d is shown in panel~(e). Here, the experimental data (red trace) is fitted with the theoretical model (green trace) to identify the number and the transmissions of the nanowire subbands, which are plotted in Fig.~\ref{fig:FigureSI_JJ1_01}f. Similarly, in Fig.~\ref{fig:FigureSI_JJ1_02}a, we illustrate the extracted transmission probabilities of the three lowest subbands in the back-gate voltage range of Fig.~\ref{fig:FigureSI_JJ1_01}c. The sum of these transmission probabilities extracted from the MAR pattern is compared to the normal-state conductance in Fig.~\ref{fig:FigureSI_JJ1_02}b.
\paragraph*{\textbf{Device 2}}
The normal-state current and conductance ($V_{\mathrm{SD}} = 10\,$mV) as a function of back-gate voltage are displayed in Figs.~\ref{fig:FigureSI_JJ2_01}a,b. While conductance plateaus are more difficult to identify than in the case of device~1, the presence of an induced superconducting gap is clear from the MAR conductance peaks and the supercurrent peak (Figs.~\ref{fig:FigureSI_JJ2_01}c,d). By fitting each line-cut of panel~(c) (just like in panel~(d)), we can extract the transmissions of the nanowire subbands across the full measurement range (Fig.~\ref{fig:FigureSI_JJ2_01}e). The closing of the superconducting gap and the suppression of the switching current with the magnetic field aligned along three perpendicular orientations are shown in Fig.~\ref{fig:FigureSI_JJ2_02} and Fig.~\ref{fig:FigureSI_JJ2_03}, respectively.
\paragraph*{\textbf{Device 3}}
Device~3 differs from the first two samples by having a significantly thinner Al shell. To protect the thin film from oxidation, the device is capped in-situ with a $20\,$nm Al$_2$O$_3$ layer. This results in a large zero-field switching current of more than $50\,$nA (Fig.~\ref{fig:FigureSI_JJ3_02}) and a critical magnetic field of $\sim 2\,$T (Fig.~\ref{fig:FigureSI_JJ3_01}).
\paragraph*{\textbf{Device 4}}
In this nano-flake device, the normal-state current manifests sharp steps and the differential conductance features quantized plateaus owing to ballistic transport in the junction (Figs.~\ref{fig:FigureSI_JJ4_01}a,b). The presence of a moderate supercurrent (Fig.~\ref{fig:FigureSI_JJ4_01}c) demonstrates that our fabrication technique can be used not only to proximitize one-dimensional nanowires, but also other types of nanostructures such as quasi-two-dimensional flakes.

\newpage
\section{Modelling of Andreev Transport}
\label{sec:MAR_model}
\subsection{Modelling of the Conductance of a Biased Josephson Junction and the Fitting Procedure}
\begin{figure}[h!]
\center
\includegraphics[width=0.5\linewidth]{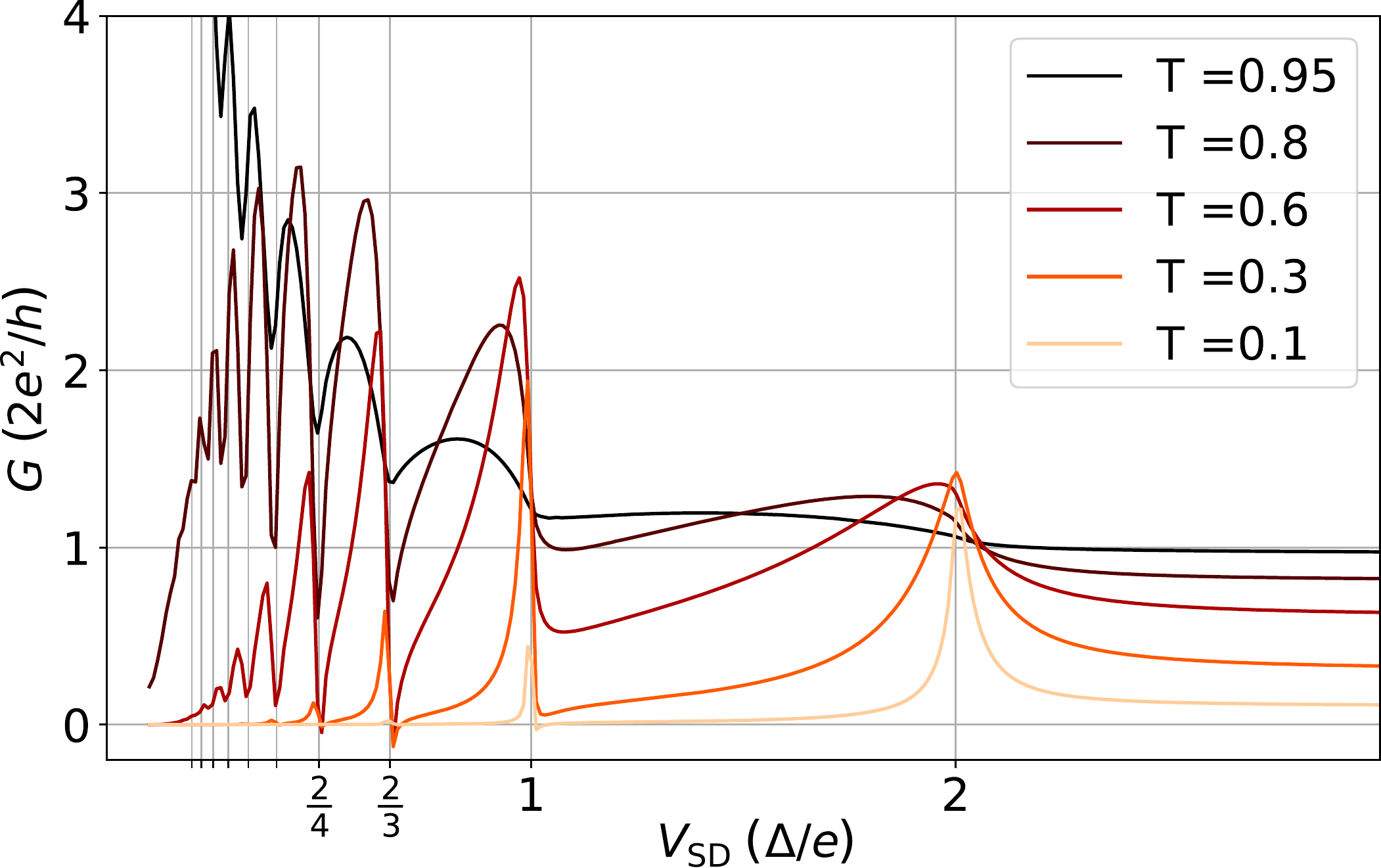}
\caption{Conductance, $G$, of a single-mode Josephson junction versus bias voltage, $V_{\mathrm{SD}}$, for five transparencies ($T$) of the normal region.}
\label{fig:mar}
\end{figure}

\noindent We calculate the conductance of a voltage-biased Josephson junction following the approach of ref.~\cite{averin_ac_1995}. In the model, we account for the electrons and holes propagating through the normal region of the junction with the transparency $T$. The quasiparticles are accelerated by the voltage $V_{\mathrm{SD}}$ applied to the structure and are Andreev reflected at the superconducting leads with the induced superconducting gap $\Delta_{\mathrm{ind}}$. The sequential Andreev reflections imprint the conductance with the subgap features appearing at $V_{\mathrm{SD}} = 2\Delta_{\mathrm{ind}}/Ne$, where $N$ is integer -- see Fig.~\ref{fig:mar}.

\noindent For the analysis of the experimental conductance traces we estimate the total conductance $G_{\mathrm{theory}}(V_{\mathrm{SD}})$ of a multimode nanowire Josephson junction as a sum of $M$ single-mode contributions resulting from the presence of $M$ modes of the transverse quantization~\cite{bardas_electron_1997}:
\begin{equation}
G_{\mathrm{theory}}\left(V_{\mathrm{SD}}\right) = \sum_{i=1}^{M}G_{i}\left(V_{\mathrm{SD}}, T_i, \Delta_{\mathrm{ind}}\right),
\end{equation} 
where $T_i$ is the transmission probability for the $i$'th mode.\\
We obtain $T_i$ and $\Delta_{\mathrm{ind}}$ (induced in the nanowire by the presence of the Al shell) by fitting the numerically calculated conductance to the experimental one by minimizing $\chi = \int[G_\mathrm{{exp}}(V_{\mathrm{SD}})-G_{\mathrm{theory}}(V_{\mathrm{SD}})]^2 dV_{\mathrm{SD}}$. $M$ is a free parameter of the fitting procedure and it is chosen as the smallest number for which at least one of the parameters $T_i$ is zero.

\subsection{Theory for Multiple Andreev Reflections in the Presence of Subgap States}
\noindent The original theory developed in ref.~\cite{averin_ac_1995} assumes a bulk superconducting density of states in the leads. To account for different properties of the two leads, especially the presence of subgap states in one of the contacts, we extend this theory as follows.\\
We consider a Josephson junction consisting of two superconducting electrodes connected through a normal scattering region. We assume that the first contact is kept at zero voltage, while the second one is biased at $V_{\mathrm{SD}}$.\\
In the normal region, adjacent to the $L$'th lead, the quasiparticle wave function takes the form,
\begin{equation}
\Psi_L = \sum_{n} \left[ \left(
  \begin{array}{c}
    A_{n}^L\\
    B_{n}^L
  \end{array} \right) e^{ikx}+
    \left(\begin{array}{c}
    C_{n}^L\\
    D_{n}^L
  \end{array} \right) e^{-ikx}
  \right]e^{-i\left[E+neV_{\mathrm{SD}}\right]t/\hbar},
\end{equation}
where $A_{n}^L$, $C_{n}^L$ ($B_{n}^L$, $D_{n}^L$) correspond to the electron (hole) amplitudes, the time dependence stems from the voltage applied to the leads and $x$ points in the direction opposite to the scattering region.\\
We describe the scattering properties of the normal region by the scattering matrix:
\begin{equation}
S_{0}=\left(\begin{array}{cc}
    r & t \\
    t & -r \\
  \end{array}\right),
\label{smat}
\end{equation}
which sets the transmission probability through the scattering region with the transmission amplitude $t=\sqrt{T}$ and the reflection amplitude $r=\sqrt{1-T}$.

\noindent We rely on the short-junction approximation and use the energy-independent $S_0$ to setup the matching conditions for the wave functions $\Psi_L$. The electron and hole coefficients are related by:
\begin{equation}
\left(\begin{array}{c}
    A_{n}^\text{I}\\
    A_{n+1}^{\text{II}}\\
  \end{array}\right) = S_{0} \left(
  \begin{array}{c}
    C_{n}^\text{I}\\
    C_{n+1}^\text{II}\\
  \end{array} \right),
\label{sle1}
\end{equation}
and
\begin{equation}
\left(\begin{array}{c}
    D_{n}^\text{I}\\
    D_{n-1}^\text{II}\\
  \end{array}\right) = S_{0} ^*\left(
  \begin{array}{c}
    B_{n}^\text{I}\\
    B_{n-1}^\text{II}\\
  \end{array} \right),
\label{slh1}
\end{equation}
respectively. The shifts of the indexes correspond to the changes of quasiparticle energies due to the bias voltage.\\
At each superconductor--normal-conductor interface we take into account the Andreev reflection:
\begin{equation}
\left(\begin{array}{c}
    C_{n}^L\\
    B_{n}^L
  \end{array}\right)
  = \left(\begin{array}{c c}
    a_{n} & 0\\
    0 & a_{n}
  \end{array} \right)
  \left(
  \begin{array}{c}
    D_{n}^L\\
    A_{n}^L
  \end{array} \right),
\label{AA}
\end{equation}
with the amplitude $a_{n}\equiv a(E+neV_{\mathrm{SD}})$, where,
\begin{small}
\begin{equation}
a\left(E\right) = \frac{1-\delta_{L,I}\Gamma\left(E\right)}{\Delta_{\mathrm{ind}}}\left\{
  \begin{array}{l l}
    E - \mathrm{sgn}\left(E\right)\sqrt{E^2-\Delta_{\mathrm{ind}}^2} & \quad |E|>\Delta_{\mathrm{ind}}\\
    E - i \sqrt{\Delta_{\mathrm{ind}}^2-E^2} & \quad |E|\leq \Delta_{\mathrm{ind}}.
  \end{array} \right.
\end{equation}
\end{small}

\noindent The Andreev reflection amplitude is modified by the factor $[1-\delta_{L,I}\Gamma(E)]$ where,
\begin{equation}
\Gamma\left(E\right) = \frac{\gamma^2}{\left(E \pm E_{\mathrm{0}}\right)^2 + \gamma^2},
\end{equation}
is the Lorentzian distribution that accounts for absorption of the quasiparticles in the subgap states (with the energy $\pm E_0$) in the first lead. We set $\gamma=4\,\upmu$eV. 

\begin{figure}[ht!]
\center
\includegraphics[width=0.5\textwidth]{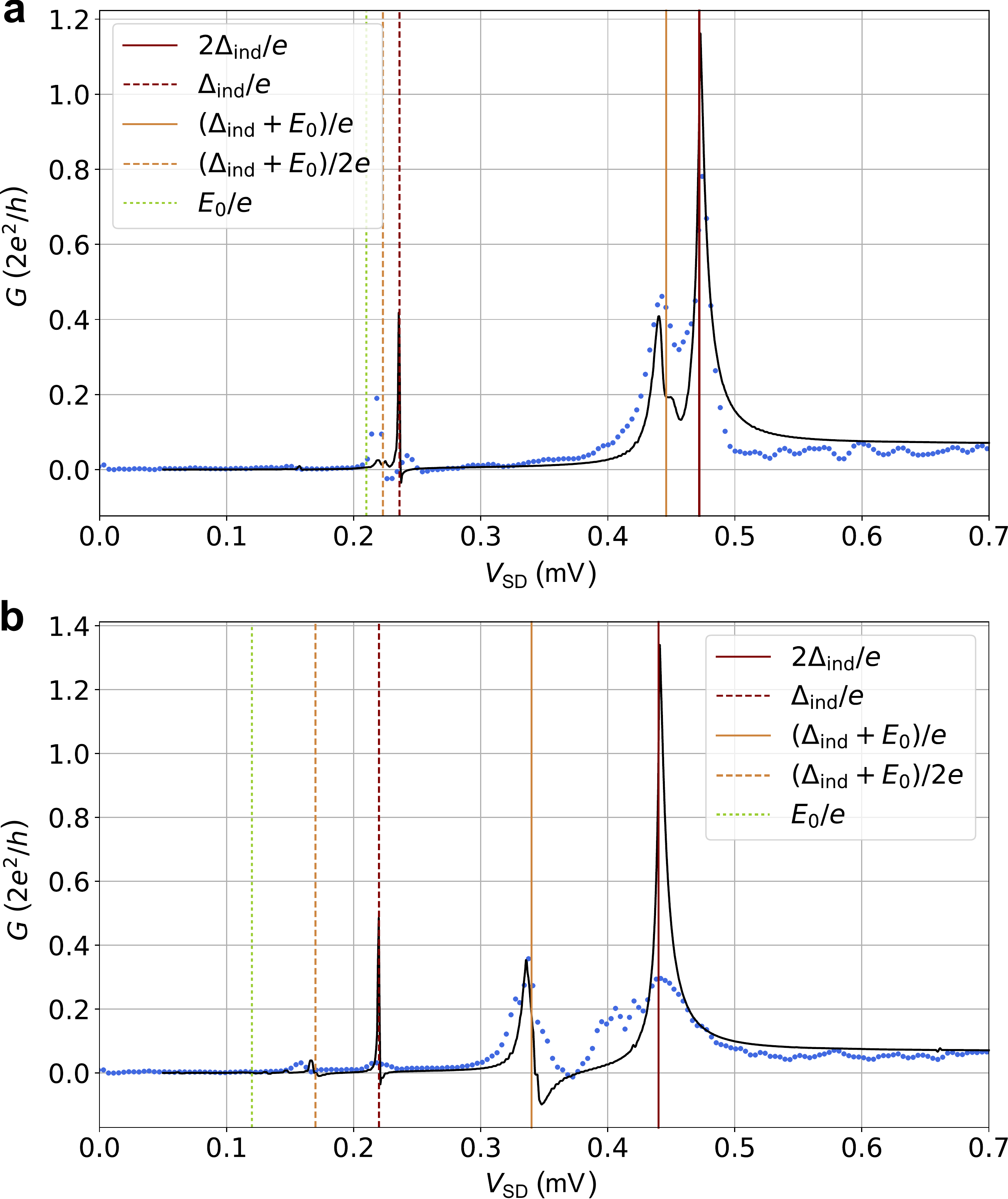}
\caption{Experimental (blue dots) and theoretical (black curves) conductance traces of a Josephson junction with the subgap states in one of the superconducting leads. \textbf{a} is for $B=0$ and \textbf{b} is for $B=0.2\,$T.}
\label{fig:subgap_mar}
\end{figure}

\noindent The electronic excitations in the normal part of the junction originate from the quasiparticles incoming from the nearby superconducting contacts. We therefore write down equation~\eqref{AA} including the quasiparticle source terms~\cite{nowak_supercurrent_2019}:
\begin{equation}
\begin{split}
\left(\begin{array}{c}
    C_{n}^{L}\\
    B_{n}^{L}
  \end{array}\right)
  =& \left(\begin{array}{c c}
    a_{n} & 0\\
    0 & a_{n}
  \end{array} \right)
  \left(
  \begin{array}{c}
    D_{n}^{L}\\
    A_{n}^{L}
  \end{array} \right)\\&+
  \left( \begin{array}{c}
    J\left(E+eV_L\right)\\
    0
  \end{array} \right)\frac{1}{\sqrt{2}}\delta_{p,e}\delta_{s,L}\kappa_{L}^+\\&+
  \left(\begin{array}{c}
    0\\
    J\left(E-eV_L\right)
  \end{array} \right)\frac{1}{\sqrt{2}}\delta_{p,h}\delta_{s,L}\kappa_{L}^-,
\end{split}
\label{AA_final}
\end{equation}
with $J(E)=\sqrt{[1-a(E)^2]F_D(E)}$, where $F_D(E,T=30\;\mathrm{mK})$ is the Fermi distribution. In equation~\eqref{AA_final} $p$ sets the injected quasiparticle type, $s$ determines the lead in which we consider the source term, and 
$\kappa_1^\pm = \delta_{n,0}$, $\kappa_2^\pm = \delta_{n,\pm1}$ keep track of the quasiparticle energy shifts due to the bias.\\
We calculate the current $I^L$ in the $L$'th lead as:
\begin{equation}
I^{L}=\sum_{\imath = -I_\text{max}}^{I_\text{max}} I^{L}_{\imath} e^{\imath V_{\mathrm{SD}} eit/\hbar},
\label{total_current_formula}
\end{equation}
with the Fourier components,
\begin{equation}
I_{\imath}^{L}=\frac{e}{\hbar\pi}\sum_{s=1,2}\sum_{p=e,h}\int_{-\infty}^{\infty}dE\sum_{n = -N_\text{max}}^{N_\text{max}}(\mathbf{U}^{L*}_{\imath+n}\mathbf{U}^L_{n}
-\mathbf{V}^{L*}_{\imath+n}\mathbf{V}^L_{n}).
\label{CII}
\end{equation}
$\mathbf{U}^L_{n}=\left(A_{n}^L,B_{n}^L\right)^T$ and $\mathbf{V}^L_{n}=\left(C_{n}^L,D_{n}^L\right)^T$ are vectors that consist of the electron and hole amplitudes. The DC current is obtained for $\imath = 0$ and subsequently used to calculate the conductance, $G = dI^L/dV_{\mathrm{SD}}$. To efficiently sample the non-uniform conductance trace we use the Adaptive package~[\onlinecite{Nijholt2019}].

\begin{table}[h]
\begin{tabular}{|c|c|c|c|}
\hline
$\bm{B_{||}}$ (T) & $\bm{\Delta_{\mathrm{ind}}\;(\upmu\mathrm{eV})}$ & $\bm{T_1}$ & $\bm{E_0\;(\upmu\mathrm{eV})}$\\
\hline\hline
0 & 236 & 0.065 & 210\\
\hline
0.2 & 220 & 0.065 & 120\\
\hline
\end{tabular}
\caption{Parameters used for the calculation of the conductance traces in Fig.~\ref{fig:subgap_mar}.}
\label{tab:subgap_parameters}
\end{table}

\noindent In Fig.~\ref{fig:subgap_mar}, we show the calculated MAR conductance traces (black curves) together with two cross-sections (blue dots) from the experimental map in Fig.~3c of the main text. We focus here on two cases: $B=0$ and $B=0.2\,$T with the parameters used for the calculations given in the first and second row of Table~\ref{tab:subgap_parameters}, respectively. The calculated traces agree qualitatively well with the data: they capture the peak positions and the overall line shape. In particular, we observe two ordinary MAR peaks at $V_{\mathrm{SD}} = 2\Delta_{\mathrm{ind}}/Ne$ with $N=1,2$ and two peaks induced by the presence of the subgap state at $V_{\mathrm{SD}} = (\Delta_{\mathrm{ind}}+E_0)/Ne$ with $N=1,2$. The increase of the magnetic field significantly alters the energy of the subgap state causing a further splitting between the MAR and the subgap-induced peaks. Nevertheless, the low-energy transport at higher magnetic fields shown in the conductance map in Fig.~3c of the main text, with multiple states detaching from the gap edge, goes beyond the approximations of our model.

\newpage
\section{Additional Transport Measurements in Normal-Metal--Superconductor Junctions}
\subsection{Calibrating the AC conductance}
\begin{figure*}[hbt!]
\centering
\includegraphics[width=0.55\linewidth]{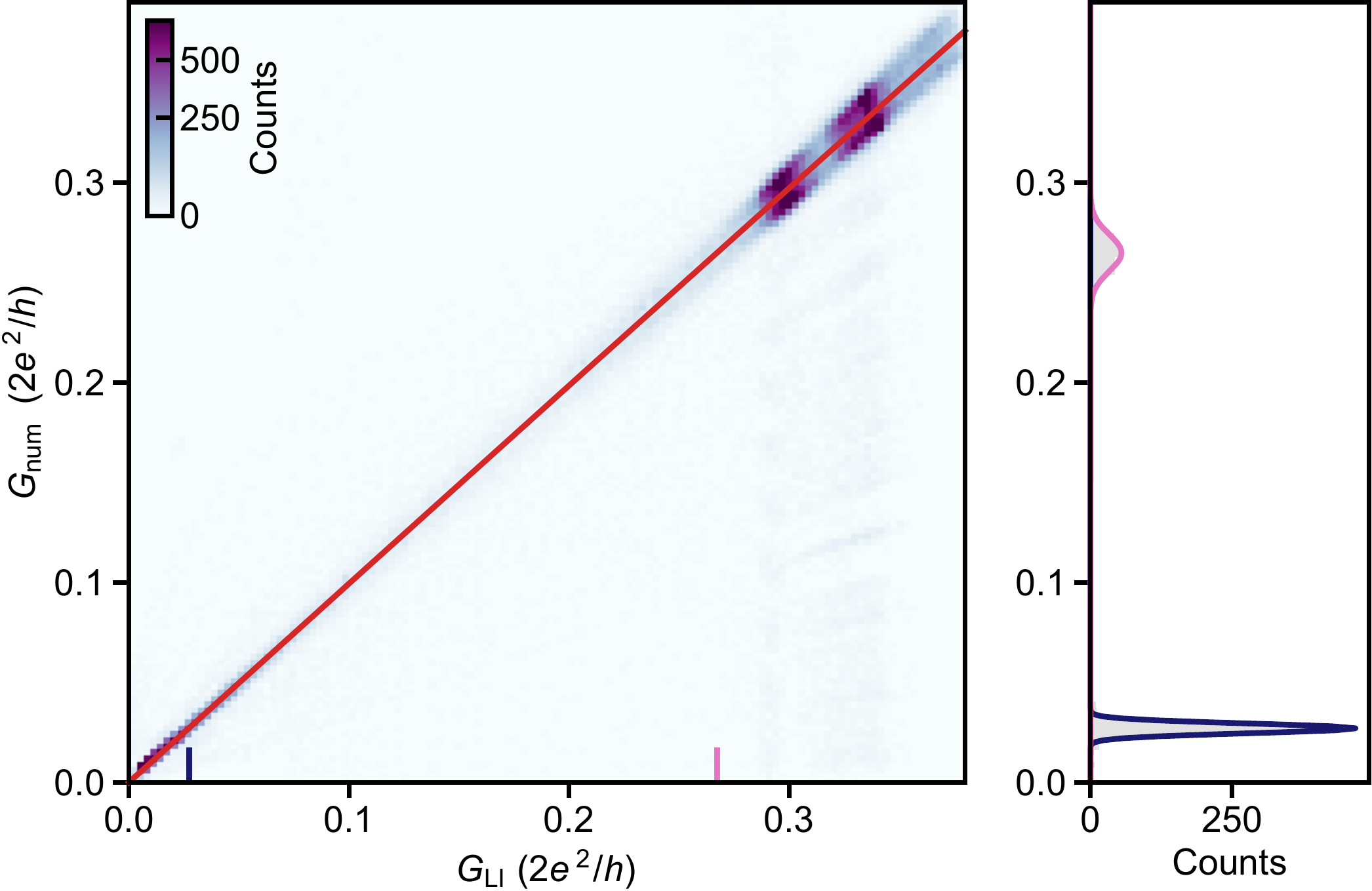}
\caption{Calibration function extracted from the conductance data of the N--S junction presented in Fig.~4 of the main text. Numerical differential conductance, $G_{\mathrm{num}}$, vs.\ AC differential conductance, $G_{\mathrm{LI}}$. Here, the lock-in frequency is $f=23\,$Hz.}
\label{fig:SI_calibration_NS}
\end{figure*}

\begin{figure*}[hbt!]
\centering
\includegraphics[width=0.55\linewidth]{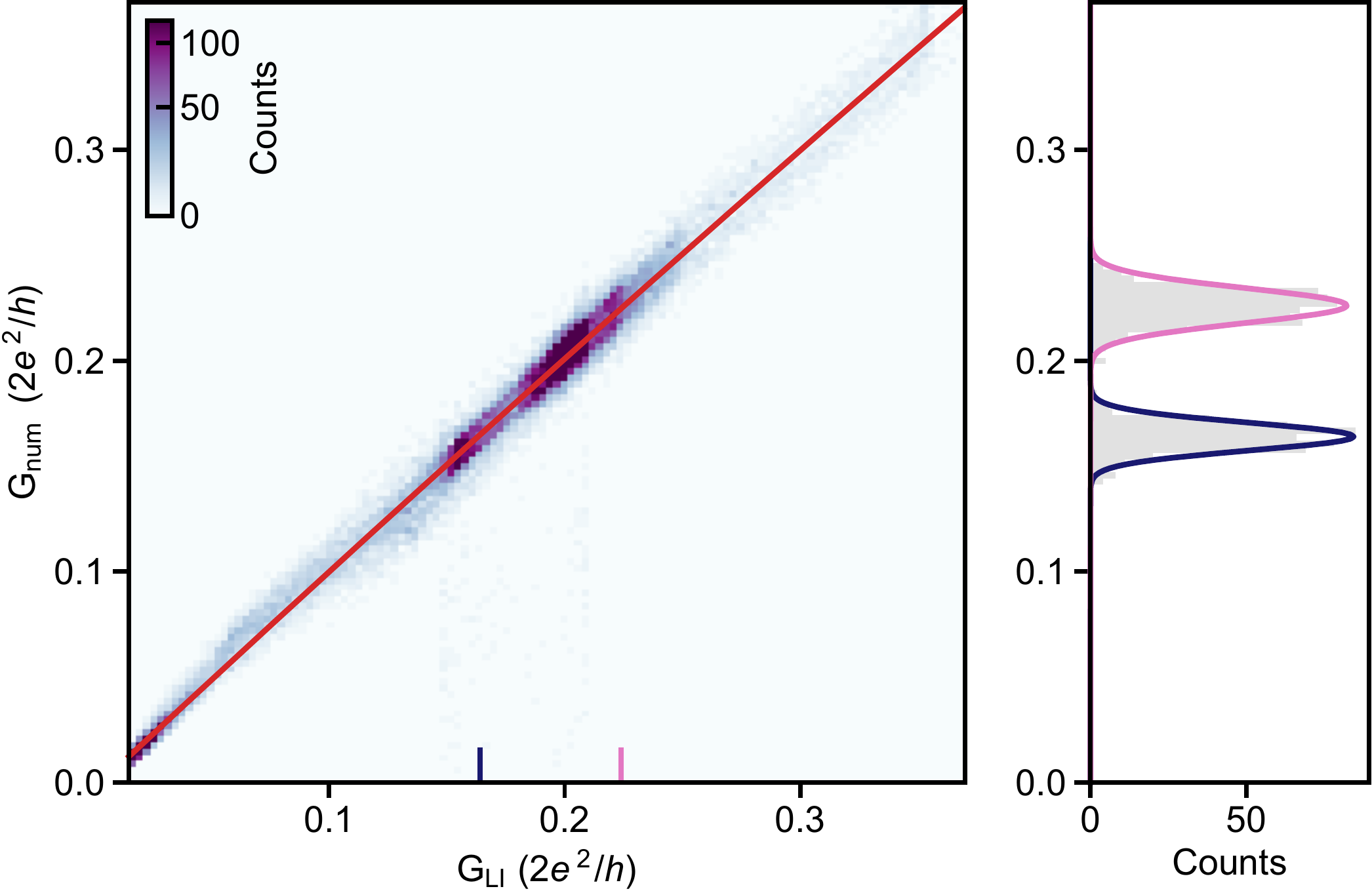}
\caption{Calibration function extracted from the conductance data of the N--S junction presented in Fig.~5e of the main text. Numerical differential conductance, $G_{\mathrm{num}}$, vs.\ AC differential conductance, $G_{\mathrm{LI}}$. Here, the lock-in frequency is $f=23\,$Hz.}
\label{fig:SI_calibration_corr_left}
\end{figure*}

\begin{figure*}[hbt!]
\centering
\includegraphics[width=1.0\linewidth]{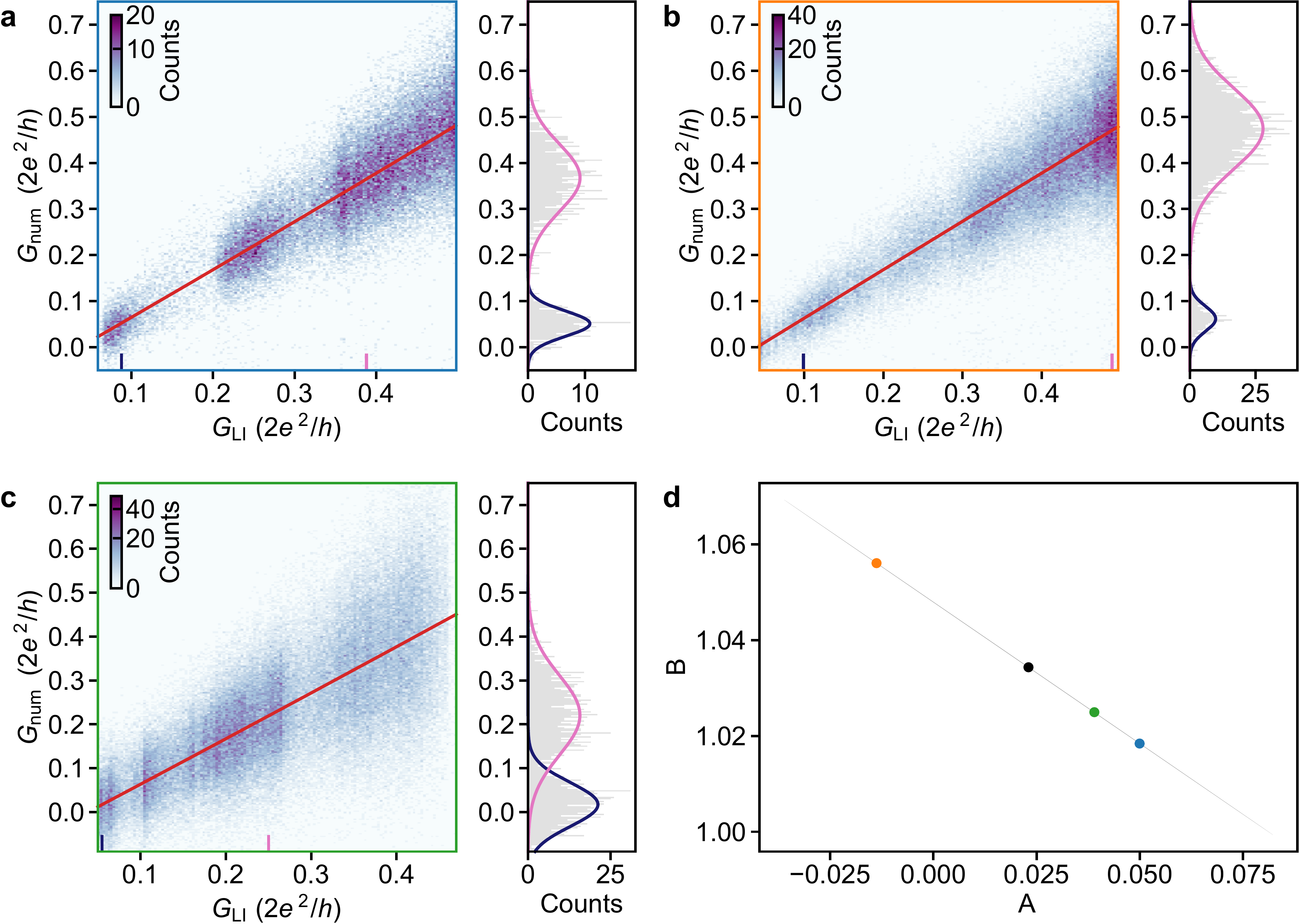}
\caption{Calibration functions extracted for the N--S junction presented in Fig.~5f of the main text. \textbf{a}-\textbf{c}~Numerical differential conductance, $G_{\mathrm{num}}$, vs.\ AC differential conductance, $G_{\mathrm{LI}}$. Each of the three panels is from separate data set. Here, the lock-in frequency is $f=72\,$Hz. \textbf{d}~Summary of the parabolic ($A$) and linear ($B$) fit parameters in panels~(\textbf{a}-\textbf{c}). The colors of the data points correspond to the axis colors of the respective panels. The black data point denotes the weighted average fit parameters, where the weights are determined by the residuals of the individual fits. The grey area designates the $95\%$ confidence interval.}
\label{fig:SI_calibration_corr_right}
\end{figure*}

\noindent The AC conductance is measured using a standard low-frequency lock-in technique. Some of the employed current-to-voltage amplifiers have been found to suffer from a relatively low bandwidth. This required a recalibration of the measured differential conductance of the N--S devices. The approach shown here is similar to a calibration procedure developed by Jouri Bommer, Guanzhong Wang and Michiel de Moor (see also guidelines on lock-in measurements by the same authors: \url{http://homepage.tudelft.nl/q40r9/lockin-meas-guide-v20200603.pdf}).\\
Fig.~\ref{fig:SI_calibration_NS} shows the raw conductance data from Fig.~4 of the main text prior to the subtraction of any series resistance. For the mapping of the lock-in conductance, $G_{\mathrm{LI}}$, to the numerical DC conductance, $G_{\mathrm{num}}$, the data is binned into a two-dimensional histogram (resolution $0.003 \cdot 2e^2/h$). Since the numerical conductance suffers from noise, we determine the centre of the distribution for each bin of $G_{\mathrm{LI}}$ by fitting a Gaussian distribution to the histogram of $G_{\mathrm{num}}$ (see right panel of Fig.~\ref{fig:SI_calibration_NS}). Data points that are more than $5$ standard deviations from the centre of the distributions are discarded as outliers. Here, the mapping yields the parametrization $G_{\mathrm{num}} = -0.016 \cdot G_{\mathrm{LI}}^2 + 0.995 \cdot G_{\mathrm{LI}}$.
~\\
Fig.~\ref{fig:SI_calibration_corr_left} shows the calibration for the left junction of the correlation device in Fig.~5e of the main text. In Fig.~\ref{fig:SI_calibration_corr_right}, the calibration is presented for the right junction of the correlation device in Fig.~5f of the main text. This is the only device that was measured at a relatively large lock-in frequency ($f=72\,$Hz). The right panels of Fig.~\ref{fig:SI_calibration_corr_left} and Figs.~\ref{fig:SI_calibration_corr_right}a-c show exemplary fits of the histograms using a Gaussian. The red traces represent the fitting by the least-squares method using the polynomial regression function $G_{\mathrm{num}} = A \cdot G_{\mathrm{LI}}^2 + B \cdot G_{\mathrm{LI}} + C$. The mapping in Fig.~\ref{fig:SI_calibration_corr_left} yields the parametrization $G_{\mathrm{num}} = -0.108 \cdot G_{\mathrm{LI}}^2 + 1.043 \cdot G_{\mathrm{LI}} - 0.003$. In Fig.~\ref{fig:SI_calibration_corr_right}, the weighted average of the fitting functions yields the mapping function $G_{\mathrm{num}} = 0.023 \cdot G_{\mathrm{LI}}^2 + 1.034 \cdot G_{\mathrm{LI}} - 0.040$, where the residuals of the individual measurements provide the weights. Fig.~\ref{fig:SI_calibration_corr_right}d summarizes the parabolic ($A$) and linear ($B$) fit parameters from Figs.~\ref{fig:SI_calibration_corr_right}a-c. The black data point indicates the weighted average of the fit parameters.

\clearpage
\newpage
\subsection{N--S Junction Spectroscopy}
\begin{figure*}[hbt!]
\centering
\includegraphics[width=0.6\linewidth]{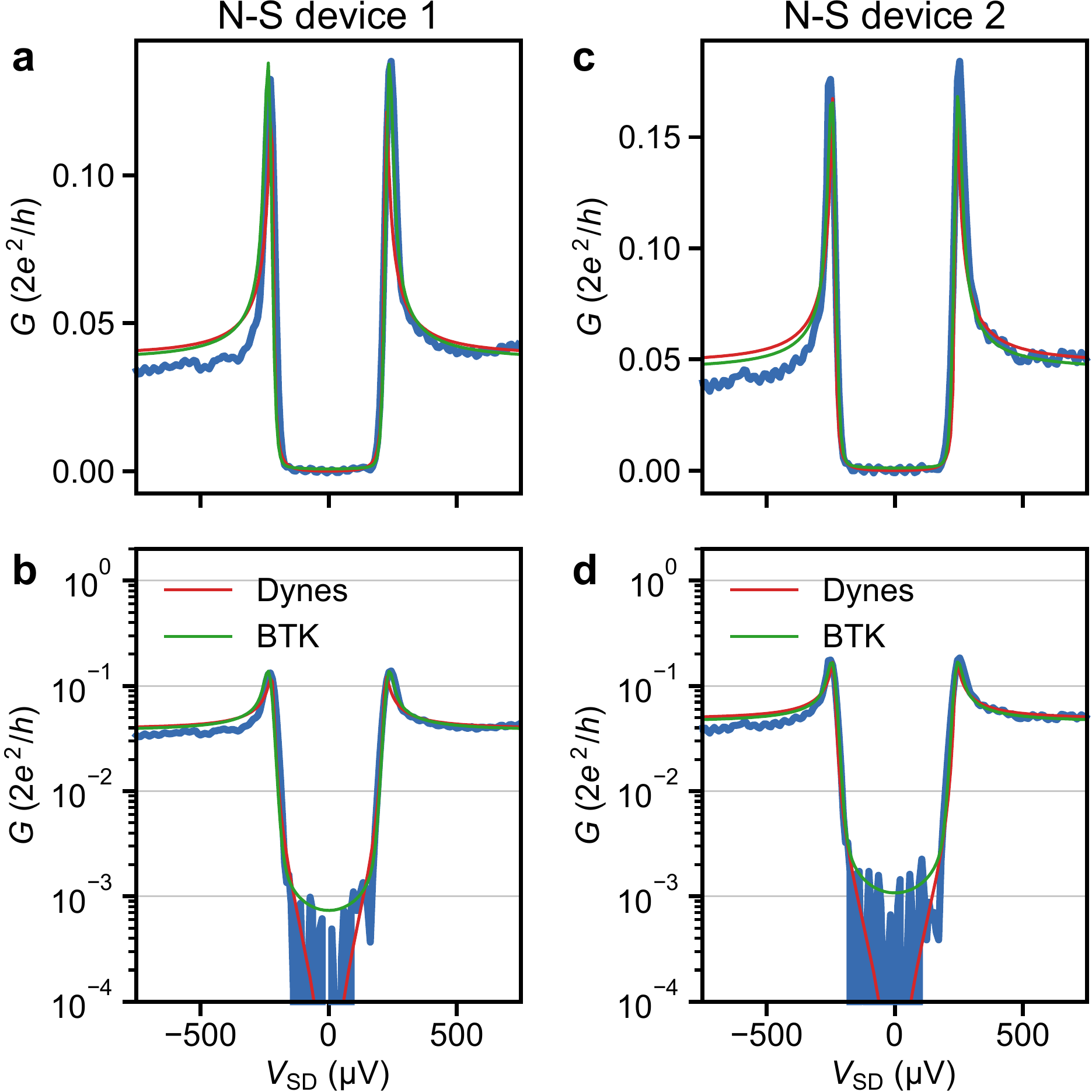}
\caption{\textbf{a}, \textbf{b}~Differential conductance vs.\ bias-voltage line-cuts from N--S device~1 (same as in Fig.~4 of the main text) on a linear scale in (\textbf{a}) and on a logarithmic scale in (\textbf{b}). Here, the tunnel-gate voltage is $V_{\mathrm{TG}}=0.530\,$V and the super-gate voltage is $V_{\mathrm{SG}}=0\,$V. \textbf{c}, \textbf{d}~Differential conductance vs.\ bias-voltage line-cuts from N--S device~2 (not presented in the main text) on a linear scale in (\textbf{c}) and on a logarithmic scale in (\textbf{d}). Here, the tunnel-gate voltage is $V_{\mathrm{TG}}=2.004\,$V and the super-gate voltage is $V_{\mathrm{SG}}=7.0\,$V. The fit of the BCS--Dynes term and of the BTK model are shown in red and green, respectively.}
\label{fig:SI_hard_gap}
\end{figure*}

\noindent Deep in the tunnelling regime the subgap conductance is strongly suppressed. As illustrated in Fig.~\ref{fig:SI_hard_gap}, the ratio of the above-gap conductance and the subgap conductance is approximately a factor of $100$. In Figs.~\ref{fig:SI_hard_gap}a,b the differential conductance line-cuts from N--S device~1 (i.e.\ the same device as in Fig.~4 of the main text) are fitted using the BCS--Dynes term (red) and the BTK model (green). The data in Figs.~\ref{fig:SI_hard_gap}c,d show line-cuts from another N--S junction (device~2), which is not presented in the main text. The fitting parameters in the BTK model are the induced gap, $\Delta_{\mathrm{ind}}$, the normal-state conductance, $G_{\mathrm{N}}$, and the temperature, $T$. For device~1 it yields an induced gap of $\Delta_{\mathrm{ind}}=231\,\upmu$eV and for device~2 the extracted gap is $\Delta_{\mathrm{ind}}=241\,\upmu$eV. In the BTK model the only effective broadening parameter is the temperature, which for both devices yields $T\approx95\,$mK.

\newpage
\subsection{Temperature Dependence of the Induced Gap}
\begin{figure*}[hbt!]
\centering
\includegraphics[width=0.6\linewidth]{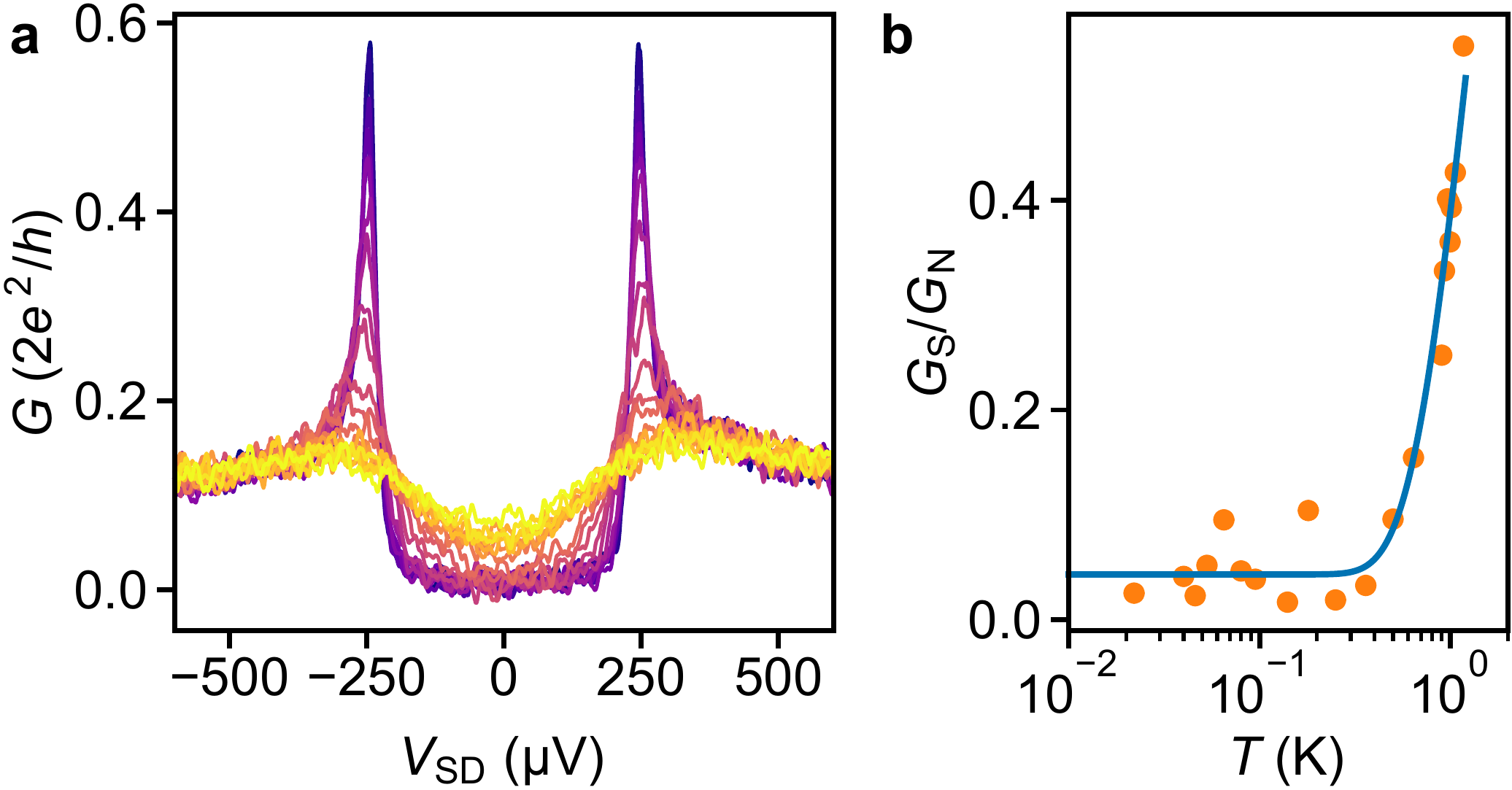}
\caption{Temperature dependence of the induced gap (N--S device~3). \textbf{a}~Tunnelling conductance vs.\ bias voltage between $T=18\,$mK (purple) and $T=1.17\,$K (yellow). \textbf{b}~Subgap conductance averaged between $V_{\mathrm{SD}}=\pm25\upmu$V ($G_{\mathrm{S}}$) divided by the normal-state conductance ($G_{\mathrm{N}}$) as a function of $T$. The blue trace is a fit to equation~\eqref{eq:N-S_Temp}.}
\label{fig:SI_Temp}
\end{figure*}

\noindent In Fig.~\ref{fig:SI_Temp}, we present the temperature dependence from another device (N--S device~3), which is not presented in the main text. In the limit $k_{\mathrm{B}}T \ll \Delta_{\mathrm{ind}}$, the subgap conductance, $G_{\mathrm{S}}$, scales with temperature, $T$, as~\cite{Tinkham1996}
\begin{equation}
  G_{\mathrm{S}}\left(V_{\mathrm{SD}}=0\right)=G_{\mathrm{N}}\sqrt{\frac{2\pi\Delta_{\mathrm{ind}}}{k_{\mathrm{B}}T}}e^{-\Delta_{\mathrm{ind}}/k_{\mathrm{B}}T},
  \label{eq:N-S_Temp}
\end{equation}
where $G_{\mathrm{N}}$ is the normal-state conductance and $k_{\mathrm{B}}$ is the Boltzmann constant. The purple trace in Fig.~\ref{fig:SI_Temp}a measured at $T=18\,$mK is well described by the BTK model with an induced gap of $\Delta_{\mathrm{ind}}=237\,\upmu$eV. This is very similar to the magnitude of the induced gap of the other two N--S devices shown in Fig.~\ref{fig:SI_hard_gap}, albeit those junctions are formed during a separate Al deposition step. The theoretical model in equation~\eqref{eq:N-S_Temp} can describe the smearing of the density of states with temperature. It yields a fit parameter of $\Delta_{\mathrm{ind}}\approx210\,\upmu$eV, which is a bit smaller than the gap directly extracted from the tunnelling spectroscopy.

\newpage
\subsection{Hard Induced Gap}
\begin{figure*}[hbt!]
\centering
\includegraphics[width=0.85\linewidth]{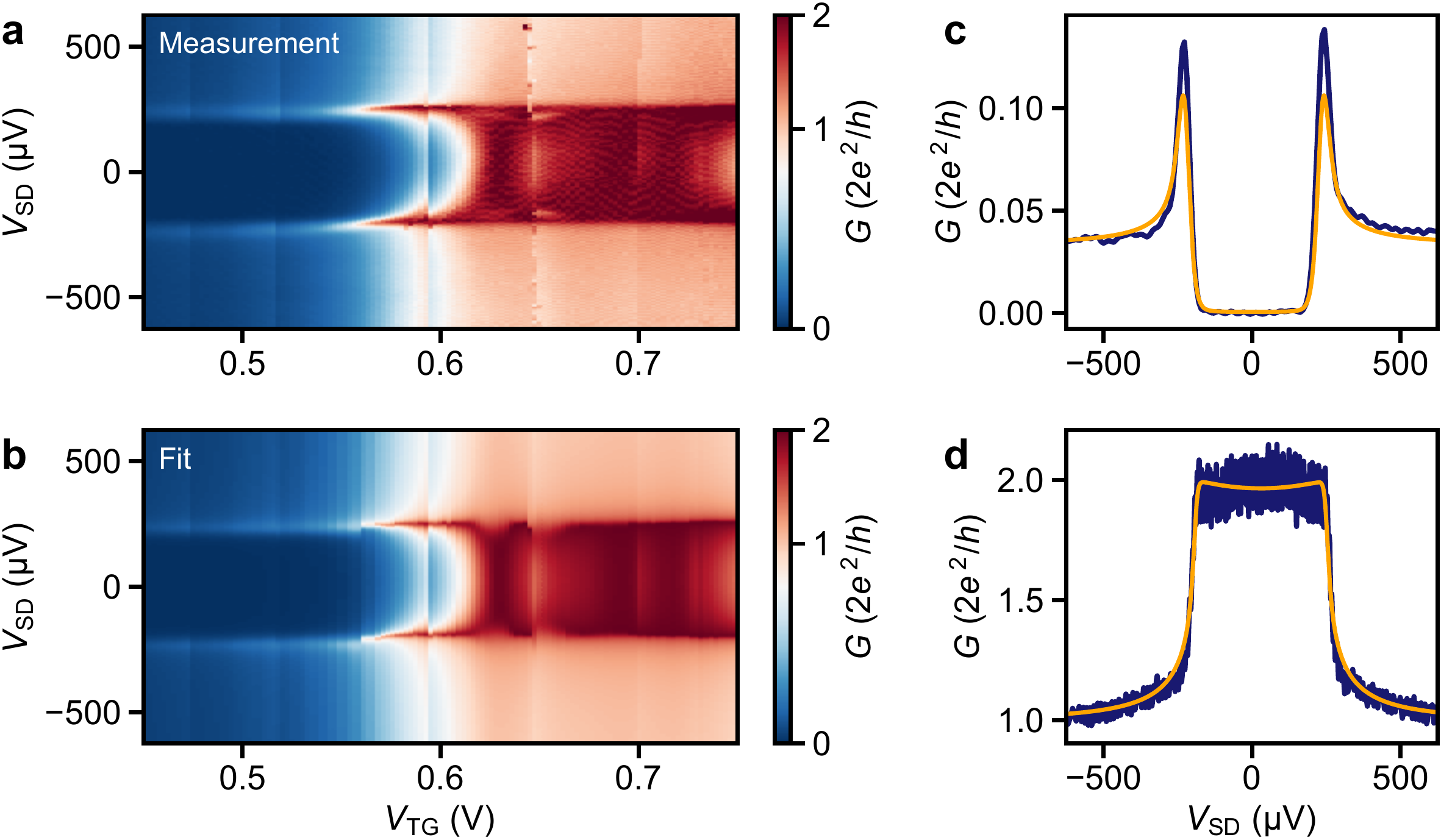}
\caption{N--S junction voltage-bias spectroscopy and the corresponding fit of the BTK model~\cite{Blonder1982} for N--S device~1. \textbf{a}~Differential conductance, $G$, as a function of source--drain voltage, $V_{\mathrm{SD}}$, and bottom tunnel-gate voltage, $V_{\mathrm{TG}}$, from Fig.~4b of the main text. \textbf{b}~Fit of the BTK model to the data set in panel~(\textbf{a}). The fit parameters include the induced gap, the temperature, and the barrier strength $Z$, wich is given by the transmission $(1+Z^2)^{-1}$. \textbf{c}, \textbf{d}~Line-cut of the data in panel~(\textbf{a}) (dark blue) at $V_{\mathrm{TG}}=0.53\,$V and at $V_{\mathrm{TG}}=0.69\,$V, respectively. The orange traces show the corresponding fits to the BTK model.}
\label{fig:SI_BTK}
\end{figure*}

\noindent In Fig.~\ref{fig:SI_BTK}, we report the fit of the BTK model to the data shown in Fig.~4b of the main text (N--S device~1). The extracted induced superconducting gap is $\Delta_{\mathrm{ind}}\sim230\,\upmu$eV.

\newpage
\subsection{Zero-Bias Peaks in the N--S Device}
\begin{figure*}[hbt!]
\centering
\includegraphics[width=0.85\linewidth]{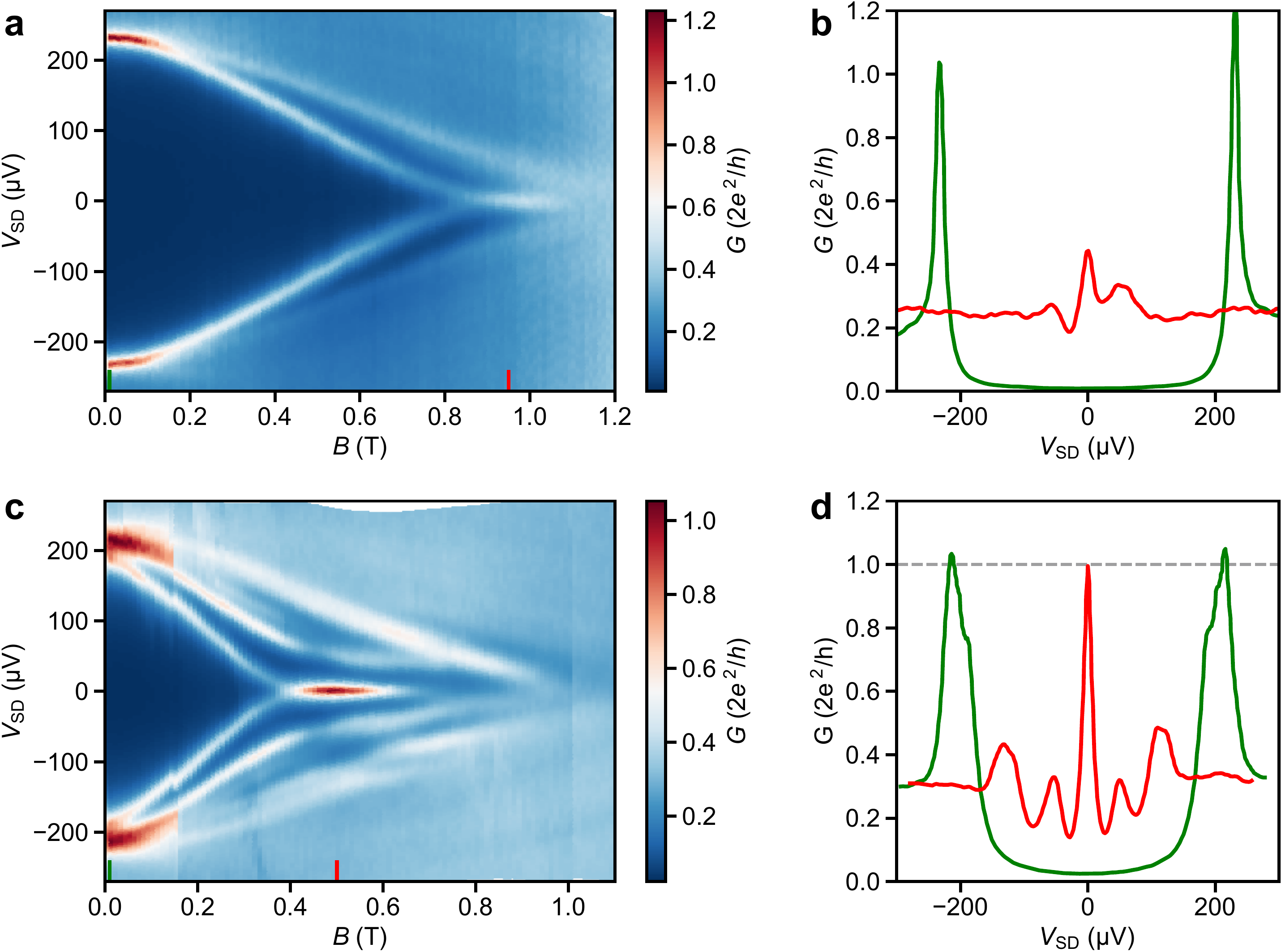}
\caption{Magnetic-field-dependent voltage-bias spectroscopy for N--S device~1 from Fig.~4 of the main text, demonstrating the formation of zero-bias peaks in the differential conductance. \textbf{a}~$G$ as a function of $V_{\mathrm{SD}}$ and $B$. The super-gate voltage is $V_{\mathrm{SG}}=7.5\,$V and the tunnel-gate voltage is $V_{\mathrm{TG}}=0.5\,$V. \textbf{b}~Line-cuts of (\textbf{a}) at the positions indicated by the two lines. \textbf{c}~$G$ as a function of $V_{\mathrm{SD}}$ and $B$. Here, $V_{\mathrm{SG}}=2.97\,$V and $V_{\mathrm{TG}}=0.417\,$V. \textbf{d}~Line-cuts of (\textbf{c}) at the positions indicated by the two lines.}
\label{fig:SI_NS1_ZBP}
\end{figure*}

\noindent In Fig.~\ref{fig:SI_NS1_ZBP}, we present additional data from the first N--S device (cf.\ Fig.~4 of the main text) in a magnetic field for two different super-gate voltages. In the main text, we present ballistic transport and pronounced Andreev enhancement for the same N--S device.

\newpage
\subsection{Zero-Bias Peaks and Super-Gate Dependence}
\begin{figure*}[hbt!]
\centering
\includegraphics[width=0.75\linewidth]{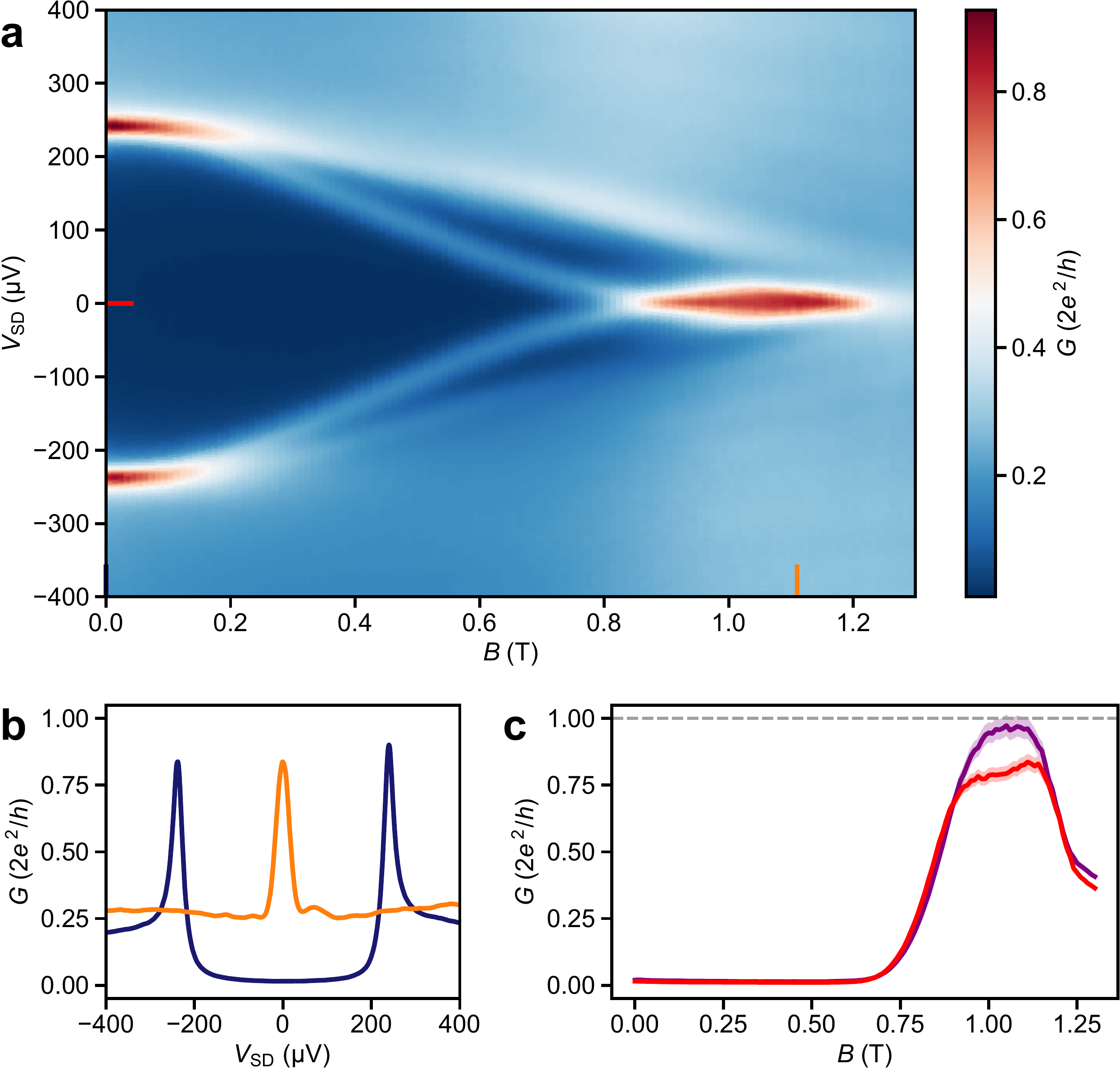}
\caption{Voltage-bias spectroscopy of a subgap state with a large zero-bias peak conductance close to $2e^2/h$ (measured at the left N--S junction of the device presented in Fig.~5 of the main text). Here, the super-gate voltage $V_{\mathrm{SG}}=0.525\,$V. \textbf{a}~Differential conductance, $G$, as a function of the bias voltage at the left terminal, $V_{\mathrm{SD}}$, and the magnetic field along the wire axis. \textbf{b}~Voltage-bias line-cut of the differential conductance at zero field (blue) and at $B=1.11\,$T (orange). \textbf{c}~$G$ vs.\ $B$ line-cuts at $V_{\mathrm{SD}}=0\,\upmu$V from panel~(\textbf{a}) (red, at $V_{\mathrm{SG}}=0.525\,$V) and from Fig.~5e (purple, at $V_{\mathrm{SG}}=0\,$V). The shaded areas behind the solid traces correspond to the variation in conductance assuming an uncertainty of $\pm0.5\,$k$\Omega$ in estimating the actual series resistance.}
\label{fig:SI_NS2_ZBP}
\end{figure*}

\noindent Additional N--S spectroscopy measurements of the left N--S junction of the device presented in Fig.~5 of the main text are shown in Fig.~\ref{fig:SI_NS2_ZBP}. Here, the voltage at the super gate -- the bottom gate controlling the electrochemical potential in the hybrid nanowire segment -- is larger ($V_{\mathrm{SG}}=0.525\,$V vs.\ $0\,$V). The differential conductance vs.\ $V_{\mathrm{SD}}$ and $B$ is depicted in Fig.~\ref{fig:SI_NS2_ZBP}a, the bias-voltage line-cut in Fig.~\ref{fig:SI_NS2_ZBP}b illustrates the pronounced zero-bias conductance peak at large magnetic fields. However, the magnitude of the ZBP conductance depends on the tuning of the tunnel-gate and super-gate voltages (cf.\ Fig.~\ref{fig:SI_NS2_ZBP}c).

\begin{figure*}[hbt!]
\centering
\includegraphics[width=0.75\linewidth]{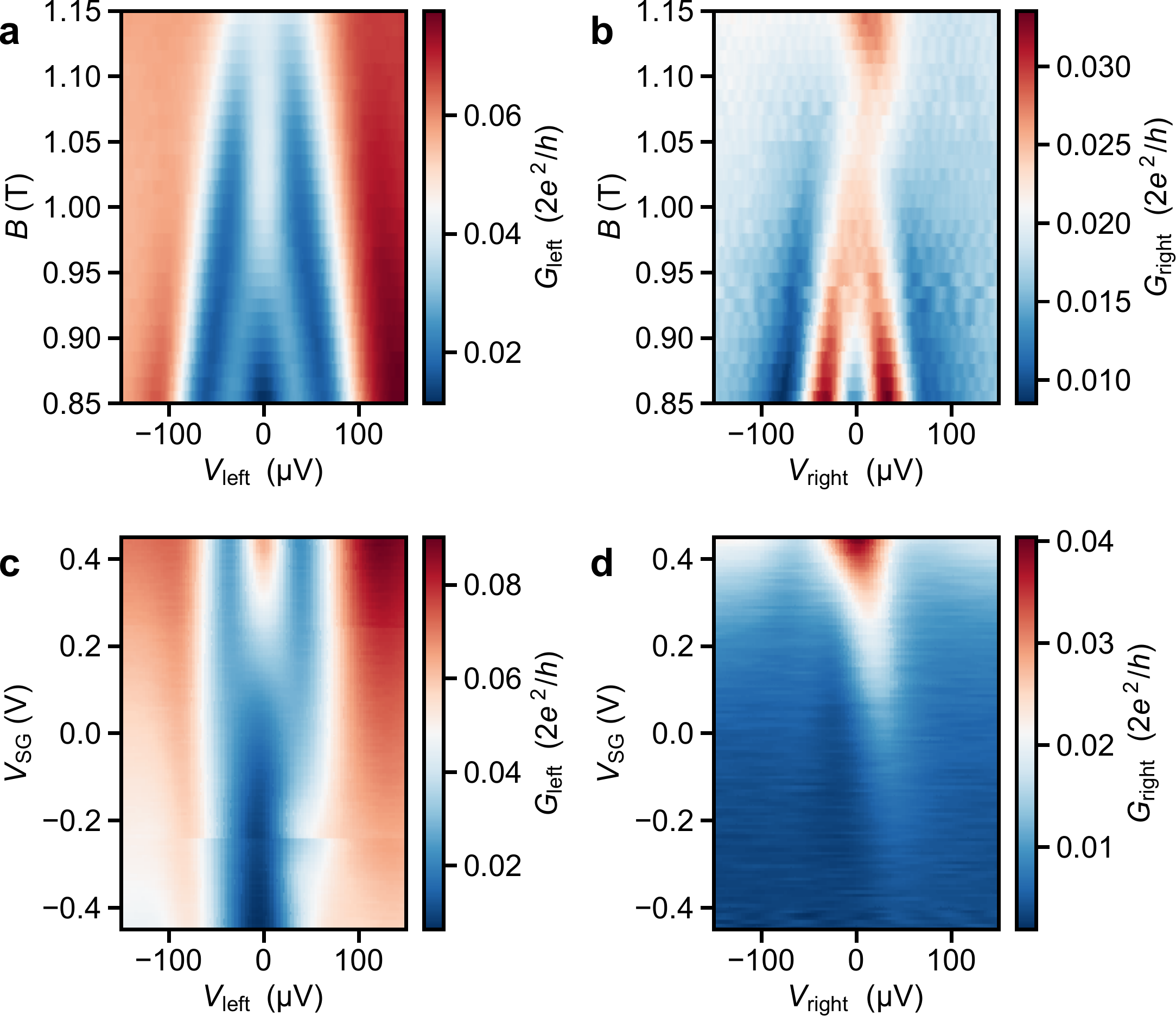}
\caption{Simultaneous appearance of zero-bias peaks on both hybrid boundaries (same device as in Fig.~5 of the main text). The two tunnel gates are set to $V_{\mathrm{TG,left}}=0.47\,$V and $V_{\mathrm{TG,right}}=0.13\,$V. \textbf{a}, \textbf{b}~Differential conductance, $G_{\mathrm{left/right}}$, as a function of magnetic field, $B$, and bias voltage at the left and right terminal, respectively. Here, the super-gate voltage $V_{\mathrm{SG}}=0.525\,$V, i.e.\ identical as for the data in Fig.~\ref{fig:SI_NS2_ZBP}a. \textbf{c}, \textbf{d}~Differential conductance, $G_{\mathrm{left/right}}$, at $B=1.0\,$T as a function of $V_{\mathrm{SG}}$ and bias voltage at the left and right terminal, respectively.}
\label{fig:NS2_sim-ZBP}
\end{figure*}

\begin{figure*}[hbt!]
\centering
\includegraphics[width=0.75\linewidth]{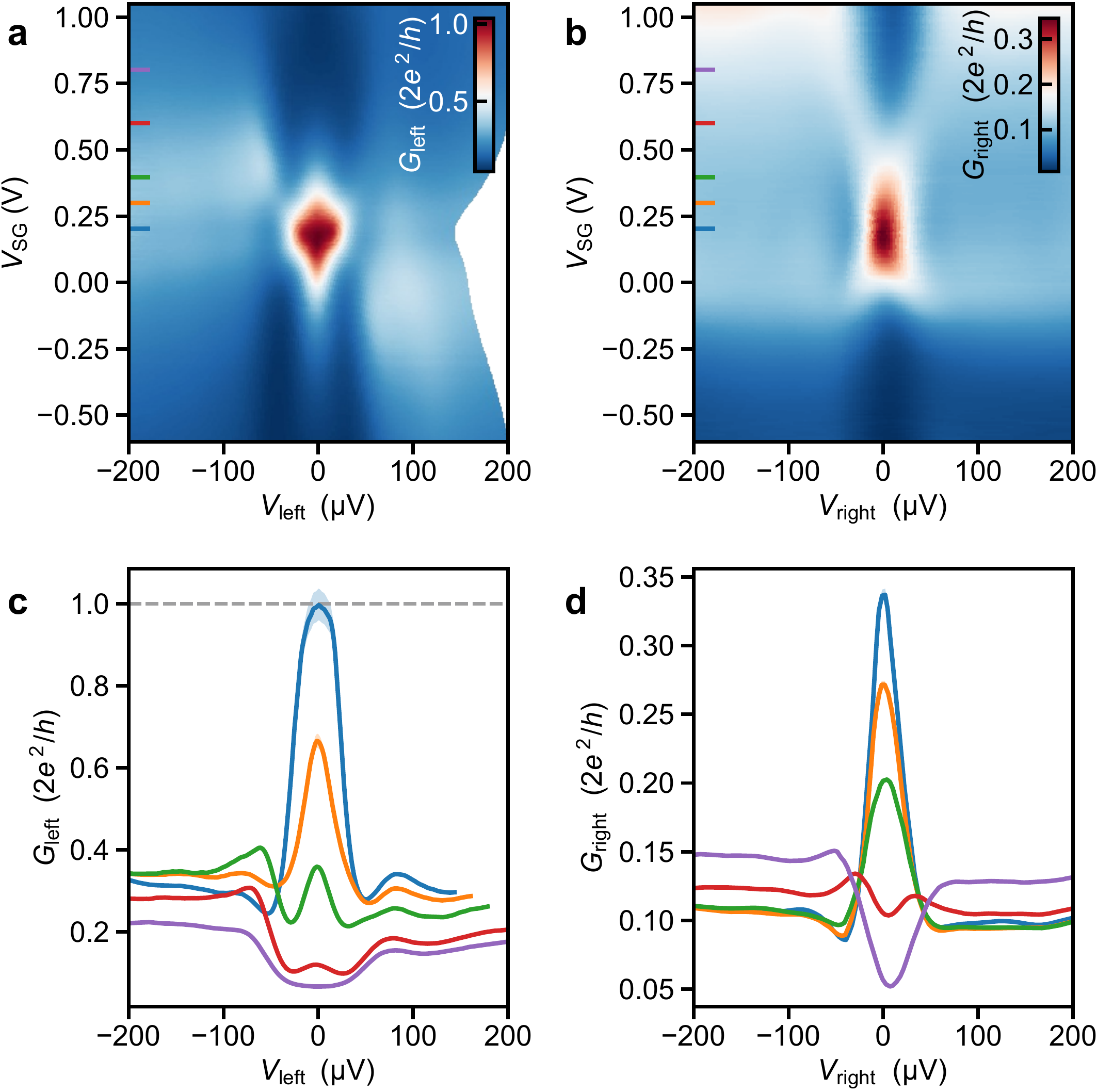}
\caption{Simultaneous appearance of zero-bias peaks on both hybrid boundaries (same device and same tunnel-gate settings as in Fig.~5 of the main text). The two tunnel gates are set to $V_{\mathrm{TG,left}}=0.52\,$V and $V_{\mathrm{TG,right}}=0.21\,$V. \textbf{a}, \textbf{b}~Differential conductance, $G_{\mathrm{left/right}}$, at $B=1.0\,$T as a function of $V_{\mathrm{SG}}$ and bias voltage at the left and right terminal, respectively. \textbf{c}, \textbf{d}~Line-cuts from panels~(\textbf{a}) and (\textbf{b}) at the values of $V_{\mathrm{SG}}$ designated by the coloured lines. The shaded areas behind the solid traces correspond to the variation in conductance assuming an uncertainty of $\pm0.5\,$k$\Omega$ in estimating the actual series resistance.}
\label{fig:SI_SG-dependence}
\end{figure*}

\newpage
\noindent In Fig.~\ref{fig:NS2_sim-ZBP} additional data from the high-field regime are presented (here $B=0.85-1.15\,$T). For the same bottom-gate settings as in Fig.~\ref{fig:SI_NS2_ZBP} we observe ZBPs that emerge concurrently on both boundaries of the superconductor--semiconductor nanowire segment (cf.\ Figs.~\ref{fig:NS2_sim-ZBP}a,b). By fixing the magnetic field at $B=1.0\,$T we can observe the evolution of the ZBPs at the left and right N--S junctions as a function of the voltage on the super gate underneath the hybrid nanowire segment (see Figs.~\ref{fig:NS2_sim-ZBP}c,d). The asymmetry in the conductance of Fig.~\ref{fig:NS2_sim-ZBP}d with respect to bias polarity is related to energy-dependent tunnel barrier transmission at the right N--S junction.\\
The concurrent evolution of the ZBPs on both N--S boundaries of the correlation device as a function of the super-gate voltage is also depicted in Fig.~\ref{fig:SI_SG-dependence} for same tunnel-gate settings as in Fig.~5 of the main text.

\clearpage
\newpage
\section{Realization of Advanced Hybrid Devices}
\begin{figure}[H]
\centering
\includegraphics[width=0.75\linewidth]{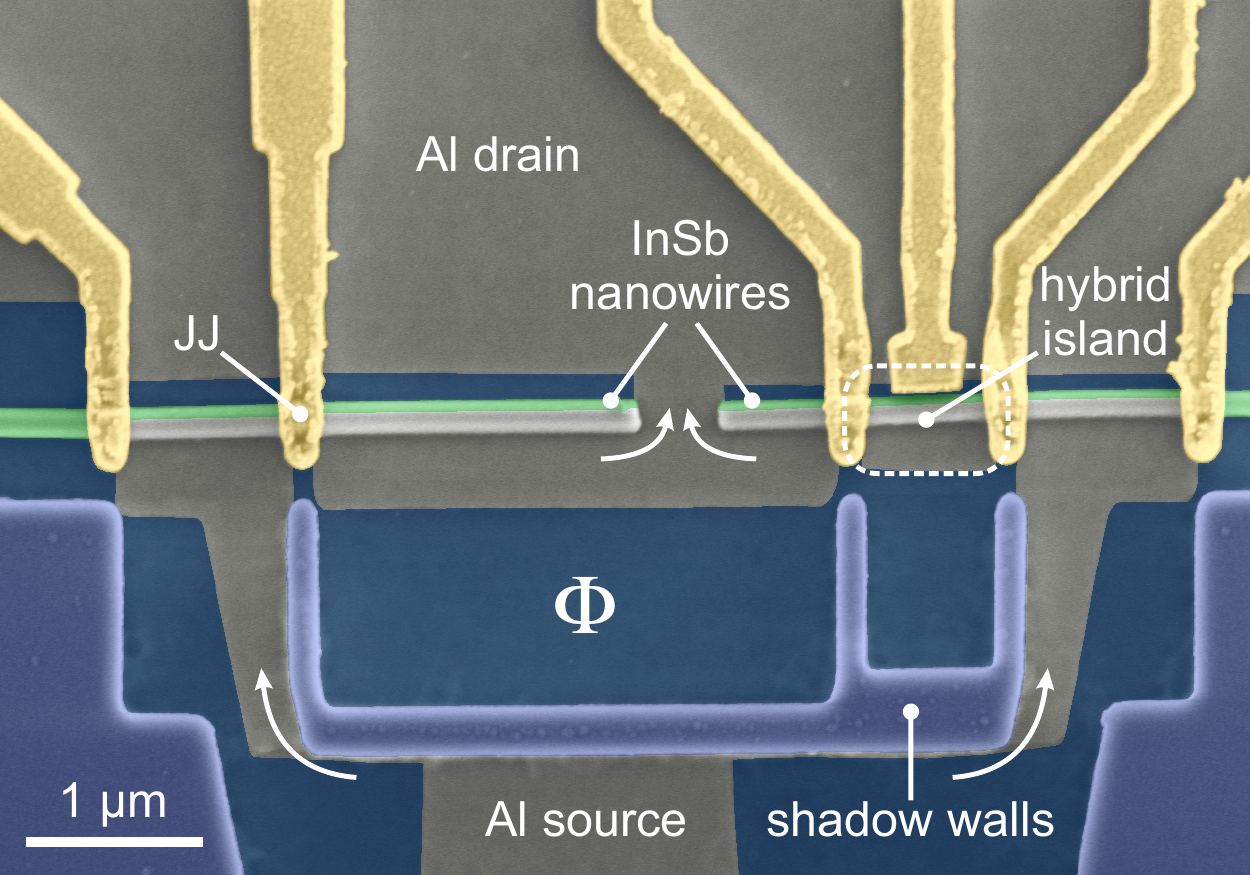}
\caption{SQUID sample formed by placing two InSb nanowires next to each other in the shadow region of the dielectric walls. Electrical current flows from source to drain via the Josephson junction (denoted as JJ) and the hybrid charge island as indicated by the white arrows. The magnetic flux threading through the SQUID loop is denoted as $\Phi$. The bottom of the SQUID loop is partly formed by the Al thin film covering the side of the central shadow wall.}
\label{fig:SQUID}
\end{figure}

\noindent In this section, we present another example of more advanced nanowire devices that can be realized using the shadow-wall technique. In the main text, we have introduced the necessary ingredients to realize the basic implementation of a topological qubit using the shadow-wall technique. In Fig.~\ref{fig:SQUID}, we show another application of the shadow-wall concept, which is intended as an experimental implementation of a theoretical proposal by Schrade and Fu~\cite{Schrade2018}. It represents a superconducting quantum interference device (SQUID) formed by two InSb nanowires (green) placed deterministically  in close vicinity of shadow walls (blue). Previous realizations of nanowire SQUIDs relied on electron-beam lithography and standard lift-off technique~\cite{Szombati2016}. Here, top gates (yellow) are fabricated to form a single Josephson junction (JJ) on the left side of the device and a superconducting island is defined by two tunnel gates and one plunger gate on the right side of the device. Source and drain electrodes are created by bonding directly to the Al film (grey) at the bottom and at the top of the SQUID loop, respectively. By utilizing shadow-wall substrates with bottom gates, this SQUID sample can be realized without any post-interface fabrication steps.

\end{bibunit}


\begin{thebibliography}{50}

\bibitem{Oreg2010}
Y.~Oreg, G.~Refael, and F.~von Oppen.
\newblock Helical liquids and \mbox{M}ajorana bound states in quantum wires.
\newblock {\em Phys. Rev. Lett.}, 105:177002, 2010.

\bibitem{Lutchyn2010}
R.~M. Lutchyn, J.~D. Sau, and S.~Das~Sarma.
\newblock Majorana fermions and a topological phase transition in
  semiconductor-superconductor heterostructures.
\newblock {\em Phys. Rev. Lett.}, 105:077001, 2010.

\bibitem{Lutchyn2018}
R.~M. Lutchyn, E.~P. A.~M. Bakkers, L.~P. Kouwenhoven, P.~Krogstrup, C.~M.
  Marcus, and Y.~Oreg.
\newblock Majorana zero modes in superconductor–semiconductor
  heterostructures.
\newblock {\em Nat. Rev. Mater.}, 3:52--68, 2018.

\bibitem{Kitaev2001}
A.~Yu. Kitaev.
\newblock Unpaired \mbox{M}ajorana fermions in quantum wires.
\newblock {\em Phys.-Uspekhi}, 44(10S):131--136, 2001.

\bibitem{Nayak2008}
C.~Nayak, S.~H. Simon, A.~Stern, M.~Freedman, and S.~Das~Sarma.
\newblock Non-\mbox{A}belian anyons and topological quantum computation.
\newblock {\em Rev. Mod. Phys.}, 80:1083--1159, Sep 2008.

\bibitem{Plugge2017}
S.~Plugge, A.~Rasmussen, R.~Egger, and K.~Flensberg.
\newblock Majorana box qubits.
\newblock {\em New J. Phys.}, 19(1):012001, 2017.

\bibitem{Karzig2017}
T.~Karzig, C.~Knapp, R.~M. Lutchyn, P.~Bonderson, M.~B. Hastings, C.~Nayak,
  J.~Alicea, K.~Flensberg, S.~Plugge, Y.~Oreg, C.~M. Marcus, and M.~H.
  Freedman.
\newblock Scalable designs for quasiparticle-poisoning-protected topological
  quantum computation with \mbox{M}ajorana zero modes.
\newblock {\em Phys. Rev. B}, 95:235305, 2017.

\bibitem{Takei2013}
S.~Takei, B.~M. Fregoso, H.-Y. Hui, A.~M. Lobos, and S.~Das~Sarma.
\newblock Soft superconducting gap in semiconductor \mbox{M}ajorana nanowires.
\newblock {\em Phys. Rev. Lett.}, 110:186803, Apr 2013.

\bibitem{Gul2017}
\"O. G\"ul, H.~Zhang, F.~K. de~Vries, J.~van Veen, K.~Zuo, V.~Mourik,
  S.~Conesa-Boj, M.~P. Nowak, D.~J. van Woerkom, M.~Quintero-Pérez, M.~C.
  Cassidy, A.~Geresdi, S.~Koelling, D.~Car, S.~R. Plissard, E.~P. A.~M.
  Bakkers, and L.~P. Kouwenhoven.
\newblock Hard superconducting gap in \mbox{InSb} nanowires.
\newblock {\em Nano Lett.}, 17(4):2690--2696, 2017.

\bibitem{Krogstrup2015}
P.~Krogstrup, N.~L.~B. Ziino, W.~Chang, S.~M. Albrecht, M.~H. Madsen,
  E.~Johnson, J.~Nyg{\aa}rd, C.~M. Marcus, and T.~S. Jespersen.
\newblock Epitaxy of semiconductor-superconductor nanowires.
\newblock {\em Nat. Mater.}, 14:400, 2015.

\bibitem{Gazibegovic2017}
S.~Gazibegovic, D.~Car, H.~Zhang, S.~C. Balk, J.~A. Logan, M.~W.~A. de~Moor,
  M.~C. Cassidy, R.~Schmits, D.~Xu, G.~Wang, P.~Krogstrup, R.~L.~M. \mbox{Op
  het Veld}, K.~Zuo, Y.~Vos, J.~Shen, D.~Bouman, B.~Shojaei, D.~Pennachio,
  J.~S. Lee, P.~J. van Veldhoven, S.~Koelling, M.~A. Verheijen, L.~P.
  Kouwenhoven, C.~J. Palmstr{\o}m, and E.~P. A.~M. Bakkers.
\newblock Epitaxy of advanced nanowire quantum devices.
\newblock {\em Nature}, 584:434--438, 2017.

\bibitem{deMoor2018}
M.~W.~A. de~Moor, J.~D.~S. Bommer, D.~Xu, G.~W. Winkler, A.~E. Antipov,
  A.~Bargerbos, G.~Wang, N.~van Loo, R.~L. M.~Op het Veld, S.~Gazibegovic,
  D.~Car, J.~A. Logan, M.~Pendharkar, J.~S. Lee, E.~P. A.~M. Bakkers, C.~J.
  Palmstr{\o}m, R.~M. Lutchyn, L.~P. Kouwenhoven, and H.~Zhang.
\newblock Electric field tunable superconductor-semiconductor coupling in
  \mbox{M}ajorana nanowires.
\newblock {\em New J. Phys.}, 20(10):103049, Oct 2018.

\bibitem{Cao2005}
J.~Cao, Q.~Wang, and H.~Dai.
\newblock Electron transport in very clean, as-grown suspended carbon
  nanotubes.
\newblock {\em Nat. Mater.}, 4:745--749, 2005.

\bibitem{Octavio1983}
M.~Octavio, M.~Tinkham, G.~E. Blonder, and T.~M. Klapwijk.
\newblock Subharmonic energy-gap structure in superconducting constrictions.
\newblock {\em Phys. Rev. B}, 27:6739--6746, 1983.

\bibitem{Nilsson2012}
H.~A. Nilsson, P.~Samuelsson, P.~Caroff, and H.~Q. Xu.
\newblock Supercurrent and multiple \mbox{A}ndreev reflections in an
  \mbox{InSb} nanowire \mbox{J}osephson junction.
\newblock {\em Nano Lett.}, 12(1):228--233, 2012.

\bibitem{Li2016}
S.~Li, N.~Kang, D.~X. Fan, L.~B. Wang, Y.~Q. Huang, P.~Caroff, and H.~Q. Xu.
\newblock Coherent charge transport in ballistic \mbox{InSb} nanowire
  \mbox{J}osephson junctions.
\newblock {\em Sci. Rep.}, 6:24822, Apr 2016.

\bibitem{Mourik2012}
V.~Mourik, K.~Zuo, S.~M. Frolov, S.~R. Plissard, E.~P. A.~M. Bakkers, and L.~P.
  Kouwenhoven.
\newblock Signatures of \mbox{M}ajorana fermions in hybrid
  superconductor-semiconductor nanowire devices.
\newblock {\em Science}, 336(6084):1003--1007, 2012.

\bibitem{Rosdahl2018}
T.~\"O. Rosdahl, A.~Vuik, M.~Kjaergaard, and A.~R. Akhmerov.
\newblock Andreev rectifier: A nonlocal conductance signature of topological
  phase transitions.
\newblock {\em Phys. Rev. B}, 97:045421, 2018.

\bibitem{Blonder1982}
G.~E. Blonder, M.~Tinkham, and T.~M. Klapwijk.
\newblock Transition from metallic to tunneling regimes in superconducting
  microconstrictions: Excess current, charge imbalance, and supercurrent
  conversion.
\newblock {\em Phys. Rev. B}, 25:4515--4532, Apr 1982.

\bibitem{Beenakker1992}
C.~W.~J. Beenakker.
\newblock Quantum transport in semiconductor-superconductor microjunctions.
\newblock {\em Phys. Rev. B}, 46:12841--12844, Nov 1992.

\bibitem{Vijay2016}
S.~Vijay and L.~Fu.
\newblock Teleportation-based quantum information processing with
  \mbox{M}ajorana zero modes.
\newblock {\em Phys. Rev. B}, 94:235446, 2016.

\bibitem{Beri2012}
B.~B\'eri and N.~R. Cooper.
\newblock Topological \mbox{K}ondo effect with \mbox{M}ajorana fermions.
\newblock {\em Phys. Rev. Lett.}, 109:156803, Oct 2012.

\bibitem{Bjergfelt2019}
M.~Bjergfelt, D.~J. Carrad, T.~Kanne, M.~Aagesen, E.~M. Fiordaliso, E.~Johnson,
  B.~Shojaei, C.~J. Palmstr{\o}m, P.~Krogstrup, T.~S. Jespersen, and
  J.~Nyg{\aa}rd.
\newblock Superconducting vanadium/indium-arsenide hybrid nanowires.
\newblock {\em Nanotechnology}, 30(29):294005, May 2019.

\bibitem{Rieger2016}
T.~Rieger, D.~Rosenbach, D.~Vakulov, S.~Heedt, {\mbox{Th}}.~Sch\"apers,
  D.~Gr\"utzmacher, and M.~I. Lepsa.
\newblock Crystal phase transformation in self-assembled \mbox{InAs} nanowire
  junctions on patterned \mbox{Si} substrates.
\newblock {\em Nano Lett.}, 16(3):1933--1941, 2016.

\bibitem{Carrad2020}
D.~J. Carrad, M.~Bjergfelt, T.~Kanne, M.~Aagesen, F.~Krizek, E.~M. Fiordaliso,
  E.~Johnson, J.~Nyg{\r{a}}rd, and T.~Sand~Jespersen.
\newblock Shadow epitaxy for in situ growth of generic
  semiconductor/superconductor hybrids.
\newblock {\em Adv. Mater.}, 32:1908411, 2020.

\bibitem{Boscherini1987}
F.~Boscherini, Y.~Shapira, C.~Capasso, C.~Aldao, M.~del Giudice, and J.~H.
  Weaver.
\newblock Exchange reaction, clustering, and surface segregation at the
  \mbox{Al}/\mbox{InSb}(110) interface.
\newblock {\em Phys. Rev. B}, 35:9580--9585, Jun 1987.

\bibitem{Thomas2019}
C.~Thomas, R.~E. Diaz, J.~H. Dycus, M.~E. Salmon, R.~E. Daniel, T.~Wang, G.~C.
  Gardner, and M.~J. Manfra.
\newblock Toward durable \mbox{Al}-\mbox{InSb} hybrid heterostructures via
  epitaxy of \mbox{2ML} interfacial \mbox{InAs} screening layers.
\newblock {\em Phys. Rev. Mater.}, 3:124202, Dec 2019.

\bibitem{Badawy2019}
G.~Badawy, S.~Gazibegovic, F.~Borsoi, S.~Heedt, C.-A. Wang, S.~Koelling, M.~A.
  Verheijen, L.~P. Kouwenhoven, and E.~P. A.~M. Bakkers.
\newblock High mobility stemless \mbox{InSb} nanowires.
\newblock {\em Nano Lett.}, 19(6):3575--3582, 2019.

\bibitem{Ambegaokar1963}
V.~Ambegaokar and A.~Baratoff.
\newblock Tunneling between superconductors.
\newblock {\em Phys. Rev. Lett.}, 10:486--489, Jun 1963.

\bibitem{Doh2005}
Y.-J. Doh, J.~A. van Dam, A.~L. Roest, E.~P. A.~M. Bakkers, L.~P. Kouwenhoven,
  and S.~De~Franceschi.
\newblock Tunable supercurrent through semiconductor nanowires.
\newblock {\em Science}, 309(5732):272--275, 2005.

\bibitem{Fulton1974}
T.~A. Fulton and L.~N. Dunkleberger.
\newblock Lifetime of the zero-voltage state in \mbox{J}osephson tunnel
  junctions.
\newblock {\em Phys. Rev. B}, 9:4760--4768, Jun 1974.

\bibitem{Tinkham1996}
M.~Tinkham.
\newblock {\em Introduction to superconductivity}.
\newblock Dover Publications, 1996.

\bibitem{Scheer1997}
E.~Scheer, P.~Joyez, D.~Esteve, C.~Urbina, and M.~H. Devoret.
\newblock Conduction channel transmissions of atomic-size aluminum contacts.
\newblock {\em Phys. Rev. Lett.}, 78:3535--3538, 1997.

\bibitem{Nijholt2016}
B.~Nijholt and A.~R. Akhmerov.
\newblock Orbital effect of magnetic field on the \mbox{M}ajorana phase
  diagram.
\newblock {\em Phys. Rev. B}, 93:235434, Jun 2016.

\bibitem{Antipov2018}
A.~E. Antipov, A.~Bargerbos, G.~W. Winkler, B.~Bauer, E.~Rossi, and R.~M.
  Lutchyn.
\newblock Effects of gate-induced electric fields on semiconductor
  \mbox{M}ajorana nanowires.
\newblock {\em Phys. Rev. X}, 8:031041, Aug 2018.

\bibitem{Abay2012}
S.~Abay, H.~Nilsson, F.~Wu, H.~Q. Xu, C.~M. Wilson, and P.~Delsing.
\newblock High critical-current superconductor-\mbox{InAs}
  nanowire-superconductor junctions.
\newblock {\em Nano Lett.}, 12(11):5622--5625, 2012.

\bibitem{Law2009}
K.~T. Law, P.~A. Lee, and T.~K. Ng.
\newblock Majorana fermion induced resonant \mbox{A}ndreev reflection.
\newblock {\em Phys. Rev. Lett.}, 103:237001, Dec 2009.

\bibitem{Zhang2017}
H.~Zhang, {\"O}.~G\"ul, S.~Conesa-Boj, M.~P. Nowak, M.~Wimmer, K.~Zuo,
  V.~Mourik, F.~K. de~Vries, J.~van Veen, M.~W.~A. de~Moor, J.~D.~S. Bommer,
  D.~J. van Woerkom, D.~Car, S.~R. Plissard, E.~P. A.~M. Bakkers,
  M.~Quintero-Pérez, M.~C. Cassidy, S.~Koelling, S.~Goswami, K.~Watanabe,
  T.~Taniguchi, and L.~P. Kouwenhoven.
\newblock Ballistic superconductivity in semiconductor nanowires.
\newblock {\em Nat. Commun.}, 8:16025, 2017.

\bibitem{Heedt2016}
S.~Heedt, A.~Manolescu, G.~A. Nemnes, W.~Prost, J.~Schubert, D.~Gr\"utzmacher,
  and {\mbox{Th}}.~Sch\"apers.
\newblock Adiabatic edge channel transport in a nanowire quantum point contact
  register.
\newblock {\em Nano Lett.}, 16(7):4569--4575, 2016.

\bibitem{Lai2019}
Y.-H. Lai, J.~D. Sau, and S.~Das~Sarma.
\newblock Presence versus absence of end-to-end nonlocal conductance
  correlations in \mbox{M}ajorana nanowires: Majorana bound states versus
  \mbox{A}ndreev bound states.
\newblock {\em Phys. Rev. B}, 100:045302, Jul 2019.

\bibitem{Nichele2017}
F.~Nichele, A.~C.~C. Drachmann, A.~M. Whiticar, E.~C.~T. O'Farrell, H.~J.
  Suominen, A.~Fornieri, T.~Wang, G.~C. Gardner, C.~Thomas, A.~T. Hatke,
  P.~Krogstrup, M.~J. Manfra, K.~Flensberg, and C.~M. Marcus.
\newblock Scaling of \mbox{M}ajorana zero-bias conductance peaks.
\newblock {\em Phys. Rev. Lett.}, 119:136803, 2017.

\bibitem{Gul2018}
\"O. G\"ul, H.~Zhang, J.~D.~S. Bommer, M.~W.~A. de~Moor, D.~Car, S.~R.
  Plissard, A.~Bakkers, E. P. A. M.~Geresdi, K.~Watanabe, T.~Taniguchi, and
  L.~P. Kouwenhoven.
\newblock Ballistic \mbox{M}ajorana nanowire devices.
\newblock {\em Nat. Nanotechnol.}, 13(3):192--197, 2018.

\bibitem{Grivnin2019}
A.~Grivnin, E.~Bor, M.~Heiblum, Y.~Oreg, and H.~Shtrikman.
\newblock Concomitant opening of a bulk-gap with an emerging possible
  \mbox{M}ajorana zero mode.
\newblock {\em Nat. Commun.}, 10:1940, 2019.

\bibitem{Chen2017}
J.~Chen, P.~Yu, J.~Stenger, M.~Hocevar, D.~Car, S.~R. Plissard, E.~P. A.~M.
  Bakkers, T.~D. Stanescu, and S.~M. Frolov.
\newblock Experimental phase diagram of zero-bias conductance peaks in
  superconductor/semiconductor nanowire devices.
\newblock {\em Sci. Adv.}, 3(9), 2017.

\bibitem{Pan2020b}
H.~Pan and S.~Das~Sarma.
\newblock Physical mechanisms for zero-bias conductance peaks in
  \mbox{M}ajorana nanowires.
\newblock {\em Phys. Rev. Research}, 2:013377, Mar 2020.

\bibitem{Anselmetti2019}
G.~L.~R. Anselmetti, E.~A. Martinez, G.~C. M\'enard, D.~Puglia, F.~K.
  Malinowski, J.~S. Lee, S.~Choi, M.~Pendharkar, C.~J. Palmstr\o{}m, C.~M.
  Marcus, L.~Casparis, and A.~P. Higginbotham.
\newblock End-to-end correlated subgap states in hybrid nanowires.
\newblock {\em Phys. Rev. B}, 100:205412, Nov 2019.

\bibitem{Yu2020}
P.~Yu, J.~Chen, M.~Gomanko, G.~Badawy, E.~P. A.~M. Bakkers, K.~Zuo, V.~Mourik,
  and S.~M. Frolov.
\newblock Non-\mbox{M}ajorana states yield nearly quantized conductance in
  superconductor-semiconductor nanowire devices.
\newblock {\em arXiv e-prints}, arXiv:2004.08583, 2020.

\bibitem{Vuik2019}
A.~Vuik, B.~Nijholt, A.~R. Akhmerov, and M.~Wimmer.
\newblock {Reproducing topological properties with quasi-\mbox{M}ajorana
  states}.
\newblock {\em SciPost Phys.}, 7:61, 2019.

\bibitem{Pan2020}
H.~Pan, W.~S. Cole, J.~D. Sau, and S.~Das~Sarma.
\newblock Generic quantized zero-bias conductance peaks in
  superconductor-semiconductor hybrid structures.
\newblock {\em Phys. Rev. B}, 101:024506, Jan 2020.

\bibitem{deJong2019}
D.~de~Jong, J.~van Veen, L.~Binci, A.~Singh, P.~Krogstrup, L.~P. Kouwenhoven,
  W.~Pfaff, and J.~D. Watson.
\newblock Rapid detection of coherent tunneling in an $\mathrm{In}\mathrm{As}$
  nanowire quantum dot through dispersive gate sensing.
\newblock {\em Phys. Rev. Applied}, 11:044061, Apr 2019.

\bibitem{Dynes1978}
R.~C. Dynes, V.~Narayanamurti, and J.~P. Garno.
\newblock Direct measurement of quasiparticle-lifetime broadening in a
  strong-coupled superconductor.
\newblock {\em Phys. Rev. Lett.}, 41:1509--1512, Nov 1978.

\bibitem{Liu2017}
C.-X. Liu, F.~Setiawan, J.~D. Sau, and S.~Das~Sarma.
\newblock Phenomenology of the soft gap, zero-bias peak, and zero-mode
  splitting in ideal \mbox{M}ajorana nanowires.
\newblock {\em Phys. Rev. B}, 96:054520, Aug 2017.

\end{thebibliography}

\begin{thebibliography}{10}

\bibitem{Flohr2011}
K.~Fl\"ohr, M.~Liebmann, K.~Sladek, H.~Y. G\"unel, R.~Frielinghaus, F.~Haas,
  C.~Meyer, H.~Hardtdegen, {\mbox{Th}}.~Sch\"apers, D.~Gr\"utzmacher, and
  M.~Morgenstern.
\newblock Manipulating \mbox{InAs} nanowires with submicrometer precision.
\newblock {\em Rev. Sci. Instrum.}, 82(11):113705, 2011.

\bibitem{Webb2015}
J.~L. Webb, J.~Knutsson, M.~Hjort, S.~Gorji~Ghalamestani, K.~A. Dick, R.~Timm,
  and A.~Mikkelsen.
\newblock Electrical and surface properties of \mbox{InAs}/\mbox{InSb}
  nanowires cleaned by atomic hydrogen.
\newblock {\em Nano Lett.}, 15(8):4865--4875, 2015.

\bibitem{Haworth2000}
L.~Haworth, J.~Lu, D.~I. Westwood, and J.~E. MacDonald.
\newblock Atomic hydrogen cleaning, nitriding and annealing \mbox{InSb} (100).
\newblock {\em Appl. Surf. Sci.}, 166(1):253 -- 258, 2000.

\bibitem{Tessler2006}
R.~Tessler, C.~Saguy, O.~Klin, S.~Greenberg, E.~Weiss, R.~Akhvlediani,
  R.~Edrei, and A.~Hoffman.
\newblock Oxide-free \mbox{InSb} (100) surfaces by molecular hydrogen cleaning.
\newblock {\em Appl. Phys. Lett.}, 88:1918--031918, Jan 2006.

\bibitem{Dong1996}
L.~Dong, R.~W. Smith, and D.~J. Srolovitz.
\newblock A two-dimensional molecular dynamica simulation of thin film growth
  by oblique deposition.
\newblock {\em J. Appl. Phys.}, 80:5682 -- 5690, 1997.

\bibitem{BARRANCO201659}
A.~Barranco, A.~Borras, A.~R. Gonzalez-Elipe, and A.~Palmero.
\newblock Perspectives on oblique angle deposition of thin films: From
  fundamentals to devices.
\newblock {\em Prog. Mater. Sci.}, 76:59 -- 153, 2016.

\bibitem{deVries2019}
F.~K. de~Vries, M.~L. Sol, S.~Gazibegovic, R.~L. M. op~het Veld, S.~C. Balk,
  D.~Car, E.~P. A.~M. Bakkers, L.~P. Kouwenhoven, and J.~Shen.
\newblock Crossed \mbox{A}ndreev reflection in \mbox{InSb} flake
  \mbox{J}osephson junctions.
\newblock {\em Phys. Rev. Research}, 1:032031, Dec 2019.

\bibitem{Gazibegovic2019}
S.~Gazibegovic, G.~Badawy, T.~L.~J. Buckers, P.~Leubner, J.~Shen, F.~K.
  de~Vries, S.~Koelling, L.~P. Kouwenhoven, M.~A. Verheijen, and E.~P. A.~M.
  Bakkers.
\newblock Bottom-up grown \mbox{2D} \mbox{InSb} nanostructures.
\newblock {\em Adv. Mater.}, 31(14):1808181, 2019.

\bibitem{averin_ac_1995}
D.~Averin and A.~Bardas.
\newblock ac \mbox{J}osephson effect in a single quantum channel.
\newblock {\em Phys. Rev. Lett.}, 75(9):1831--1834, Aug 1995.

\bibitem{bardas_electron_1997}
A.~Bardas and D.~V. Averin.
\newblock Electron transport in mesoscopic disordered
  superconductor--normal-metal--superconductor junctions.
\newblock {\em Phys. Rev. B}, 56(14):R8518--R8521, Oct 1997.

\bibitem{nowak_supercurrent_2019}
M.~P. Nowak, M.~Wimmer, and A.~R. Akhmerov.
\newblock Supercurrent carried by nonequilibrium quasiparticles in a
  multiterminal {Josephson} junction.
\newblock {\em Phys. Rev. B}, 99(7):075416, Feb 2019.

\bibitem{Nijholt2019}
B.~Nijholt, J.~Weston, J.~Hoofwijk, and A.~Akhmerov.
\newblock \textit{Adaptive}: parallel active learning of mathematical
  functions, 10.5281/zenodo.3475095, 2019.

\bibitem{Tinkham1996}
M.~Tinkham.
\newblock {\em Introduction to superconductivity}.
\newblock Dover Publications, 1996.

\bibitem{Blonder1982}
G.~E. Blonder, M.~Tinkham, and T.~M. Klapwijk.
\newblock Transition from metallic to tunneling regimes in superconducting
  microconstrictions: Excess current, charge imbalance, and supercurrent
  conversion.
\newblock {\em Phys. Rev. B}, 25:4515--4532, Apr 1982.

\bibitem{Schrade2018}
C.~Schrade and L.~Fu.
\newblock Andreev or \mbox{M}ajorana, \mbox{C}ooper finds out.
\newblock {\em arXiv e-prints arXiv:1809.06370}, Sep 2018.

\bibitem{Szombati2016}
D.~B. Szombati, S.~Nadj-Perge, D.~Car, S.~R. Plissard, E.~P. A.~M. Bakkers, and
  L.~P. Kouwenhoven.
\newblock Josephson $\varphi_0$-junction in nanowire quantum dots.
\newblock {\em Nat. Phy.}, 12:568--572, 2016.

\end{thebibliography}
\end{document}